\documentclass[11pt]{article}                 
\usepackage{graphicx}
\usepackage{authblk}
\usepackage{mathptmx}      
\usepackage{latexsym}
\usepackage[dvipsnames]{xcolor}
\usepackage[utf8]{inputenc}
\usepackage{soul}
\usepackage{graphicx}
\usepackage{amsmath}
\usepackage{mathtools}
\usepackage{breqn}

\usepackage{amsthm}
\usepackage{authblk}

\usepackage{amsfonts}
\usepackage{enumerate}
\usepackage{verbatim}
\usepackage{listings}
\usepackage{tabularx}
\usepackage{blkarray}
\usepackage{graphicx}
\usepackage{verbatim}
\usepackage{array}
\usepackage[refpages]{gloss}
\usepackage{amssymb}
\usepackage{courier}
\usepackage{float}
\usepackage{rotating}
\usepackage{color}
\usepackage{comment}
\usepackage[compatibility=false]{caption}
\usepackage{standalone}
\usepackage{tikz}
\usepackage{tcolorbox}
\newtheorem{theorem}{Theorem}
\newtheorem{proposition}{Proposition}
\usepackage[title]{appendix}
\usetikzlibrary{arrows.meta, positioning, quotes}
\usetikzlibrary{decorations.pathreplacing}
\DeclareMathOperator*{\Ex}{\mathrm{E}}
\DeclareMathOperator*{\Var}{\text{Var}}
\newcommand{\Mod}[1]{\ (\mathrm{mod}\ #1)}

  \usepackage{pgfplots}
  \pgfplotsset{compat=newest}
  \usetikzlibrary{plotmarks}
  \usetikzlibrary{arrows.meta}
  \usepgfplotslibrary{patchplots}
  \usepackage{grffile}
\usepackage{hyperref}

\usepackage{caption}
\usepackage{subcaption}
\usepackage{natbib}
\setcitestyle{authoryear,open={(},close={)}} 
\providecommand{\keywords}[1]
{
  \small	
  \textbf{\textit{Keywords---}} #1
}
\begin{document}

\title{Finding analytical approximations for discrete, stochastic, individual-based models of ecology
}
%



\author[1]{Linnéa Gyllingberg} 
\author[2]{ David J.T. Sumpter}   
    \author[3]{$\mathrm{\mathring{A}}$ke Brännström }

\affil[1]{Department of Mathematics, Uppsala University, Uppsala, Sweden}
\affil[2]{Department of Information Technology, Uppsala University, Uppsala, Sweden}
\affil[3]{Department of Mathematics and Mathematical Statistics, Umeå University, Umeå, Sweden}




\date{}

\maketitle

\begin{abstract}
Discrete time, spatially extended models play an important role in ecology, modelling population dynamics of species ranging from micro-organisms to birds. An important question is how 'bottom up', individual-based models can be approximated by ’top down’ models of dynamics. Here, we study a class of spatially explicit individual-based models with contest competition: where species compete for space in local cells and then disperse to nearby cells. We start by describing simulations of the model, which exhibit large-scale discrete oscillations and characterise these oscillations by measuring spatial correlations. We then develop two new approximate descriptions of the resulting spatial population dynamics. The first is based on local interactions of the individuals and allows us to give a difference equation approximation of the system over small dispersal distances. The second approximates the long-range interactions of the individual-based model. These approximations capture demographic stochasticity from the individual-based model and show that dispersal stabilizes population dynamics. We calculate extinction probability for the individual-based model and show convergence between the local approximation and the non-spatial global approximation of the individual-based model as dispersal distance and population size simultaneously tend to infinity. Our results provide new approximate analytical descriptions of a complex bottom-up model and deepen understanding of spatial population dynamics.
\end{abstract}

\section{Introduction}
\label{intro}
Spatial structure plays an important role in ecology  \citep{deangelis2017spatially} and numerous models have been developed and explored to advance understanding of spatial population dynamics  \citep{pacala1985neighborhood,   nowak1992evolutionary, ermentrout1993cellular, boerlijst1993evolutionary, pacala1994limiting, hanski1994metapopulation, wu1997patch, iwasa1998allelopathy,  durrett1999stochastic, keeling2000reinterpreting, deangelis2018individual}. The importance of using spatially explicit and discrete models was highlighted by \cite{durrett1994importance} in a seminal paper that elucidated differences in outcomes between plausible modelling assumptions: discrete individuals and explicit space were shown to be critically important for capturing salient features of ecological systems. The importance of spatial structure has been manifested in many other ways as well. One well studied aspect is the effect of dispersal on population dynamics. For example,  \cite{gyllenberg1993does} and \cite{hastings1993complex} show through simple models that dispersal between two metapopulations stabilizes population dynamics.

Among the different ways to model spatial population dynamics, spatially explicit indi\-vidual-based models stand out as the arguably most realistic approach. These type of models are often referred to as ‘bottom-up’, since they explicitly represent individual actions and interactions. This contrasts with the more classical ’top-down’ approach, where the population dynamics is modelled through partial or ordinary differential or difference equations. There are many benefits of individual-based models. One is the ease by which spatial structure can be represented. Another advantage, from a biological perspective, is that observations of individual behaviour and interactions can be included. However, individual-based models are rarely analytically tractable and often computationally demanding. Therefore, it is important to study the relationship between individual-based models and analytical, top-down models of populations dynamics. 

Techniques that have been developed for this purpose include reaction-diffusion approximations \citep{oelschlager1989derivation, stevens1997aggregation, isaacson2021reaction, isaacson2022mean}, the method of moments \citep{bolker1997using, dieckmann2000geometry, murrell2004moment, ovaskainen2014general, surendran2018spatial,bordj2022moment} and pair approximations \citep{sato2000pair, van2000pair}. All these methods have advantages and drawbacks. For example, pair approximation methods work in epidemiological settings, but perform poorly when reproduction and dispersal take place on different spatial scales. Also, both pair approximations and the general method of moments fail to capture large scale patterns that can emerge through local interactions.  Reaction-diffusion approximations are useful for deriving limit dynamics of individual-based models with continuous time and space \citep{oelschlager1989derivation}, and there are rigorous results for many of these models.  Recently, \cite{patterson2020probabilistic} developed a new method demonstrating convergence to a mean-field limit for an ecologically motivated discrete-space, continuous-time stochastic individual-based model. These methods are essentially techniques for approximating interacting particle systems and spatial point processes, in which a continuous time approximation is possible.

Here, we are interested in a different class of individual-based models: what \cite{berec2002techniques} refer to as Individual-based models or D-space,  D-time IBMs.  These models are motivated by, for example, empirical studies of many insects \citep{wilson1971insect, costa2006other},  birds \citep{brown1996coloniality}, microorganisms \citep{hibbing2010bacterial, hall2004bacterial}, and other species the population that are distributed across resource sites and have a lifecycle involving discrete phases of: (1) competition for resource sites, (2) reproduction by one or more of these individuals and (3) dispersal of surviving offspring to new potential resource sites.  In this case, competition and reproduction happen on the same time-scale and, as a result, classical spatial methods like pair-approximations and reaction-diffusion approximations do not apply to models of these systems. 

The simplest way to approximate spatial population dynamics in discrete-time discrete-space models is to neglect spatial structure altogether, which is sometimes referred to as a mean field approximation \citep{morozov2012spatially}. If the population is assumed to be well mixed, the approximation describes how the mean population changes over time through a set of differential or difference equations. A large class of discrete-time individual-based models, in which individuals from one or more species compete for resources in discrete sites but mix uniformly in space, have been studied exhaustively \citep{johansson2003local, brannstrom2005role, anazawa2014individual, royama2012analytical, anazawa2009bottom}. However, these studies do not extend to the spatial case when offspring are more likely to disperse to nearby sites and, in this setting, the task of deriving suitable approximations through first principles becomes more difficult. The discrete nature of time and space in these site-based models together with potentially long dispersal distances limit the applicability of existing spatial approximation techniques.  

As a first effort to describe the spatial population dynamics of discrete-time, discrete-space, site-based models, Bränn\-ström and Sumpter (2005a) introduced a new approximation method called the coupled map lattice approximation. The idea is to divide the lattice which individuals inhabit into nine or more sublattices and decouple the individual interactions within each sublattice from the dispersal between each sublattice. An advantage of their approach is the ability to disentangle the stochastic and deterministic parts of individual interactions and dispersal, but a downside is that it leads to a large system of difference equations which is not necessarily easier to analyse than the individual-based model itself. \cite{brannstrom2005coupled} attempted to reduce the model to single analytical attractable equation, but in doing so they needed to make several simplifying assumptions that are not necessarily justified for the full system.  We complement the previous work of \cite{brannstrom2005coupled} by studying and characterizing the spatial dynamics that can occur in discrete-time, discrete-space site-based models.  


In this paper we start to ask the question about what methods might be a way in to capture the dynamics of D-space,  D-time IBMs, by studying one specific example. The paper is organized as follows. We start in Section \ref{sec:IBM} by describing a simple discrete time, discrete space biologically motivated individual-based model with local dispersal and reproduction. The qualitative behaviour of the model is studied through bifurcation plots and the spatial statistics of the model is analysed, to gain insight into the spatial patterns of the individual-based model.  From this, we continue in Section \ref{sec:approx} where we derive three different approximations: first we derive a global approximation (also known as a mean field approximation), reducing the individual-based model to a one-dimensional discrete dynamical system.  Based on the insight of the local cluster sizes and the spatial distribution of the individual-based model in Section \ref{sec:spatstat}, we derive a 'local correlation approximation' where the dispersal length is taken into account. The local correlation approximation is also given by a one-dimensional discrete dynamical system, but with the dispersal length included as a parameter in the model. We then derive a long-range dispersal approximation, in the form of a two-dimensional coupled map lattice.  We continue to analyse these three models in Section \ref{sec:analysis}, where we first present bifurcation plots for the three approximations, to compare the qualitative behaviour of the approximations with the individual-based model. We then carry out a stability analysis of the global approximation and local correlation approximation and then calculate a 'general extinction probability' of the individual-based model to explain parts of the discrepancies in stability between the global approximation and the individual-based model.  Finally, we show that the local correlation approximation converges to the global approximation in the limit of large dispersal distances.  

\section{The individual-based model}

\label{sec:IBM}
\subsection{Model description}

We consider a model where individuals live on a $D \times D$ lattice with cyclic boundary conditions. There a thus a total of $n = D^2$ resource sites. We divide the life cycle of the individuals into a competition phase, a reproduction phase and a dispersal phase. Let $X_{ij}^t$ be a random variable denoting the number of individuals at site $(i,j)$ at time $t$. At the \textbf{competition phase} there may be more than one individual at a particular site. In order to capture competition for resources we define a survival function $\phi : \mathbb{N} \rightarrow \{0,1\}$ that acts on $X_{ij}^t$ giving a new site population $\phi(X_{ij}^{t})$. The survival function maps the number of individuals at a given site to 0 if the site is over exploited or 1 if there is a survivor. This captures over-crowding: if there are too many individuals at a resource site, they cannot all reproduce. 

After competition the next phase is \textbf{reproduction}. The surviving individual at a site (if there is one) produces $R_{ij}^t$ offspring, where $R_{ij}^t$ is a random variable with  expectation $\Ex[R_{ij}^t]=r$ for all $i$, $j$ and $t$.  
We consider discrete generations, so after the reproduction phase the parents die, resulting in $R_{ij}^t \cdot \phi(X_{ij}^{t})$ individuals at each site $(i,j)$.  These offspring then \textbf{disperse} to a random site in a $(2s+1) \times (2s+1)$ Moore neighbourhood,  with uniform probability, resulting in a new site count $X_{ij}^{t+1}$.  The Moore neighbourhood of site $(i,j)$ with range $s$, $N(i,j,s)$, is given by the set of sites $(i', j')$ for which $|i-i'|\leq s$ and $|j-j'| \leq s$. Since the lattice has cyclic boundary conditions, individuals leaving the lattice in one end, end up on the opposite side, i.e. two sites $(i,j)$ and $(i',j')$ are identical if  $i \Mod D = i' \Mod D$ and $j \Mod D = j' \Mod D$. The life cycle is complete and the process starts over again. 

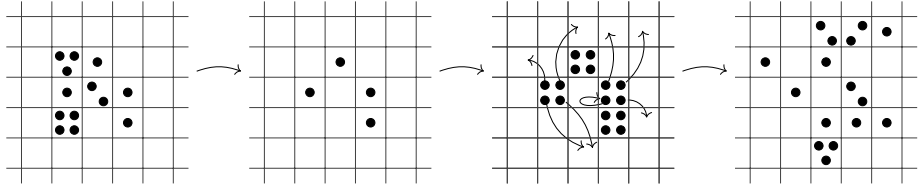
\begin{figure}[H]
  \definecolor{mycolor}{RGB}{238,233,233}
\definecolor{mygreen}{RGB}{0,0,0}
\definecolor{myred}{RGB}{0,0,0}
\begin{tikzpicture}

\node (1) at (0.5, 3.75) {};
\node (2) at (4.5, 3.75) {};
\node (3) at (8.5, 3.75) {};
 \draw[step=0.5cm,color=darkgray] (.75,.75) grid (3.75,3.75);


\filldraw[fill=mygreen, draw=black, very thin] (2.25,2.75) circle (2pt);

\filldraw[fill=myred, draw=black, very thin] (2.16,2.35) circle (2pt);
\filldraw[fill=myred, draw=black, very thin] (2.35,2.1) circle (2pt);

\filldraw[fill=myred, draw=black, very thin] (2.75,1.75) circle (2pt);

\filldraw[fill=mygreen, draw=black, very thin] (2.75,2.25) circle (2pt);

\filldraw[fill=mygreen, draw=black, very thin] (1.63,2.85) circle (2pt);
\filldraw[fill=mygreen, draw=black, very thin] (1.87,2.85) circle (2pt);
\filldraw[fill=myred, draw=black, very thin] (1.75,2.6) circle (2pt);

\filldraw[fill=myred, draw=black, very thin] (1.75,2.25) circle (2pt);

\filldraw[fill=myred, draw=black, very thin] (1.63,1.87) circle (2pt);
\filldraw[fill=myred, draw=black, very thin] (1.87,1.87) circle (2pt);
\filldraw[fill=mygreen, draw=black, very thin] (1.63,1.63) circle (2pt);
\filldraw[fill=mygreen, draw=black, very thin] (1.87,1.63) circle (2pt);

\draw[step=0.5cm,color=darkgray] (4.75,.75) grid (7.75,3.75);

\filldraw[fill=mygreen, draw=black, very thin](6.25,2.75) circle (2pt);

\filldraw[fill=myred, draw=black, very thin] (6.75,1.75) circle (2pt);

\filldraw[fill=mygreen, draw=black, very thin] (6.75,2.25) circle (2pt);

\filldraw[fill=myred, draw=black, very thin] (5.75,2.25) circle (2pt);

\draw[step=0.5cm,color=darkgray] (8.75,.75) grid (11.75,3.75);

\filldraw[fill=mygreen, draw=black, very thin](10.12,2.87) circle (2pt);
\filldraw[fill=mygreen, draw=black, very thin](10.12,2.63) circle (2pt);
\filldraw[fill=mygreen, draw=black, very thin](10.37,2.87) circle (2pt);
\filldraw[fill=mygreen, draw=black, very thin](10.37,2.63) circle (2pt);

\filldraw[fill=myred, draw=black, very thin] (10.62,1.87) circle (2pt);
\filldraw[fill=myred, draw=black, very thin] (10.62,1.63) circle (2pt);
\filldraw[fill=myred, draw=black, very thin] (10.87,1.87) circle (2pt);
\filldraw[fill=myred, draw=black, very thin] (10.87,1.63) circle (2pt);

\filldraw[fill=mygreen, draw=black, very thin] (10.62,2.12) circle (2pt);
\filldraw[fill=mygreen, draw=black, very thin] (10.62,2.37) circle (2pt);
\filldraw[fill=mygreen, draw=black, very thin] (10.87,2.12) circle (2pt);
\filldraw[fill=mygreen, draw=black, very thin] (10.87,2.37) circle (2pt);

\filldraw[fill=myred, draw=black, very thin] (9.62,2.12) circle (2pt);
\filldraw[fill=myred, draw=black, very thin] (9.62,2.37) circle (2pt);
\filldraw[fill=myred, draw=black, very thin] (9.87,2.12) circle (2pt);
\filldraw[fill=myred, draw=black, very thin] (9.87,2.37) circle (2pt);

\node (1) at (10.62, 2.3) {};
\node (2) at (10.62, 3.37) {};

\node (3) at (9.62, 2.2) {};
\node (4) at (10.4, 1.3) {};

\node (5) at (10.7, 2.1) {};
\node (6) at (10.9, 1.2) {};

\node (7) at (9.83, 2.2) {};
\node (8) at (10.43, 1.2) {};

\node (9) at (9.6, 2.3) {};
\node (10) at (9.2, 2.8) {};

\node (11) at (9.94, 2.3) {};
\node (12) at (10.3, 3.4) {};

\node (13) at (10.85, 2.12) {};
\node (14) at (11.3, 1.7) {};

\node (15) at (10.82, 2.3) {};
\node (16) at (11.2, 3.4) {};

\draw [->] (1) edge[bend right=20] node [left] {} (2);
\draw [->] (3) edge[bend right=30] node [left] {} (4);

\draw [->] (5) edge[loop left] node [left] {} (5);

\draw [->] (7) edge[bend left=20] node [left] {} (8);
\draw [->] (9) edge[bend right=45] node [left] {} (10);

\draw [->] (11) edge[bend left=45] node [left] {} (12);

\draw [->] (13) edge[bend left=45] node [left] {} (14);

\draw [->] (15) edge[bend right=30] node [left] {} (16);

\draw[step=0.5cm,color=darkgray] (8.75,.75) grid (11.75,3.75);

\draw[step=0.5cm,color=darkgray] (12.75,.75) grid (15.75,3.75);

\filldraw[fill=myred, draw=black, very thin](14.13,1.37) circle (2pt);
\filldraw[fill=myred, draw=black, very thin] (14.37,1.37) circle (2pt);
\filldraw[fill=mygreen, draw=black, very thin] (14.25,1.13) circle (2pt);

\filldraw[fill=myred, draw=black, very thin] (14.16, 3.35) circle (2pt);
\filldraw[fill=mygreen, draw=black, very thin] (14.35,3.1) circle (2pt);

\filldraw[fill=mygreen, draw=black, very thin] (14.66, 3.1) circle (2pt);
\filldraw[fill=mygreen, draw=black, very thin] (14.85,3.35) circle (2pt);

\filldraw[fill=myred, draw=black, very thin] (14.66, 2.35) circle (2pt);
\filldraw[fill=mygreen, draw=black, very thin] (14.85,2.1) circle (2pt);

\filldraw[fill=myred, draw=black, very thin] (13.25,2.75) circle (2pt);

\filldraw[fill=myred, draw=black, very thin] (14.25,2.75) circle (2pt);

\filldraw[fill=myred, draw=black, very thin] (14.75,1.75) circle (2pt);

\filldraw[fill=myred, draw=black, very thin] (14.25,1.75) circle (2pt);

\filldraw[fill=mygreen, draw=black, very thin] (13.75,2.25) circle (2pt);

\filldraw[fill=mygreen, draw=black, very thin] (15.25,3.25) circle (2pt);

\filldraw[fill=mygreen, draw=black, very thin] (15.25,1.75) circle (2pt);

\node (17) at (3.75, 2.55) {};
\node (18) at (4.75, 2.55) {};

\draw [->] (17) edge[bend left=20] node [left] {} (18);

\node (19) at (7.75, 2.55) {};
\node (20) at (8.75, 2.55) {};

\draw [->] (19) edge[bend left=20] node [left] {} (20);

\node (21) at (11.75, 2.55) {};
\node (22) at (12.75, 2.55) {};

\draw [->] (21) edge[bend left=20] node [left] {} (22);


\end{tikzpicture} 
  \caption{An illustration of the individual-based model. Each box represents a site, and individuals are represented by black dots. After the competition phase, only the individuals that are alone in their boxes survive. They then reproduce $r=4$ offspring that disperse in an $s=2$ range (arrows for the dispersal illustrated for just two sites). We count the population in the third state. }
  \label{fig_IBM_ill}	
\end{figure}

There are several ways to define the survival function and \cite{brannstrom2005role} examine the population dynamics for different survival functions for the non-spatial case. In our simulations of this individual-based model,  however, we will define the survival function as
\begin{equation} \label{eq:scramble_phi}
\phi(k) = \begin{cases} 1  &\mbox{if }  k = 1 \\ 
0, & \text{otherwise}, \end{cases} 
\end{equation}
which gives a classic scramble competition model, as first described by \cite{nicholson1954outline}.  In our simulations in the main article, we also assume $R_{ij}^t$ to be equal to a constant $r$ for all $i$, $j$ and $t$,  and not random variable. In Supplementary material S3 we simulate the model for $R_{ij}^t \sim \operatorname{Bin}(r/q,q)$, for different values of $q$. Figure \ref{fig_IBM_ill} illustrates the individual-based model with scramble competition for $r=4$ and $s=2$. Individuals that are alone in a site survive, whereas in sites with more than one individual, all individuals die. The surviving individuals then produce $r=4$ offspring each, that disperse to a randomly chosen site with uniform probability in a $(2s+1) \times (2s+1)$ Moore neighbourhood.

\subsection{Simulations}

\begin{figure}[H]
\subcaptionbox{}
{\includegraphics[scale=0.445]{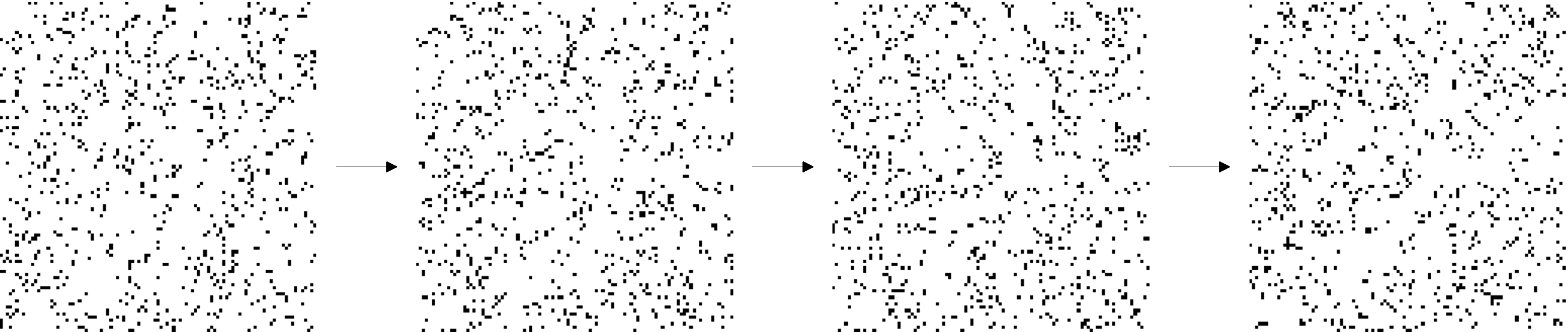} \label{snap_s1}}
\subcaptionbox{}
{\includegraphics[scale=0.445]{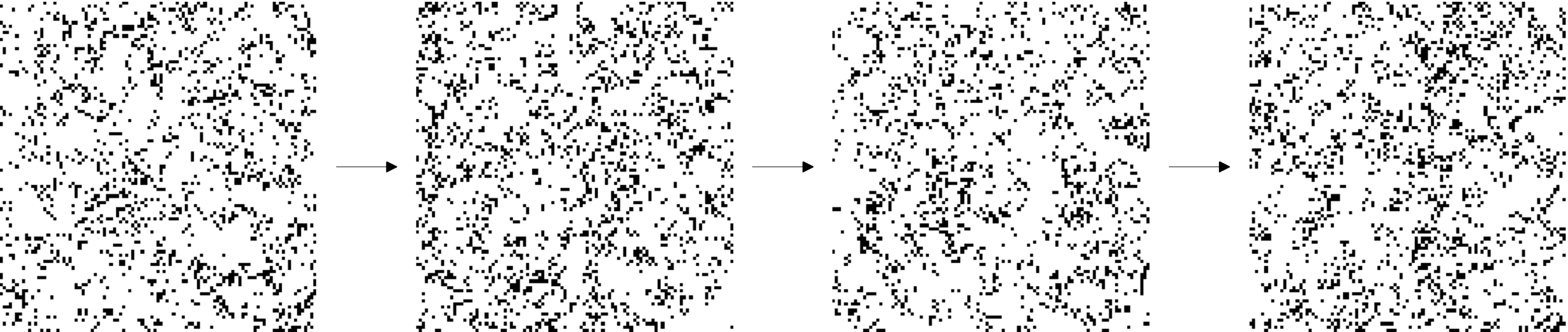} \label{snap_s2}}
\subcaptionbox{}
{\includegraphics[scale=0.445]{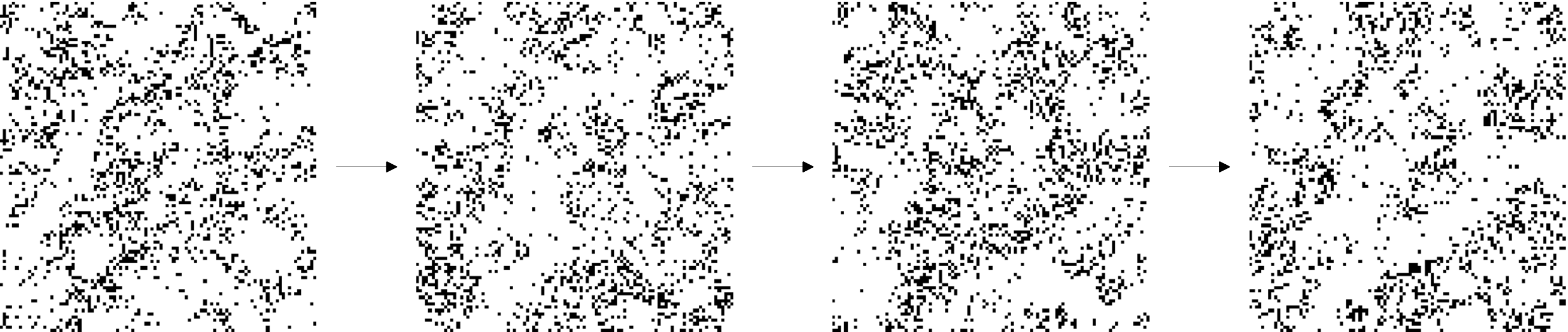} \label{snap_s3}}
\subcaptionbox{}
{\includegraphics[scale=0.445]{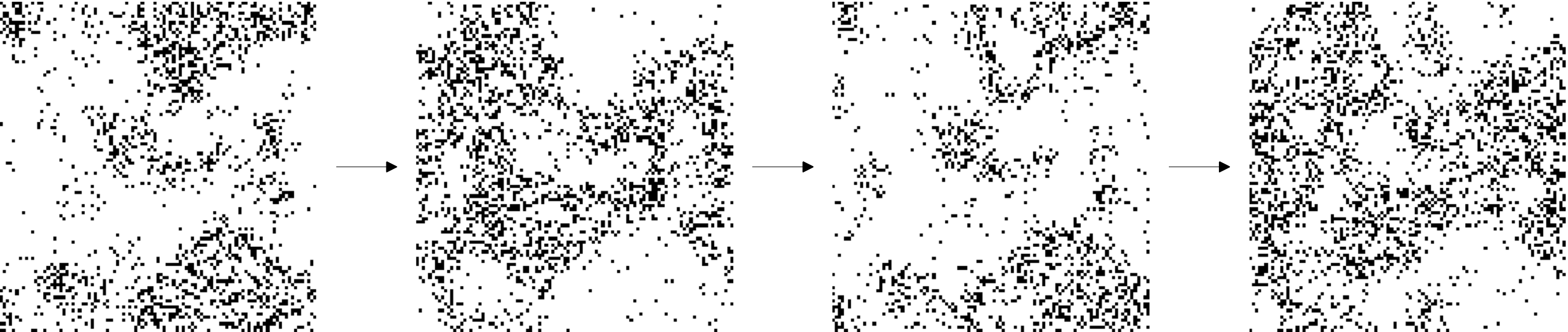} \label{snap_s5}}
\subcaptionbox{}
{\includegraphics[scale=0.445]{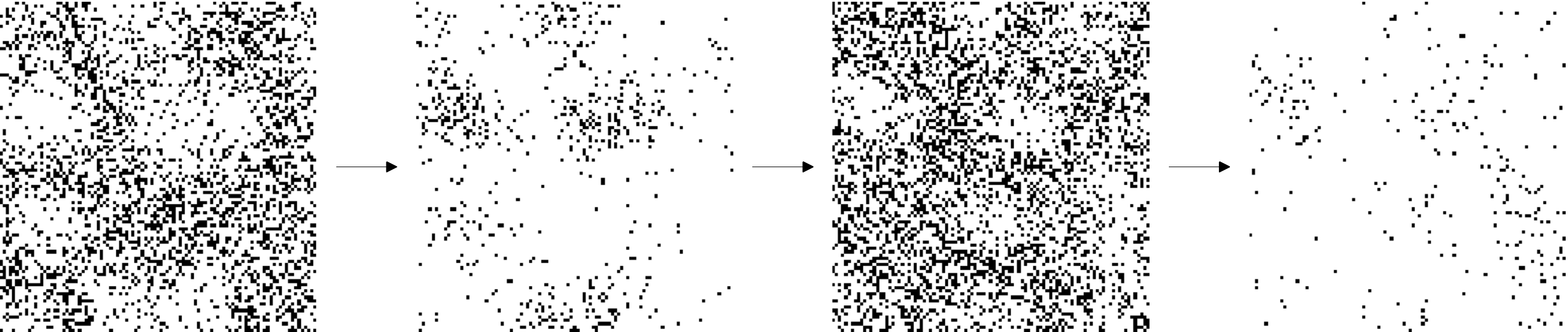} \label{snap_s10}}
\subcaptionbox{}
{\includegraphics[scale=0.445]{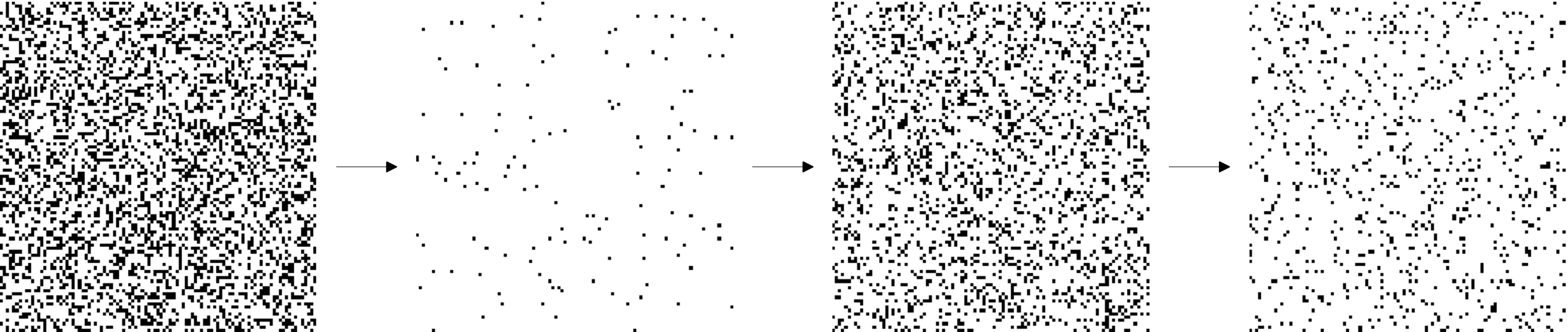} \label{snap_s50}}
\caption{Four sequential snapshots for the individual-based model on a $101 \times 101$ sized lattice when $r=20$ and (a) $s=1$ (very local dispersal), (b) $s=2$, (c) $s=3$, (d) $s=5$, (e) $s=10$ and (f) $s=50$ (global dispersal). The snapshots show the population after competition, thus there are at most 1 individual in each site.  \label{fig:snapshots}} 

\end{figure}

The model has a rich set of dynamics depending on the parameter values $r$ and $s$. Figure \ref{fig:snapshots} shows how the spatial distribution of individuals on a $101 \times 101$ lattice evolves over time for four time steps for different values of $s$, when $r=20$.
When $s=1$, there is no clear spatial clustering, and between generations the population is relatively stable. As $s$ increases to 2 and 3 (Figure \ref{fig:snapshots} (b)-(c)), we can distinguish spatial clusters. While the overall population size is still reasonably stable over time, there are local fluctuations. Patches of a size roughly $(2s+1) \times (2s+1)$, i.e. the same size as the dispersal range emerge creating a checkerboard effect. In Figure \ref{fig:snapshots} (d) and (e), where $s=5$ and $s=10$ respectively, the global population is the global population is alternatingly large and small and fluctuates in the four consecutive snapshots. The last figure, Figure \ref{fig:snapshots} (f), shows time evolution of the population for $s=50$. Dispersal is now global, i.e. individuals disperse uniformly to any site on the lattice, since $s=(D-1)/2=50$ for $D=101$. The figure displays chaotic oscillations between generations and there is no spatial clustering. 

\begin{figure}[H] \centering
\subcaptionbox*{}{\includegraphics[scale=0.48]{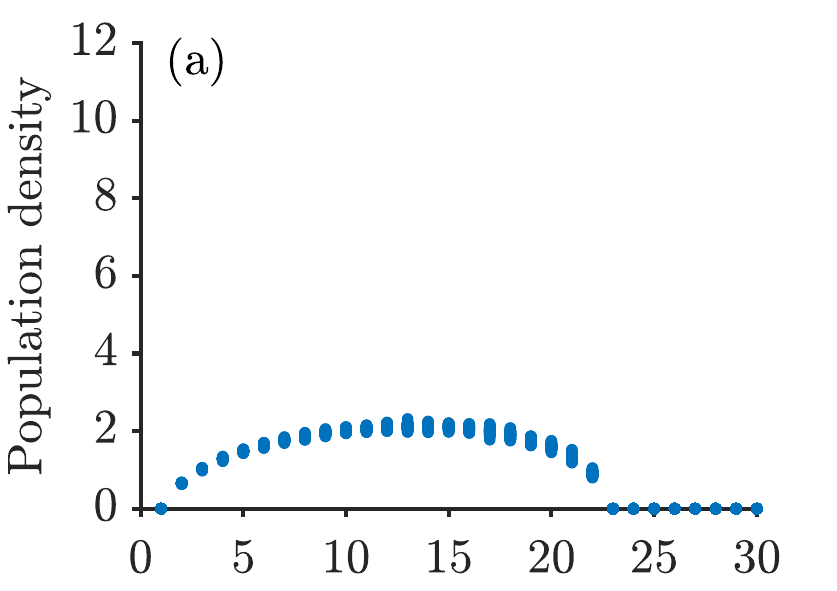}}
\subcaptionbox*{}{\includegraphics[scale=0.48]{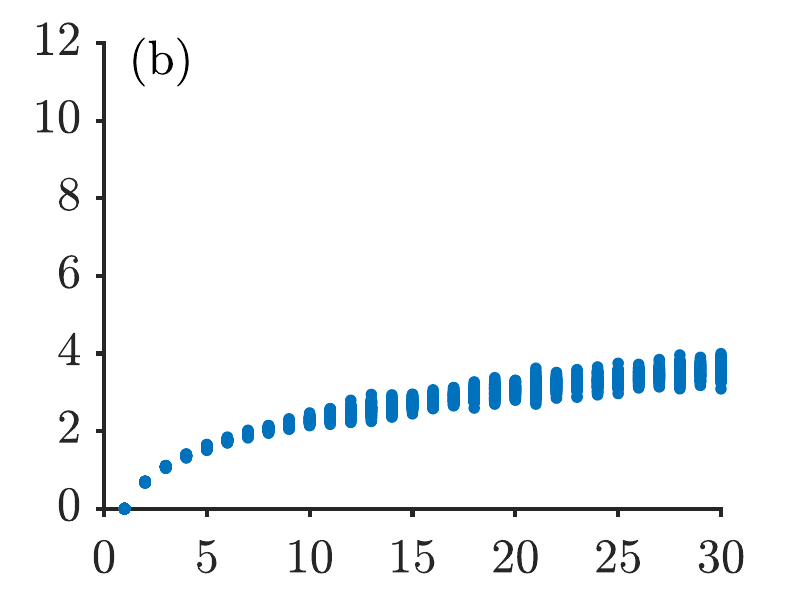}}
\subcaptionbox*{}{\includegraphics[scale=0.48]{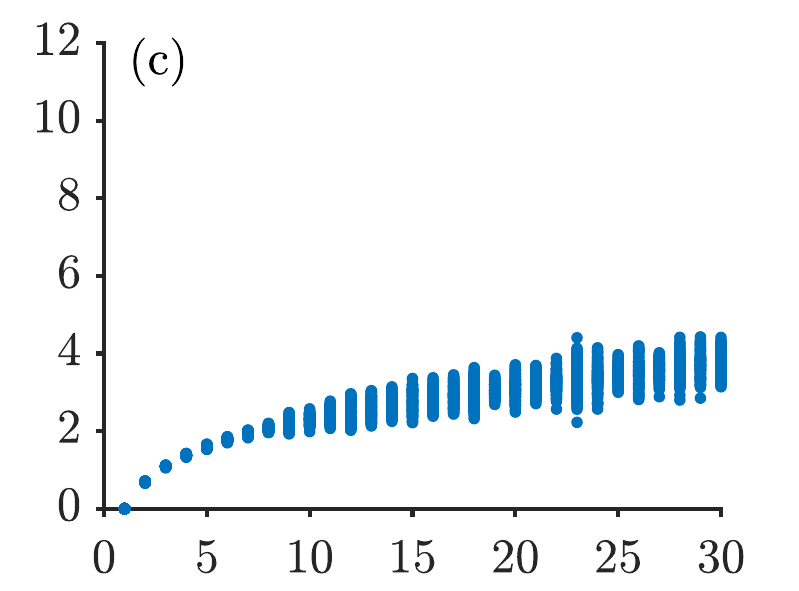}}
\subcaptionbox*{}{\includegraphics[scale=0.48]{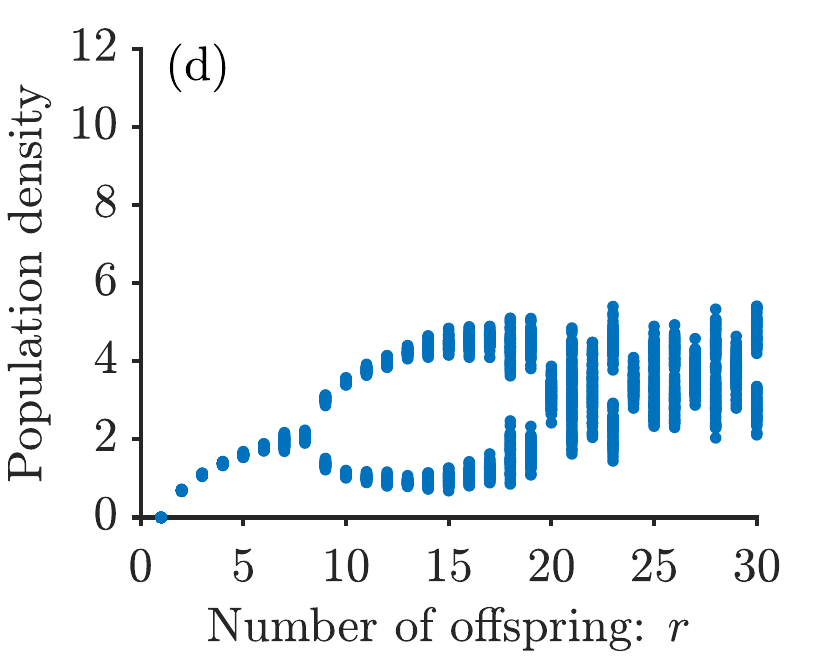}}
\subcaptionbox*{}{\includegraphics[scale=0.48]{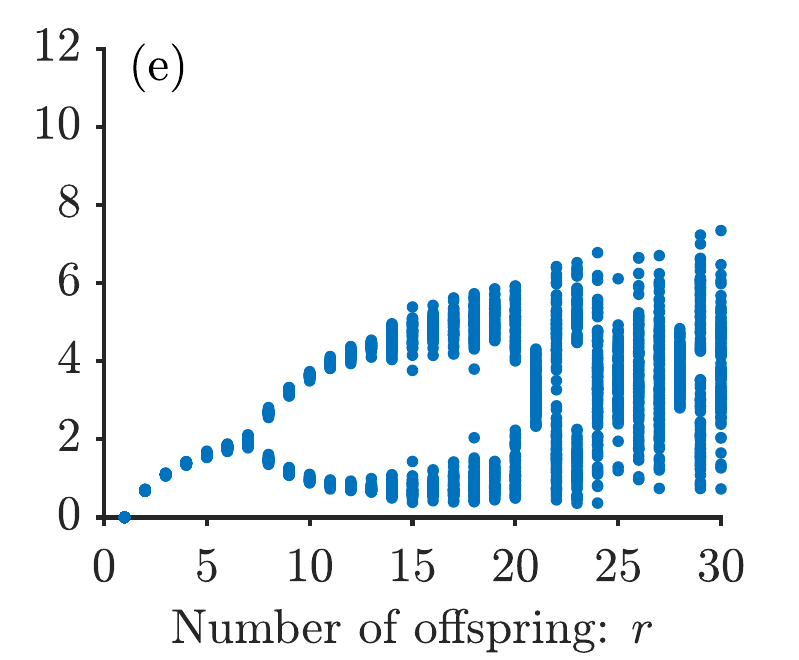}}
\subcaptionbox*{}{\includegraphics[scale=0.48]{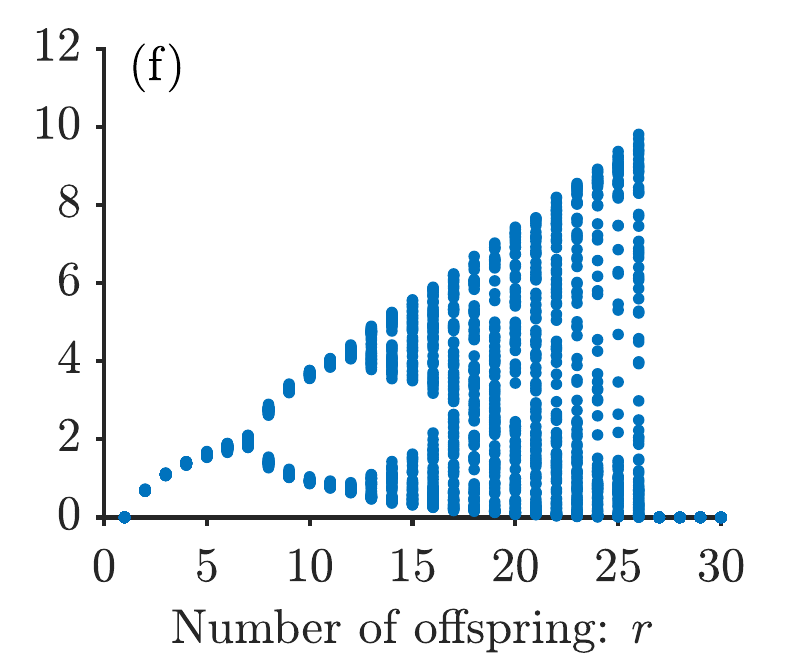}}
\caption{Bifurcation diagram for the individual-based model, with $\phi(k)$ given as equation \eqref{eq:scramble_phi}. In each subplot we vary $r$ and simulate the model for 5000 time steps and plot last 500 time steps. The population density is given by the number of individuals divided by the number of resource sites, and is counted after reproduction. We give results for (a) $s=1$ (very local dispersal) (b) $s=2$, (c) $s=3$, (d) $s=5$, (e) $s=10$ and (f) $s=50$  (global dispersal).  \label{fig:bifmodel}}
\end{figure}

Further simulations are summarised in bifurcation plots shown in Figure \ref{fig:bifmodel}. Here we varied $r$, simulated the model for 5000 time steps, and the plot shows the outcome for the last 500 time steps.  In Figure \ref{fig:bifmodel} (a), we see how very local dispersal ($s=1$) typically gives rise to a stable population. This population reaches a maximum at around $r \approx 15$, and it dies out at $r=23$. Figure \ref{fig:bifmodel} (b)-(c) shows a stable population for $s=2$ and $s=3$. Increasing $s$ to 5, there is a period doubling when $r=9$, and for $s=10$, the period doubling bifurcation occurs already at $r=8$, which we see in Figure \ref{fig:bifmodel} (d) and (e). When dispersal is global the population is stable for $r<e^2 ( \approx 7.39)$, and undergoes a period doubling bifurcation after that, leading to an oscillating population size between generations when $8<r< 14$. When $r>15$ the population oscillates chaotically.  In summary,   for all values of $s$, the population is stable for small reproductive rates, but for higher values of $r$, the population goes from stable to oscillating to chaotic by increasing $s$. 

In Supplementary material S3 we look at what happens to the dynamics if the reproductive rate is not constant,  but instead the reproductive rate at each site and time step,  $R_{ij}^t$, are i.i.d with $R_{ij}^t \sim \operatorname{Bin}(r/q,q)$, by producing bifurcation plots for different values of $s$ and $q$. We see that the overall behaviour is not affected by different values of $q$.

\begin{figure}[H]
\subcaptionbox{}{\includegraphics[scale=0.445]{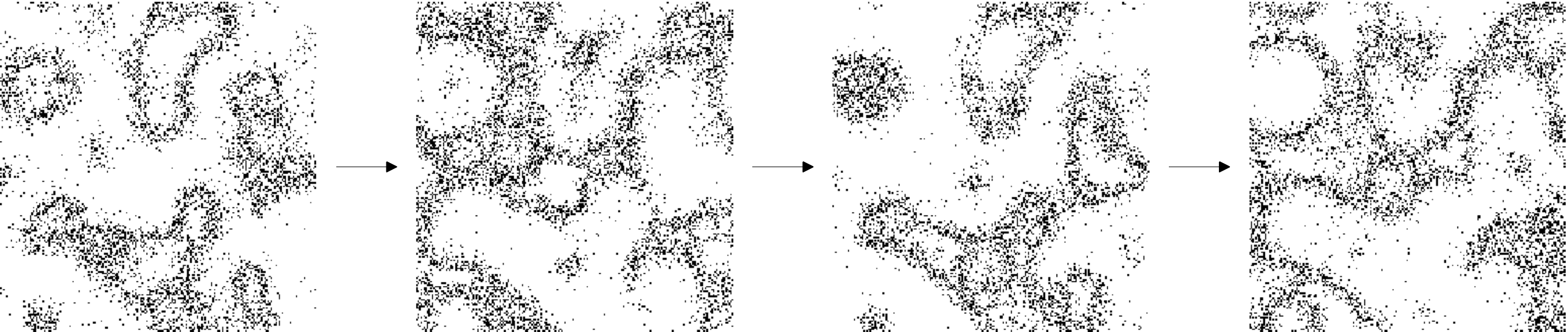}}
\subcaptionbox{}{\includegraphics[scale=0.445]{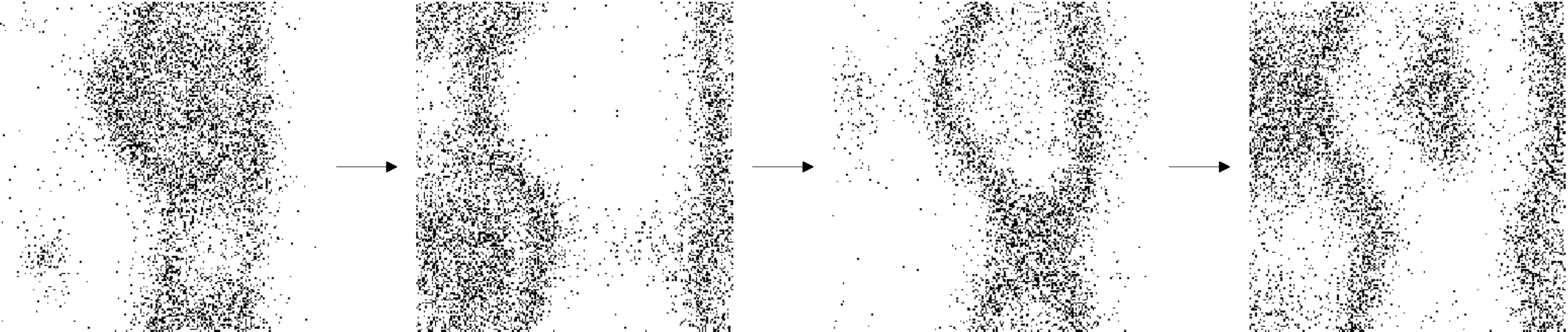}}
\subcaptionbox{}{\includegraphics[scale=0.445]{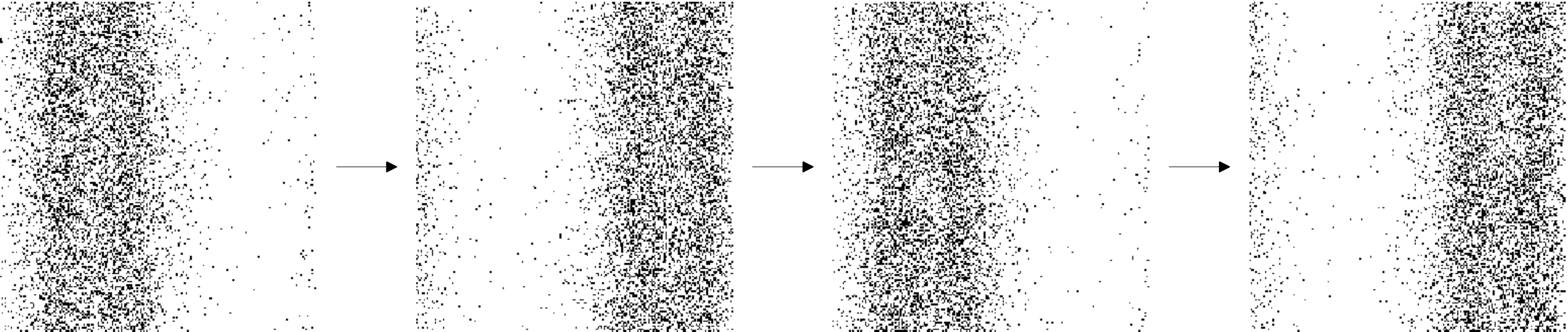}}
\caption{Four sequential snapshots for the individual-based model on a $201 \times 201$ lattice when $r=30$ and (a) $s=10$, (b) $s=20$,  and (c) $s=45$.  The snapshots show the population after competition, thus there are at most 1 individual in each site. }
 \label{fig:D201_snapshot}
\end{figure}

Increasing the lattice size to $201 \times 201$ and the reproductive rate to $r=30$, the individual-based model produces interesting patterns, which is seen in Figure \ref{fig:D201_snapshot}. When $s=10$, we see oscillating ring patterns, resembling reaction-diffusion patterns. As we increase $s$ to 20, the population dynamics instead behave as travelling waves. When $s=45$ there is a band formation jumping between the two halves of the lattice in each consecutive snapshot. In Figure \ref{fig:D201_snapshot} (c) the band formation is vertical, but the band formation can also be horizontal.  

\begin{figure}[H]
\centering
\subcaptionbox*{}{\includegraphics[scale=0.5]{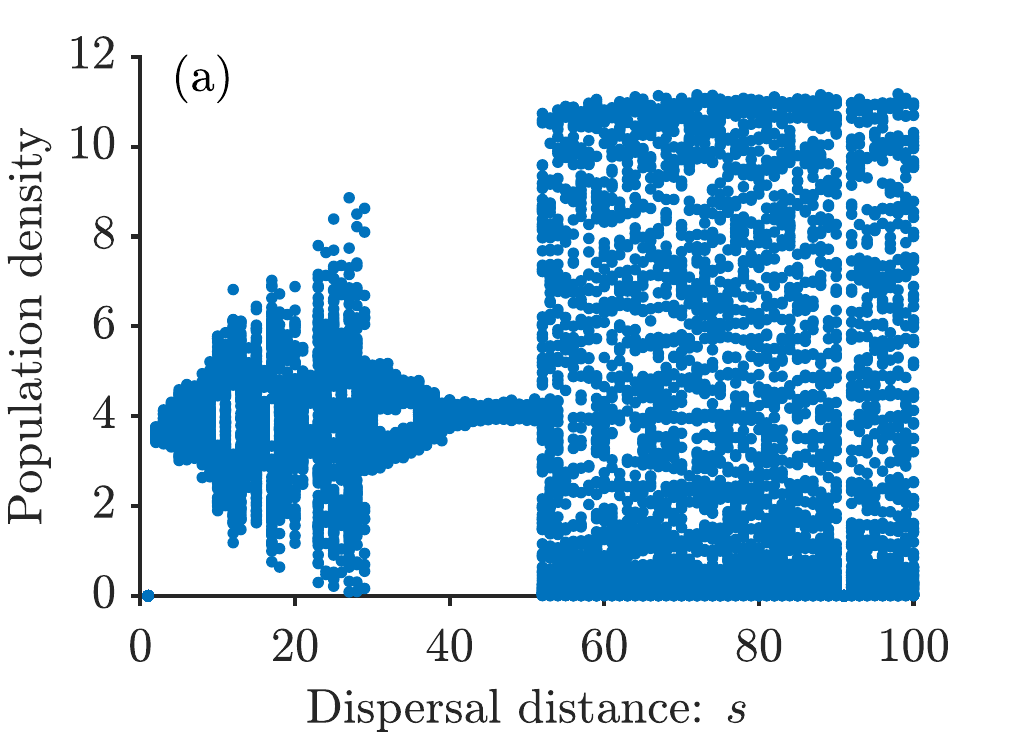}}
\subcaptionbox*{}{\includegraphics[scale=0.5]{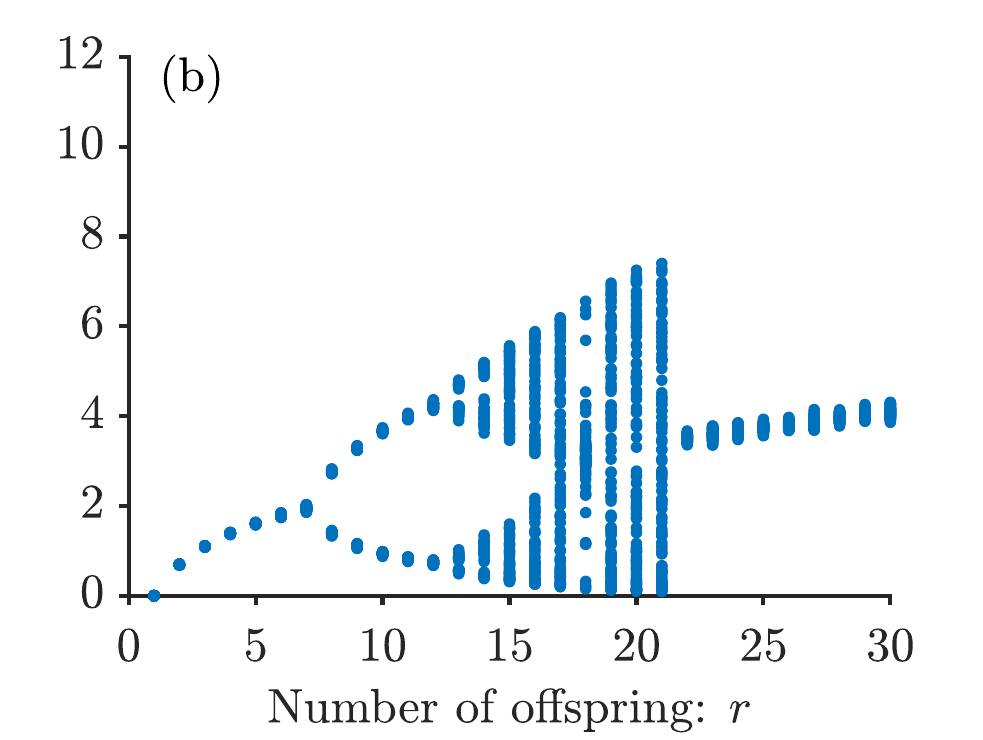}}
\caption{(a) Bifurcation diagram for the individual-based model, with $\phi(k)$ given as equation \eqref{eq:scramble_phi}, $r=30$ and $s$ as the bifurcation parameter.  For each value of $s$, we simulate the model for 5000 time steps and plot last 500 time steps.  In (b),  $s=45$ and $r$ is varied,  and for each value of $r$ we simulate the model for 5000 time steps, and the plot shows the outcome for the last 500 time steps. The population density is given by the number of individuals divided by the number of resource sites, and is counted after reproduction.} \label{fig:IBM_bif_s}
\end{figure}

In the bifurcation diagram in Figure \ref{fig:IBM_bif_s} (a) ,we fixed $r$ to 30 and varied $s$, simulated the model for 5000 time steps, and the plot shows the outcome for the last 500 time steps. From this bifurcation diagram, we can conclude that the band formation seen in Figure \ref{fig:D201_snapshot} (c) gives rise to stable population dynamics. For $s=10$, the population dynamics seem stable and for $s=20$ the population is fluctuating.  In the bifurcation diagram in Figure \ref{fig:IBM_bif_s} (b),  we instead fixed $s$ to 45 and varied $r$, simulated the model for 5000 time steps, and the plot shows the outcome for the last 500 time steps.  We see that the population is stable for $r<8$, and exhibits a period doubling bifurcations after that.  At $r=22$, the population becomes stable again.  We can thus see that long-range, but not global, dispersal has a stabilizing effect on the population dynamics for high reproductive rates. 

In essence, the snapshots of the model together with the bifurcation plots, show that our model displays behaviour which are inherently discrete in time: the population density display chaotic behaviour for some values of $r$ and $s$, and there are large-scale discrete oscillations in the spatial distribution of the population, rather than waves that move smoothly.

\subsection{Spatial statistics} \label{sec:spatstat}

The snapshots of the individual-based model, together with the bifurcation diagrams, demonstrate that spatial structure strongly influence the population dynamics. In Section \ref{sec:approx} we will make different assumptions of the spatial distribution of the individual-based model, in order to find analytical approximations of the model. Thus, understanding the spatial structure and scales of the individual-based model is valuable for the approximations done in \ref{sec:approx}. In the supplementary material S1, we use Moran's $I$ (which measures spatial correlation) to show that spatial clustering changes with dispersal distance.  This, in itself, does not give us any information of the scales of the spatial clusters. A better measure for detecting spatial pattern scales is the four-term local quadrat variance measure, referred to as 4TLQV. The idea behind 4TLQV is to look at a block of size  $b \times b$ and then compare this block to three adjacent blocks. This is done by summing all the individuals in the first block, multiply by $-3$ and then sum all individuals of the other three blocks. Since any of the four blocks can be used for comparison, there are four possible ways to do this calculation, and the 4TLQV is calculated by taking the average of the squared values of these four possibilities (see the Figure S3 for an illustration of this). The peaks in 4TLQV as a function of $b$ can be interpreted as the spatial scale of the pattern. We used the PySSaGE package for Python, to calculate the 4TQLV for 100 consecutive snapshots of 10 realisations of the individual-based model for various values $s$ (i.e. 1000 snapshots for each parameter set) \citep{Rosenberg2021}.

\begin{figure}[H]
\subcaptionbox*{}{\includegraphics[scale=0.47]{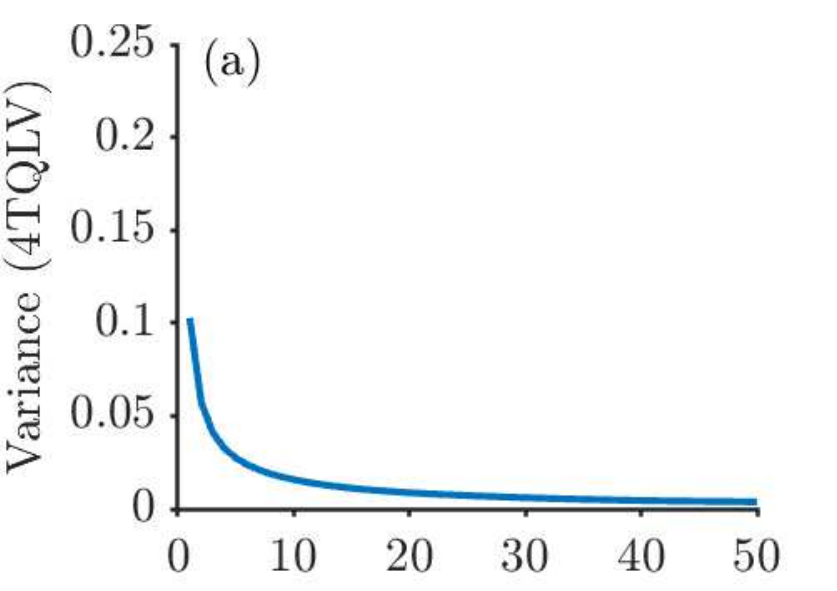}} \hspace{-1.42cm}
\subcaptionbox*{}{\includegraphics[scale=0.47]{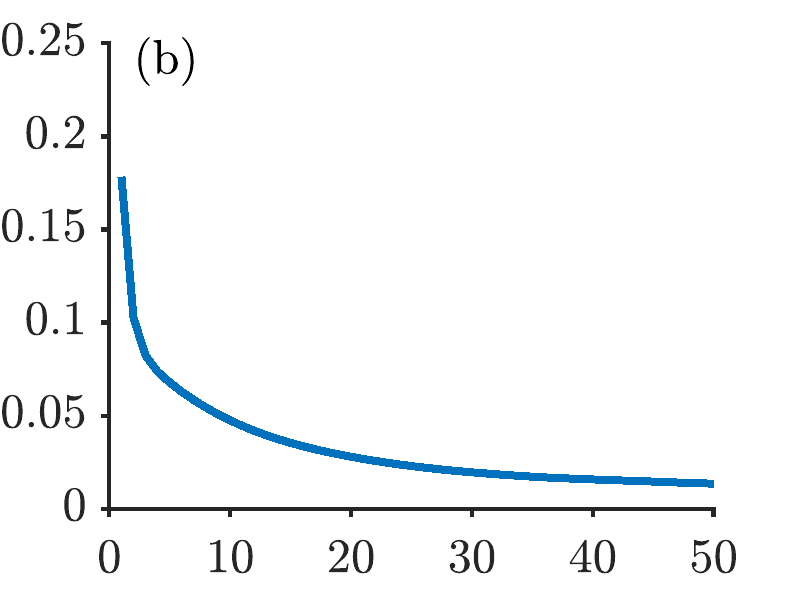}} \hspace{-1.8cm}
\subcaptionbox*{}{\includegraphics[scale=0.47]{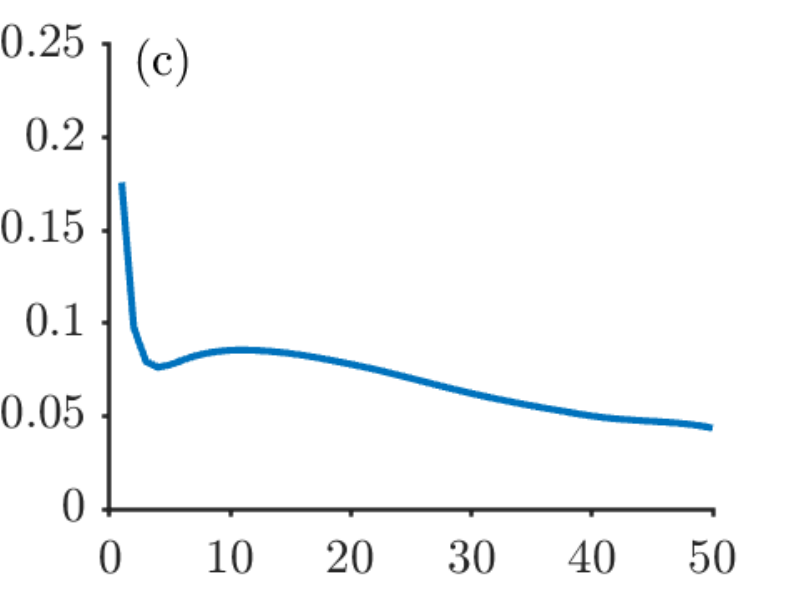}}
\subcaptionbox*{}{\includegraphics[scale=0.47]{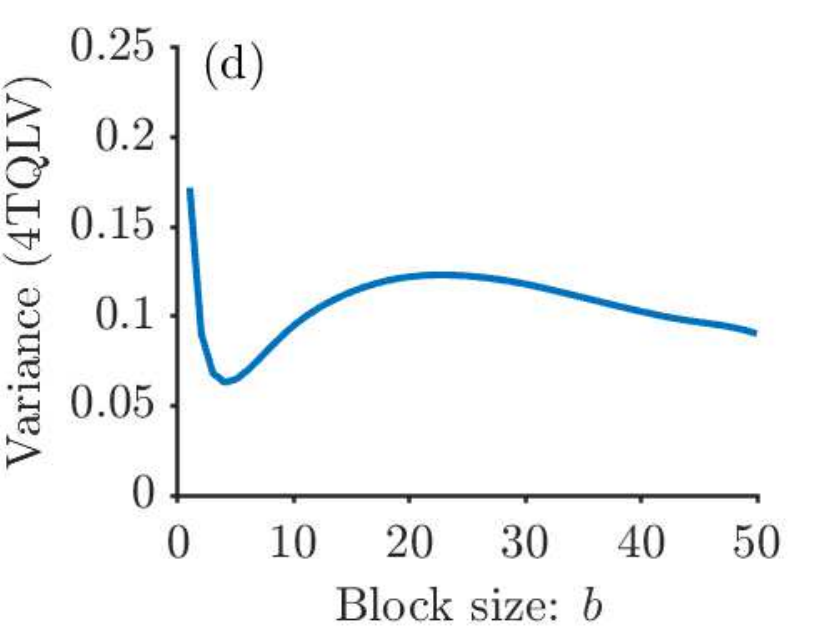}} \hspace{0.4cm}
\subcaptionbox*{}{\includegraphics[scale=0.47]{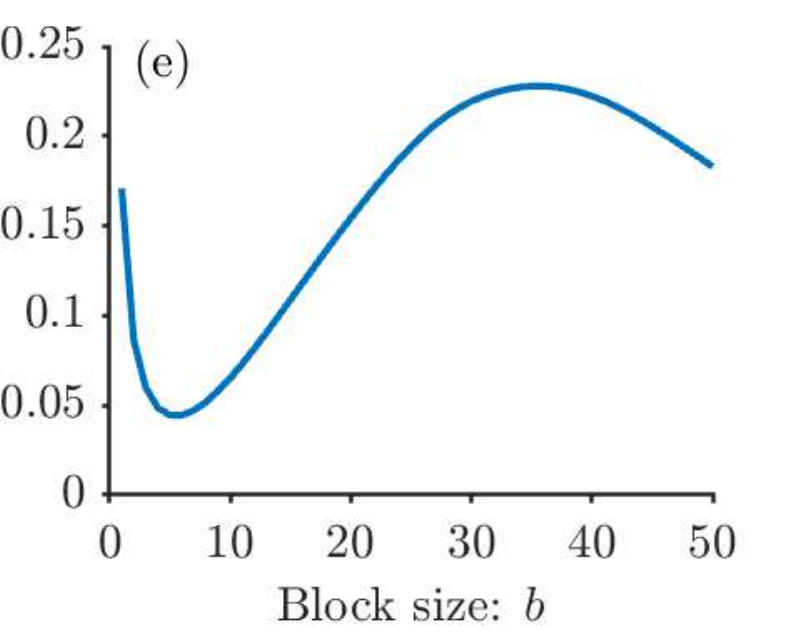}} \hspace{0.4cm}
\subcaptionbox*{}{\includegraphics[scale=0.47]{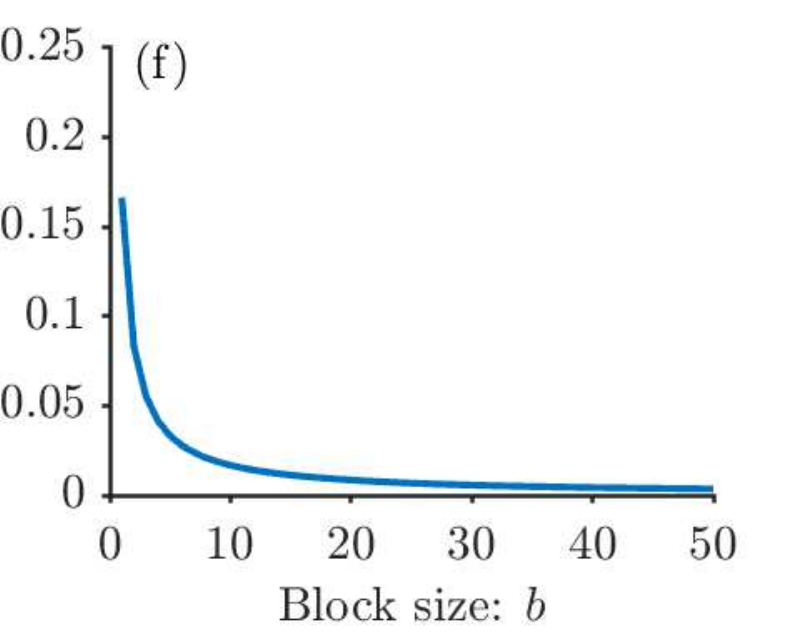}}
\caption{The average 4TLQV variance against the block size $b$ for 1000 snapshots from 10 realisations of the individual-based model simulated on a $101 \times 101$ grid when  $r=20$ and (a) $s=1$, (b) $s=2$, (c) $s=3$, (d) $s=5$, (e) $s=10$, and (f) $s=50$.\label{fig:spatial_stat_D101}}
\end{figure}

In Figure \ref{fig:spatial_stat_D101} we see the average 4TQLV of 100 consecutive snapshots from 10 realisations (1000 snapshots in total) for different values of $s$ when $r=20$ as a function of the block size, $b$. We note that for $s=1$ and $s=2$, the peak in variance is already at $b=1$, suggesting that there is no clear spatial structure. For $s=3$, there is a peak around $b=9$. As $s$ increases, the value of the 4TLQV peak also increases.  However, when $s=50$ the peak is now at $b=1$, suggesting there is no clear spatial structure. 

\begin{figure}[H]
\subcaptionbox*{}{\includegraphics[scale=0.47]{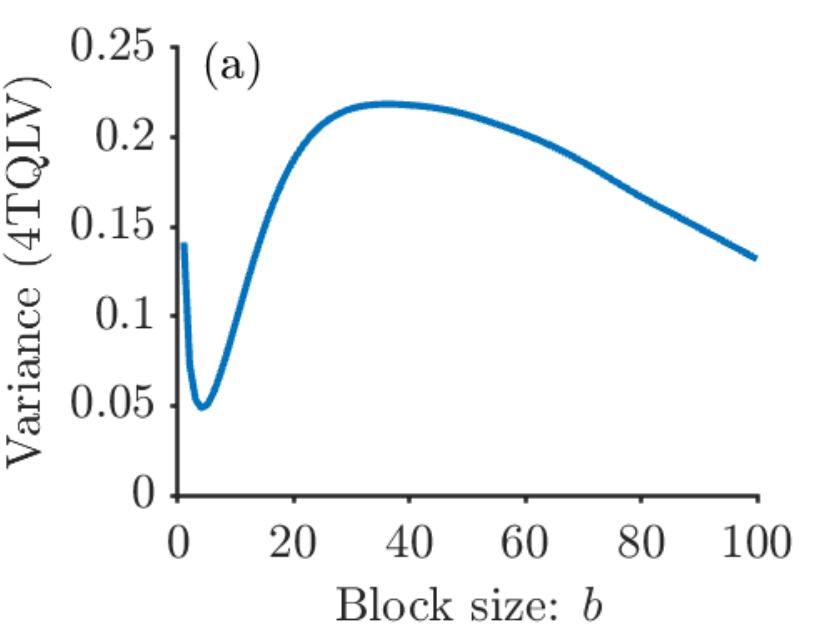}}
\subcaptionbox*{}{\includegraphics[scale=0.47]{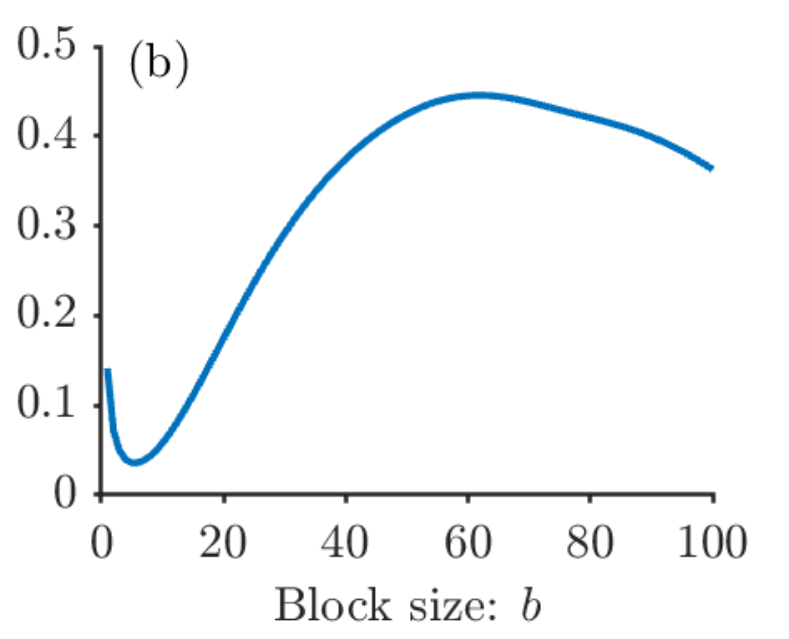}}
\subcaptionbox*{}{\includegraphics[scale=0.47]{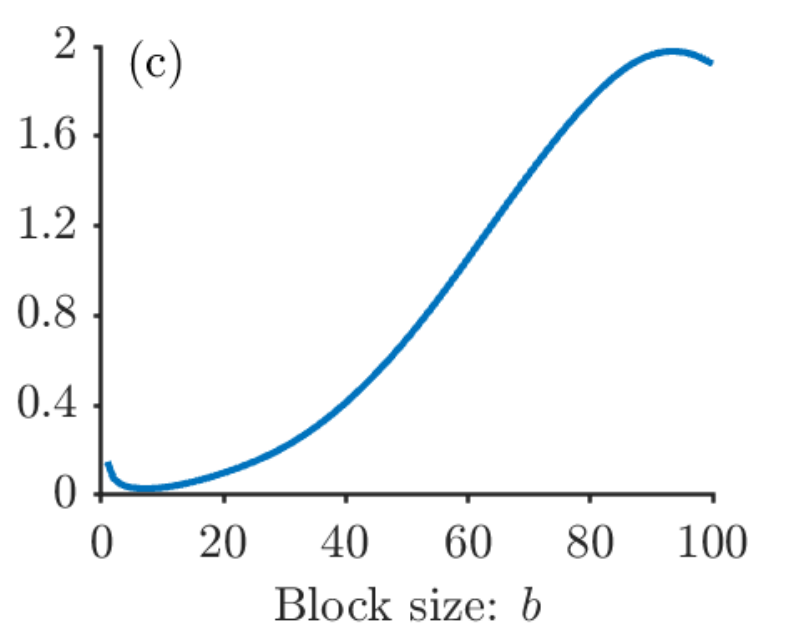}}
\caption{The average 4TLQV variance against the block size $b$ for 1000 snapshots from 10 realisations of the individual-based model simulated on a $201 \times 201$ grid when $r=30$ (a) $s=10$, (b) $s=20$, and  (c) $s=45$. \label{fig:spatial_stat_D201}}
\end{figure} 
In Figure \ref{fig:spatial_stat_D201}, the average 4TLQV variance is plotted for different values of $s$ when $r=30$ and the lattice size is $201 \times 201$. We see that when $s=10$, there is a peak around $b=30$. For $s=20$, there are there is a peak around 60, and for $s=45$, there is a peak at $b=90$. 
What should also be noted, but is not shown here, is that for all parameter sets with clear spatial patterns, the variance of the 4TLQV increases as $b$ increases, meaning that the spatial patterns sizes varies between time step.  From these plots, we can conclude that if there is a spatial pattern, the size of the pattern is always larger than $2s+1$. This is not surprising, since the individuals disperse into a $(2s+1)\times (2s+1)$ blocks and thus compete with other individuals within that range, the spatial patterns should be at least of that size.

\section{Approximation}
\label{sec:approx}

As we saw in the previous section, our model, which is very simple to state, produces a rich variety of dynamics. The aim of this article is to look at various ways to approximate the simulation model using an appropriate dynamical system and then to use this approximation to understand why these dynamics arise. In this section we describe three such approximations.

\subsection{Global approximation}

We start with the model described in Section \ref{sec:IBM} with scramble competition described by equation \eqref{eq:scramble_phi}. Let $x_t$ be the density of the population at time $t$, where $t=0, 1, 2,... $ is the time after the competition phase, i.e. after the survival function $\phi(X_{ij}^t)$ has been applied to every resource site. The expected population density at time $t+1$ given the population density $x_t$ at time $t$ can be written as: 
\begin{equation}
x_{t+1} = \Ex (X_{t+1}| X_t=x_t) = \frac{1}{D^2} \sum_{k=1}^{\infty}  \sum_{i=1}^{D}   \sum_{j=1}^{D} P(X_{ij}^t =k| X_t=x_t) \cdot \phi(k). \label{eq:full}
\end{equation} 
Here $\phi(k)$ is the survival function and $P(X_{ij}^t =k |X_t=x_t)$ is the probability to find $k$ individuals in a site  $(i,j)$ at time $t$, given that the population density is $x_t$. The question of approximating the population dynamics becomes a question of finding a good approximation $p_k$ of $P(X_{ij}^t =k)$, which utilises as much information as possible about the correlations between neighbouring sites. 
Thus $p_k$ is an approximation of $P(X_{ij}^t =k)$, which is not dependent on a specific site $(i,j)$. From this, we can reduce equation \eqref{eq:full} to
\begin{equation}
x_{t+1}=\frac{1}{D^2} \sum_{k=1}^{\infty}  \sum_{i=1}^{D}   \sum_{j=1}^{D} p_k \cdot \phi(k) =\sum_{k=1}^{\infty} p_k \cdot \phi(k),
\end{equation}
i.e. $p_k$ is site independent.

One possible way to approximate $p_k$, is to assume that the offspring disperse independently to any site on lattice with uniform probability. When the lattice is infinitely large, $X_{ij}^t$ is Poisson distributed with mean $r x_t$ for all $i$ and $j$, in which case
\begin{equation}
p_k= P(X_{ij}^t =k) =\frac{e^{-r x_t}(r x_t)^k}{k!},
\end{equation}
giving the following population dynamics
\begin{equation}
x_{t+1}=\sum_{k=1}^{\infty} \frac{e^{-rx_t} (rx_t)^k}{k!}\cdot \phi(k).
\end{equation}
In this case, the entire population dynamics for the scramble model is
\begin{equation}
x_{t+1}= rx_t e^{- r x_t}. \label{eq:meanfieldprevious}
\end{equation}
This equation, sometimes called the mean field equation \citep{morozov2012spatially},  we refer to as the global approximation equation of the individual-based model. This is the same equation as derived by \cite{sumpter2001relating} and \cite{brannstrom2005coupled} but with a change of variable $x_t' = r x_t$ to reflect the fact that we measure the population after reproduction. This global approximation works well for individual-based models with large dispersal, i.e. when $s$ is close to $(D-1)/2$. 
\newline

\subsection{Local correlation approximation}
\begin{figure}[H]
\center
  \definecolor{mycolor}{RGB}{238,233,233}
\begin{tikzpicture}


\node (1) at (4.0, 8.5) {\Large{(a)}};
\node (2) at (14.0, 8.5) {\Large{(b)}};
\node (3) at (24.0, 8.5) {\Large{(c)}};

\draw[step=1cm,black, very thin] (0,0) grid (8,8);

\draw [line width=0.3mm] (0,0) -- (8,0) -- (8,8) -- (0,8) -- (0,0);

\draw[thick,->] (0.31,0.31) -- (8.5,0.31)node[anchor=west] {$i$};
\draw[thick,->] (0.31,0.31) -- (0.3,8.5) node[anchor=south] {$j$};
;

\draw [fill=black, opacity=0.5] (3,2) -- (3,3) -- (4,3) -- (4,2) -- (3,2);
\filldraw[fill=black, draw=black, very thin] (2.5,3.5) circle (3pt);
\filldraw[fill=black, draw=black, very thin] (2.5,1.5) circle (3pt);
\filldraw[fill=black, draw=black, very thin] (4.5,2.5) circle (3pt);
\filldraw[fill=black, draw=black, very thin] (4.5,3.5) circle (3pt);
\filldraw[fill=black, draw=black, very thin] (3.5,2.5) circle (3pt);

\draw [fill=gray, opacity=0.5] (2,1) -- (2,4) -- (5,4) -- (5,1) -- (2,1);


\draw[step=1cm,black, very thin] (10,0) grid (18,8);

\draw [line width=0.3mm] (10,0) -- (18,0) -- (18,8) -- (10,8) -- (10,0);

\draw[thick,->] (10.31,0.31) -- (18.5,0.31)node[anchor=west] {$i$};
\draw[thick,->] (10.31,0.31) -- (10.3,8.5) node[anchor=south] {$j$};

\draw [fill=black, opacity=0.5] (13,2) -- (13,3) -- (14,3) -- (14,2) -- (13,2);
\filldraw[fill=black, draw=black, very thin] (12.25,3.25) circle (3pt);
\filldraw[fill=black, draw=black, very thin] (12.75,3.25) circle (3pt);
\filldraw[fill=black, draw=black, very thin] (12.25,3.75) circle (3pt);
\filldraw[fill=black, draw=black, very thin] (12.75,3.75) circle (3pt);

\filldraw[fill=black, draw=black, very thin] (12.25,1.25) circle (3pt);
\filldraw[fill=black, draw=black, very thin] (12.75,1.25) circle (3pt);
\filldraw[fill=black, draw=black, very thin] (12.25,1.75) circle (3pt);
\filldraw[fill=black, draw=black, very thin] (12.75,1.75) circle (3pt);

\filldraw[fill=black, draw=black, very thin] (14.25,2.25) circle (3pt);
\filldraw[fill=black, draw=black, very thin] (14.25,2.75) circle (3pt);
\filldraw[fill=black, draw=black, very thin] (14.75,2.25) circle (3pt);
\filldraw[fill=black, draw=black, very thin] (14.75,2.75) circle (3pt);

\filldraw[fill=black, draw=black, very thin] (14.25,3.25) circle (3pt);
\filldraw[fill=black, draw=black, very thin] (14.25,3.75) circle (3pt);
\filldraw[fill=black, draw=black, very thin] (14.75,3.25) circle (3pt);
\filldraw[fill=black, draw=black, very thin] (14.75,3.75) circle (3pt);

\filldraw[fill=black, draw=black, very thin] (13.25,2.25) circle (3pt);
\filldraw[fill=black, draw=black, very thin] (13.25,2.75) circle (3pt);
\filldraw[fill=black, draw=black, very thin] (13.75,2.25) circle (3pt);
\filldraw[fill=black, draw=black, very thin] (13.75,2.75) circle (3pt);

\draw [fill=gray, opacity=0.5] (12,1) -- (12,4) -- (15,4) -- (15,1) -- (12,1);


\draw[step=1cm,black, very thin] (20,0) grid (28,8);

\draw [line width=0.3mm] (20,0) -- (28,0) -- (28,8) -- (20,8) -- (20,0);

\draw[thick,->] (20.31,0.31) -- (28.5,0.31)node[anchor=west] {$i$};
\draw[thick,->] (20.31,0.31) -- (20.3,8.5) node[anchor=south] {$j$};

\draw [fill=black, opacity=0.5] (23,2) -- (23,3) -- (24,3) -- (24,2) -- (23,2);
\filldraw[fill=black, draw=black, very thin] (21.5,3.5) circle (3pt);
\filldraw[fill=black, draw=black, very thin] (22.5,4.5) circle (3pt);
\filldraw[fill=black, draw=black, very thin] (23.5,4.5) circle (3pt);
\filldraw[fill=black, draw=black, very thin] (24.5,4.5) circle (3pt);

\filldraw[fill=black, draw=black, very thin] (22.5,1.5) circle (3pt);
\filldraw[fill=black, draw=black, very thin] (23.5,1.5) circle (3pt);
\filldraw[fill=black, draw=black, very thin] (24.5,1.5) circle (3pt);

\filldraw[fill=black, draw=black, very thin] (22.5,2.5) circle (3pt);

\filldraw[fill=black, draw=black, very thin] (25.5,2.5) circle (3pt);

\filldraw[fill=black, draw=black, very thin] (23.25,2.75) circle (3pt);
\filldraw[fill=black, draw=black, very thin] (23.75,2.75) circle (3pt);
\filldraw[fill=black, draw=black, very thin] (23.5,2.25) circle (3pt);

\filldraw[fill=black, draw=black, very thin] (24.25,3.25) circle (3pt);
\filldraw[fill=black, draw=black, very thin] (24.75,3.75) circle (3pt);

\filldraw[fill=black, draw=black, very thin] (24.75,2.25) circle (3pt);
\filldraw[fill=black, draw=black, very thin] (24.25,2.75) circle (3pt);

\filldraw[fill=black, draw=black, very thin] (25.25,3.25) circle (3pt);
\filldraw[fill=black, draw=black, very thin] (25.75,3.75) circle (3pt);

\filldraw[fill=black, draw=black, very thin] (21.75,1.25) circle (3pt);
\filldraw[fill=black, draw=black, very thin] (21.25,1.75) circle (3pt);

\draw [fill=gray, opacity=0.5] (22,1) -- (22,4) -- (25,4) -- (25,1) -- (22,1);


\end{tikzpicture} 
\caption{An illustration of how $p_k$ is calculated for the local correlation approximation when $s=1$. The focal site is dark grey, while its neighbours within distance $s=1$ are light grey. The figure shows how individuals are (a) distributed around the focal site initially, where the number of parents is $F=5$, which in general is given by $F \sim \operatorname{Bin} ((2s+1)^2, x_t)$.  In (b) the individuals have each produced $r=4$ offspring, meaning there is $r \cdot F = 4 \cdot 5 = 20$ individuals within the neighbourhood of the focal site. In (c) the offspring have dispersed. The event that a particular offspring $j$ will disperse into the focal site, $O_j$ has probability  $1/(2s+1)^2$. The number of offspring landing in the focal site is then given by $Y = \sum_{j=1}^{F r} O_j= \sum_{j=1}^{20} O_j$, which in this example is equal to $3$, ($Y=3$).}
 \label{fig:tikz_approx}
\end{figure}
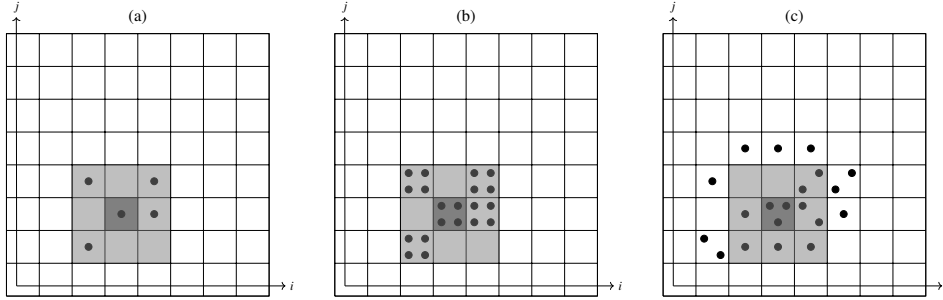

From the simulations of the individual-based model, we saw that even though the scale of the spatial pattern is the same for very local dispersal ($s=1$ and $s=2$) and global dispersal, the population dynamics is vastly different: when dispersal is global, the population dynamics is chaotic, whereas the population dynamics is always stable for local dispersal. This suggests, that for local dispersal,  ($s << D$), we need to take local correlations into account. To do this, we adopt an approach that looks at correlations which arise between dispersal and competition. Consider a particular lattice site, which we call the focal site. In order to better account for local correlations we look in the $(2s+1) \times (2s+1)$ Moore neighbourhood around the focal site. Let $F$ be the total number of parents in the $(2s+1) \times(2s+1)$ Moore neighbourhood of the focal site (see Figure \ref{fig:tikz_approx}). 

For this approximation we assume -- as we also did in our approximation of full independence between the sites -- that the population is uniformly distributed at this stage with population density $x_t$.  The spatial statistics in Section \ref{sec:spatstat} together with the supplementary material S1, showed that this assumption is justified not just for global dispersal, but also for very local dispersal (i.e.  when $s$ is small).  Now,  if  $s << D$,  we can assume that each site in the neighbourhood of the focal site has an independent probability $p=x_t$ containing an individual, meaning that $F \sim \operatorname{Bin} ((2s+1)^2, x_t)$.
In this new approximation, our aim is to calculate the number of offspring of all of the parents in the neighbourhood of the focal site that disperse to the focal site. We denote this as a random variable $Y$. The probability that a particular offspring of a particular parent within the $(2s+1) \times(2s+1)$ Moore neighbourhood ends up in the focal site is $1/(2s+1)^2$. We can thus denote the event that this offspring $j$ lands on the focal site as a Bernoulli random variable, $O_j$ with parameter $1/(2s+1)^2$. The dispersal of individuals is independent and thus $Y$ is defined by the sum
\begin{equation}
Y = \sum_{j=1}^{F \cdot r} O_j \label{eq:sum}
\end{equation}
or, equivalently, $Y\sim \operatorname{Bin}(F\cdot r, \frac{1}{(2s+1)^2})$. 

Using the law of total probability we utilise equation \eqref{eq:sum} to write 
\begin{equation}
p_k = P(Y = k) = \sum_{i=1}^{(2s+1)^2} P(Y=k | F=i) P(F=i).
\end{equation}
The two terms in this convolution are, respectively, 
\begin{equation}
P(Y=k | F=i) = \binom{ir}{k} \frac{1}{(2s+1)^{2k}}(1- \frac{1}{(2s+1)^2})^{ir-k},
\end{equation}
and
\begin{equation}
P(F=i)= \binom{(2s+1)^2}{i}x_t^i (1-x_t)^{(2s+1)^2-i}.
\end{equation} 
This gives the following explicit expression for $p_k$ 
\begin{equation}
p_k= \sum_{i=1}^{(2s+1)^2}\binom{(2s+1)^2}{i}x_t^i (1-x_t)^{(2s+1)^2-i}\binom{ir}{k}\frac{1}{(2s+1)^{2k}} \left(1- \frac{1}{(2s+1)^2} \right)^{ir-k}.
\end{equation}
With the approximation that all sites on the lattice experience the same dynamics as the focal site, the population dynamics are then given by
\begin{equation}
x_{t+1}=\sum_{k=1}^{\infty} p_k \cdot \phi(k).
\end{equation}
For scramble competition, where $\phi(k)$ is defined as equation \eqref{eq:scramble_phi}, the expression becomes
\begin{eqnarray}
x_{t+1}  & = \sum_{i=1}^{(2s+1)^2} \binom{(2s+1)^2}{i}x_t^i(1-x_t)^{(2s+1)^2-i} i r \frac{1}{(2s+1)^2}(1- \frac{1}{(2s+1)^2})^{ir-1} . 
\end{eqnarray}
By identifying the last factor as a derivative, the expression can be simplified to 
 \begin{eqnarray}
x_{t+1}  & =rx_t\left(1-\frac{1}{(2s+1)^2}\right)^{r-1}\left(x_t \left(1-\frac{1}{(2s+1)^2}\right)^{r}-x_t+1\right)^{(2s+1)^2-1} = f(x_t).
\label{mainapprox}
\end{eqnarray}
We now have a better approximation of the individual-based model that, unlike the global approximation equation (equation \eqref{eq:meanfieldprevious}), takes in the spatial properties of the individual-based model, and has the dispersal distance, $s$, as a parameter.  (See Supplementary material S4 for a detailed derivation.)

Now, if $R_{ij}^t$ is not constant, but $R_{ij}^t$ are i.i.d. random variables with $R_{ij}^t \sim \operatorname{Bin} (r/q,q)$,  the number of offspring the parents will produce will be given by
\begin{equation}
S_F=\sum_i^F R_i
\end{equation}
where $R_i\sim \operatorname{Bin}(r/q,q)$.
Thus, 
\begin{equation}
Y = \sum_{j=1}^{S_F} O_j =\sum_{j=1}^{\sum_i^F R_i} O_j .
\end{equation}
Using probability generating functions (see Supplementary material S4),  we can find a closed form of $p_1=P(Y=1)$, which will be given by 
%

\begin{dmath} 
p_1=rx_t\left(\left(1-\frac{1}{(2s+1)^2}\right)q-q+1\right)^{r/q-1}\\
\left(x_t \left(\left(1-\frac{1}{(2s+1)^2}\right)q-q+1\right)^{r/q}-x_t+1\right)^{(2s+1)^2-1}.
\end{dmath}
Thus, when the reproductive rate, $R_{ij}^t$,  is random with $R_{ij}^t \sim \operatorname{Bin} (r/q,q)$, the population dynamics is given by

\begin{dmath} \label{eq:popdyn_rand}
x_{t+1}=rx_t\left(\left(1-\frac{1}{(2s+1)^2}\right)q-q+1\right)^{r/q-1}\\
\left(x_t \left(\left(1-\frac{1}{(2s+1)^2}\right)q-q+1\right)^{r/q}-x_t+1\right)^{(2s+1)^2-1}.
\end{dmath}

In the rest of the paper, we will focus on the analysis of model with deterministic $r$, but in Supplementary material S4, derive and analyse the local correlation approximation with random $R$.

\subsection{Long-range dispersal approximation}

The bifurcation diagram in Figure \ref{fig:IBM_bif_s} shows that for long-range dispersal, i.e., $s \approx D/2$, the population density stabilizes. The four consecutive snapshots of the individual-based model for $s=45$ in Figure \ref{fig:D201_snapshot} (c) suggests that this happens when the population propagates as a band over the lattice, i.e., when the spatial pattern scale is around $D/2$, as seen in Figure \ref{fig:spatial_stat_D201} (c). Even though there is a clear global spatial structure in terms of the band formation, calculating the Moran's $I$ within each band, suggest that the population is close to well mixed locally. 

\begin{figure}[H] 
\centering
\definecolor{mycolor}{RGB}{238,233,233}
\begin{tikzpicture}

\draw[step=0.25cm,mycolor,very thin] (0,0) grid (4,4);

\draw [line width=0.3mm] (2,0) -- (2,4);
\draw [line width=0.3mm] (0,0) -- (4,0) -- (4,4) -- (0,4) -- (0,0);

\node (1) at (1, 2) {$P_1$};
\node (2) at (3, 2) {$P_2$};

\draw[thick,->] (0.31,0.31) -- (4.5,0.31)node[anchor=west] {$i$};
\draw[thick,->] (0.31,0.31) -- (0.3,4.5) node[anchor=south] {$j$};

\draw [fill=gray, opacity=0.2] (1.5,0) -- (1.5,4) -- (2,4) -- (2,0) -- (1.5,0);

\draw [fill=gray, opacity=0.2] (2.5,0) -- (2.5,4) -- (2,4) -- (2,0) -- (2.5,0);

\draw [fill=black] (1.5,0.5) -- (1.5,0.75) -- (1.75,0.75) -- (1.75, 0.5) -- (1.5,0.5);

\draw [fill=gray, opacity=0.5] (1,0) -- (1,1.25) -- (2.25,1.25) -- (2.25,0) -- (1,0);

\node (3) at (1.8, 2.75) {$\delta$};

\node (4) at (2.2, 1.63) {$\delta$};

\draw [->] (2) edge[bend right=45] node [left] {} (1);
\draw [->] (1) edge[bend right=45] node [right] {} (2);

\end{tikzpicture} 
\caption{Division of the lattice into two patches. The light grey area consists of the sites where dispersal to the other patch is possible, when the dispersal distance is $s=2$. The dark grey area shows the sites reachable by dispersal from the black site when $s=2$.}
\label{fig:patches}
\end{figure}
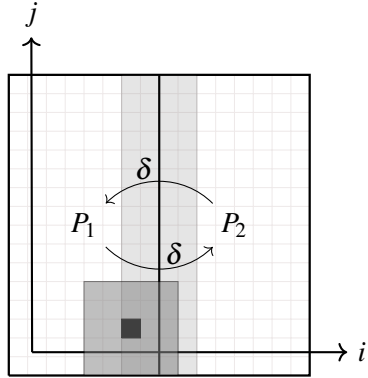

To capture this observation, we propose a long-range dispersal approximation, which divides the lattice into two patches, $P_1$ and $P_2$,  with population size $x_1$ and $x_2$ respectively, with dispersal probability $\delta$ between the patches, as seen in Figure \ref{fig:patches}. In this approximation, the population is assumed large and well mixed within each patch, allowing us to describe the population as a coupled version of the global approximation (Eq. \eqref{eq:meanfieldprevious}), i.e.
\begin{equation}
\begin{split}
x_{1,t}= (1-\delta) r e^{-rx_{1,t}}+\delta re^{-rx_{2,t}} \\
x_{2,t} = (1-\delta) r e^{-rx_{2,t}}+\delta re^{-rx_{1,t}}.
\end{split} \label{eq:twopatch}
\end{equation}

We now want to determine reasonable values of the dispersal probability $\delta$ as a function of the dispersal distance $s$, by reasoning from the individual-based model. For simplicity, we assume that an individual in $P_1$ enters $P_2$ only when it crosses the border in the middle. When crossing the other borders, it re-enters $P_1$. Since $P_1$ and $P_2$ are identical except for being laterally reversed, it suffices to determine the dispersal probability from $P_1$ to $P_2$ to find $\delta$. To do this, we start by indexing the sites by $(i,j)$, as in Figure \ref{fig:patches}.  In $P_1$ we have $1 \leq i \leq D/2$ and $1 \leq j \leq D$. 
The dispersal probability between $P_1$ and $P_2$ will be given by the probability that an individual $k$ is situated at site $(i,j)$ times the probability that it disperses to patch $P_2$, given that it is at site $(i,j)$, summed over all sites in $P_1$: 

\begin{equation}
\begin{split}
\delta &=\sum\limits_{(i,j)\in P_1}P(k \text{ at site } (i,j) | \text{ } k \text{ in } P_1)\cdot P(k \text{ disperse to } P_2 | \text{ } k \text{ at site } (i,j)).
\end{split}
\end{equation}

Assuming that within each patch the population is well mixed, meaning that individuals are uniformly distributed over all sites, we have that $P(k \text{ at site } (i,j) | \text{ } k \text{ in } P_1)$ is given by $\frac{1}{D^2/2}$.

To determine $P(k \text{ disperse to } P_2 | k \text{ at site } (i,j))$, we want to find the number of sites in $P_1$ where $P_2$ is reachable by dispersal. This obviously depends on $s$.  We have that 
\[P(k \text{ disperse to } P_2 | \text{ at site } (i,j))=\frac{(s+1)-i}{2s+1},
\]
 when $i<s$, and $0$ when $i>s$.

Thus, we have
\begin{equation}
\begin{split}
\delta &=\sum\limits_{(i,j)\in P_1}P(k \text{ at site } (i,j) | k \text{ in } P_1)\cdot P(k \text{ disperse to } P_2 | k \text{ at site } (i,j)) \\
&=\sum\limits_{j=1}^{D}\sum\limits_{i=1}^{D/2} \frac{1}{D^2/2} P(k \text{ disperse to } P_2 | k \text{ at site } (i,j)) \\
&= D \sum\limits_{i=1}^{D/4} \frac{1}{D^2/2} P(k \text{ disperse to } P_2 | k \text{ at site } (i,j)) = 2D \sum\limits_{i=1}^{D/4} \frac{(s+1)-i}{2s+1}  \\
& = \frac{s(s+1)}{D(2s+1)}.
\end{split} 
\end{equation}

We now have three different approximations that describe the population dynamics for the individual-based model. For the rest of the paper we will analyse equation \eqref{mainapprox}, \eqref{eq:twopatch} and \eqref{eq:meanfieldprevious} and compare it with the individual-based model described in Section \ref{sec:IBM}.


\section{Analysis}
\label{sec:analysis}
In Section \ref{sec:approx} we have derived three different approximations of the individual-based model described in Section \ref{sec:IBM}. In this section we discuss how these approximations are able to describe the properties of the individual-based model and the connections between the different approximations. 

\subsection{Bifurcations} \label{sec:analysis_bif}

\begin{figure}[H]
\centering
\subcaptionbox*{}{\includegraphics[scale=0.48]{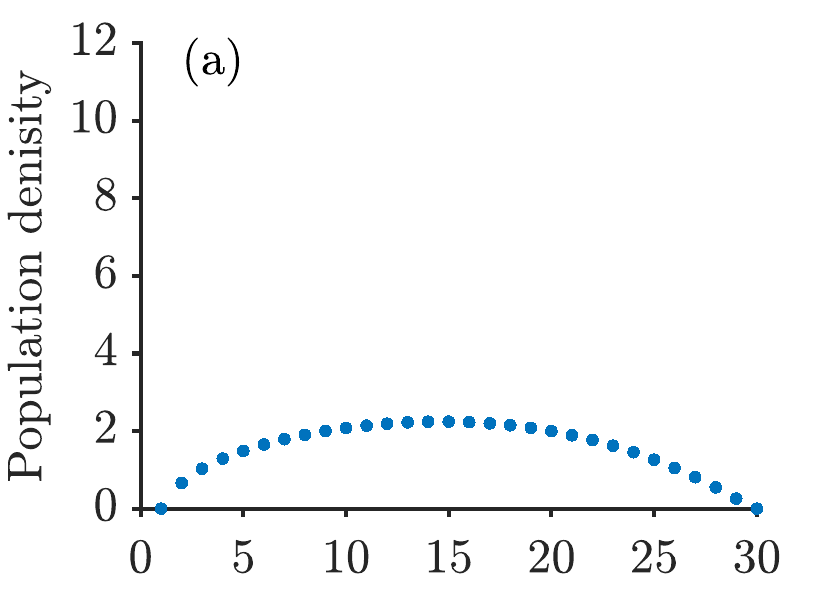}}
\subcaptionbox*{}{\includegraphics[scale=0.48]{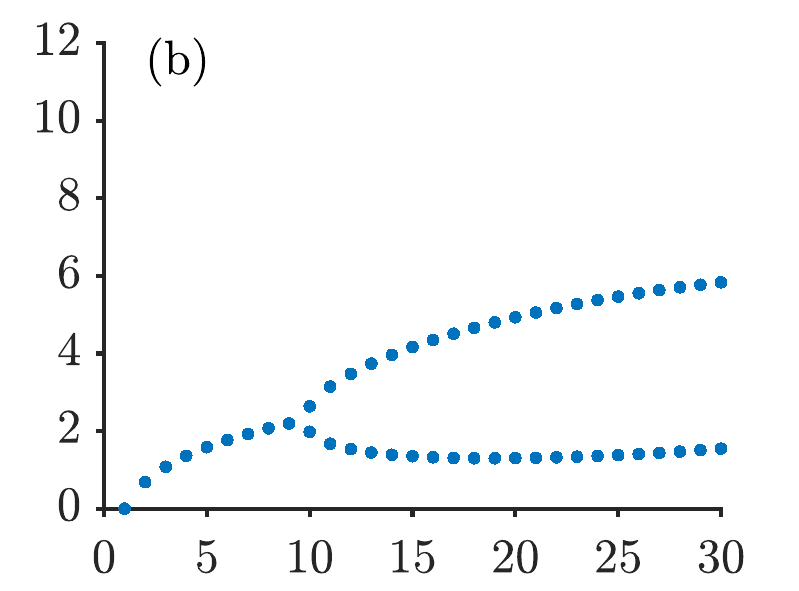}}
\subcaptionbox*{}{\includegraphics[scale=0.48]{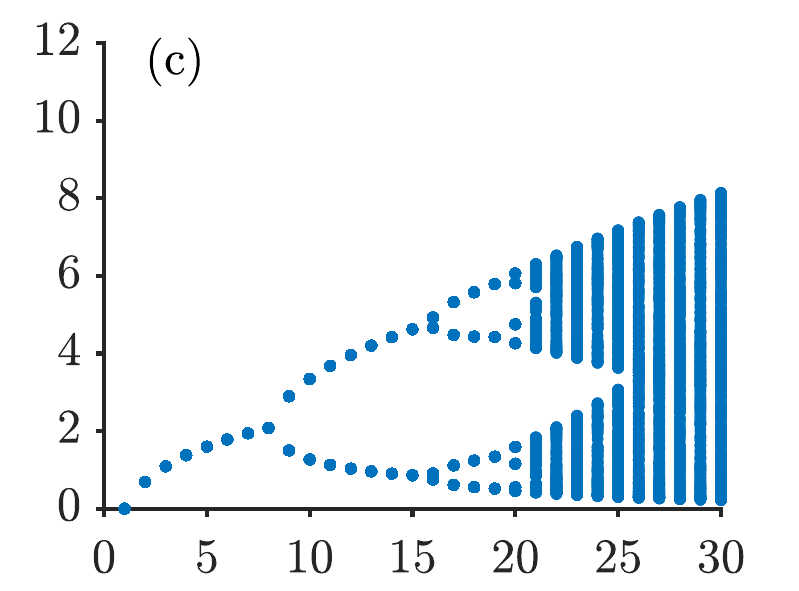}}
\subcaptionbox*{}{\includegraphics[scale=0.48]{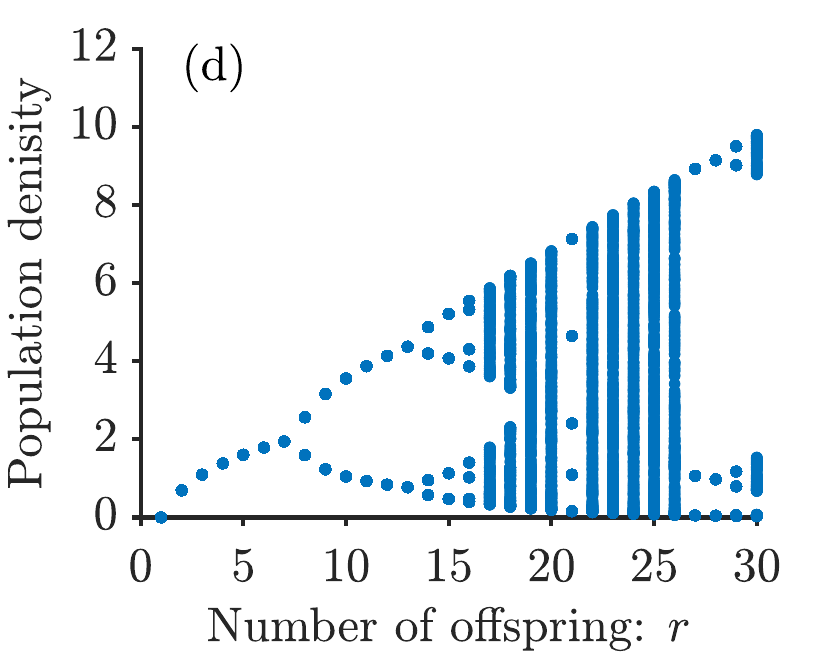}}
\subcaptionbox*{}{\includegraphics[scale=0.48]{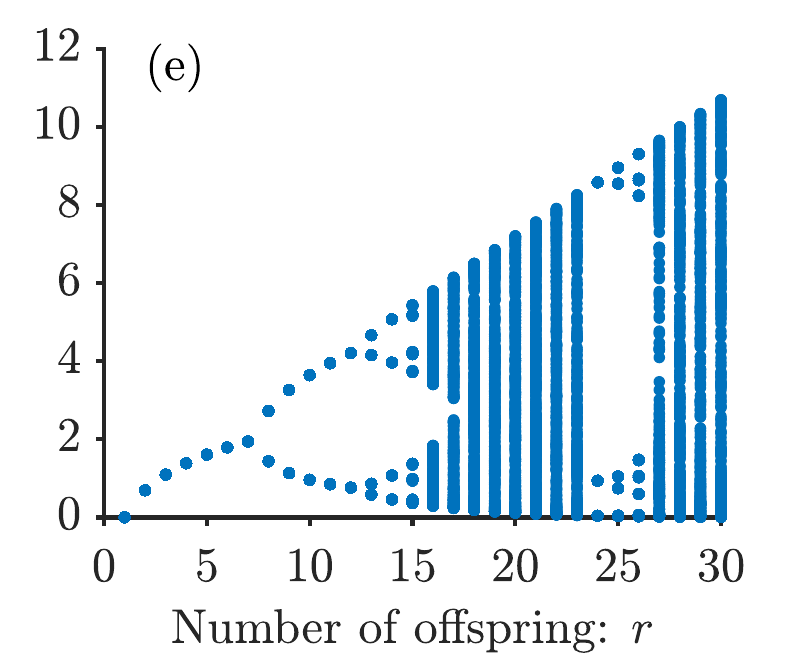}}
\subcaptionbox*{}{\includegraphics[scale=0.48]{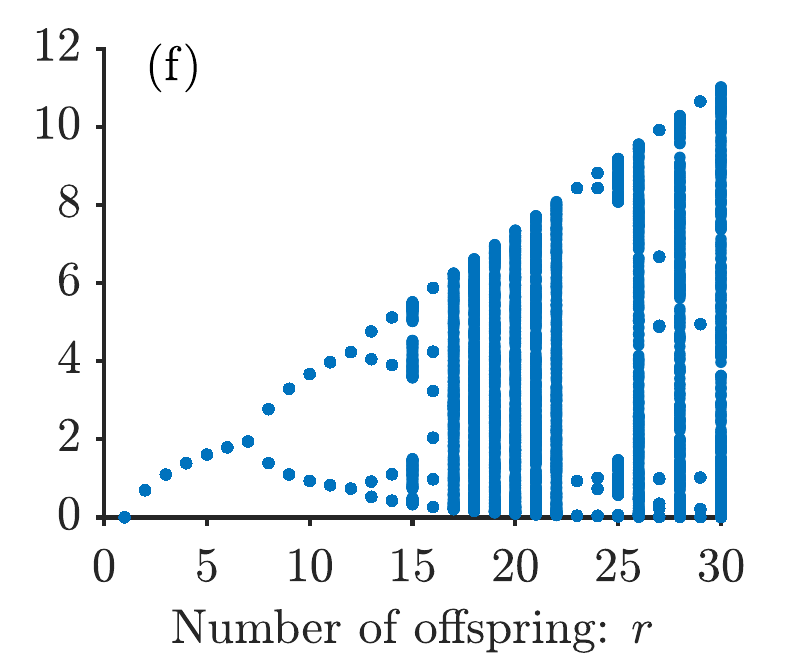}}
\caption{Bifurcation diagram for the local correlation approximation, given in equation \eqref{mainapprox}. In each subplot we vary $r$ and iterate through $x_t$ for 5000 time steps and plot last 500 time steps. We give results for (a) $s=1$ (very local dispersal) (b) $s=2$, (c) $s=3$, (d) $s=5$, (e) $s=10$ and (f) $s=50$ (close to global dispersal). \label{fig:bifapprox} }
\end{figure}

As for the individual-based model in Figure \ref{fig:bifmodel}, we produced bifurcation plots for the local correlation approximation by iterating equation \eqref{mainapprox} for different values of $r$, which can be seen in Figure \ref{fig:bifapprox}. When dispersal is very local ($s=1$), our approximation displays quantitatively the same behaviour as the individual-based model: the population is stable and at most around 2 (when $r \approx 15$). The only difference between the individual-based model and the local correlation approximation is that the population goes extinct already at $r=23$ for the individual-based model, whereas it goes extinct for $r=30$ for the approximation. In Figure \ref{fig:bifapprox} (b), we see that there is a period doubling bifurcation when $r=10$, for $s=2$, in contrary to the individual-based model, where the population is stable at least until $r=30$ for $s=2$. Increasing $s$ to 3 (Figure \ref{fig:bifapprox} (c)), the approximation model undergoes a period-doubling route to chaos: the model exhibits period doubling first at $r=8$ and then at $r=16$, and becomes chaotic at $r=26$. Increasing $s$ further, the period doubling takes place earlier and earlier: in Figure \ref{fig:bifapprox} (d) we see that for $s=5$, the first period doubling appears at $r=8$ and the second at $r=13$. 
For $s=10$, the dynamics of the approximation is more or less the same as for $s=50$. Quantitatively, the population density is higher for the local correlation approximation, than for the individual-based model, when $s>2$. In summary, the local approximation seem to approach the global approximation quicker than the individual-based model approaches the well-mixed case. 
In Supplementary material S4 we show bifurcation plots for the local correlation approximation for stochastic reproductive rates, i.e. equation \eqref{eq:popdyn_rand}. We see in the plots that over all qualitative behaviour is not affected by the value of $q$, just as for the bifurcation plots of the individual-based model with stochastic reproductive rate. 

\begin{figure}[H]
\centering
\subcaptionbox*{}{\includegraphics[scale=0.5]{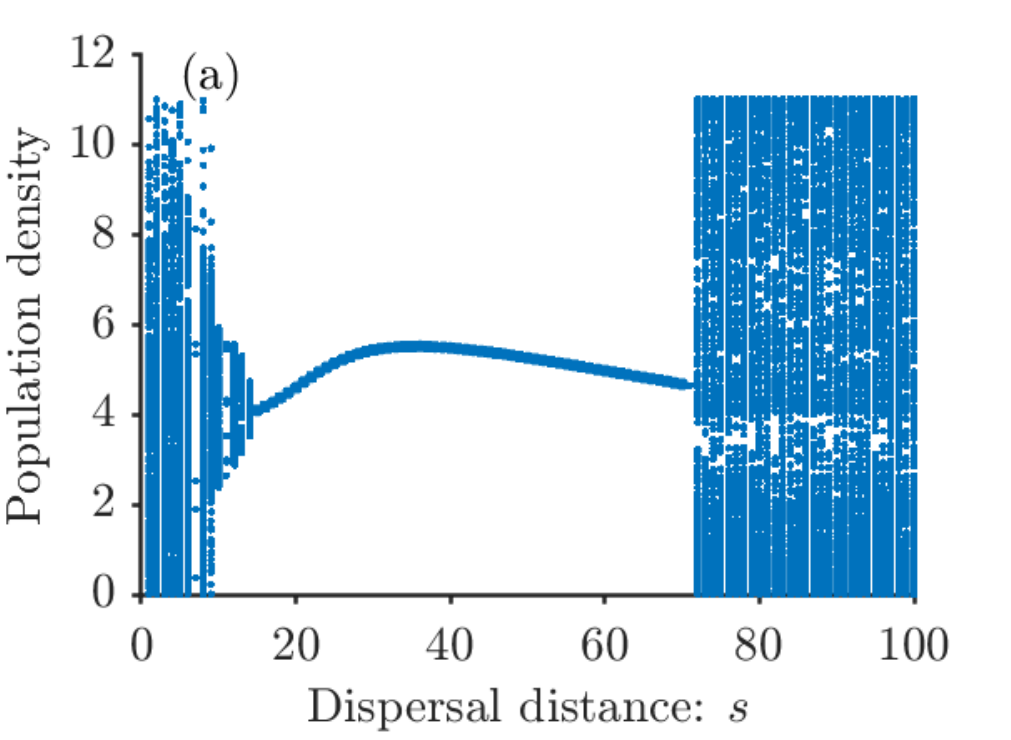}}
\subcaptionbox*{}{\includegraphics[scale=0.5]{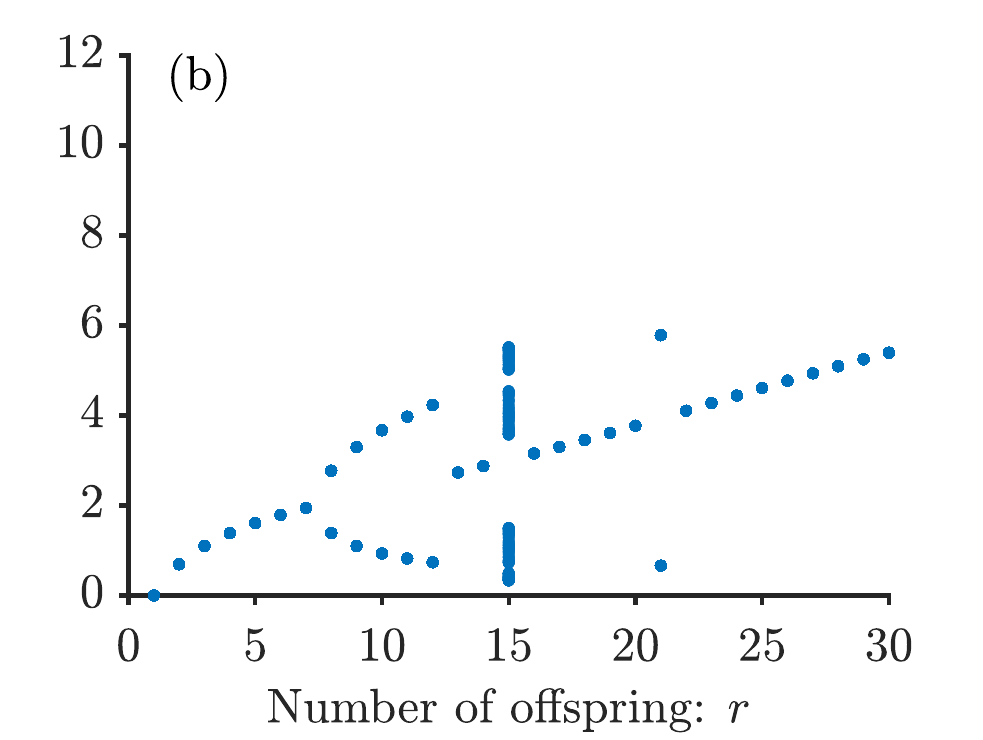}}
\caption{Bifurcation diagram for the long-range dispersal approximation, given in equation \eqref{eq:twopatch}.  In (a) we fix $r$ to 30,  set $D=201$, vary $s$, and iterate through $x_{1,t}$ and $x_{2,t}$ for 5000 time steps and plot last 500 time steps of ($x_{1,t}$+$x_{2,t}$)/2.  In (b) we fix $s$ to 45 and $D$ to 201, vary $r$, and iterate through $x_{1,t}$ and $x_{2,t}$ for 5000 time steps and plot last 500 time steps of ($x_{1,t}$+$x_{2,t}$)/2.  \label{fig:bif-twopatch}}

\end{figure}

For the long range dispersal approximation,  we produced bifurcation plots with both $s$ and $r$ as the bifurcation parameter.  In Figure \ref{fig:bif-twopatch} (a), we see that for $s$ between 40 and 50,  the approximation has both quantitatively and qualitatively the same behaviour as the individual-based model: the population is stable and has a size around 4. However,  in contrast to the individual-based model, the long-range dispersal approximation exhibits a stable population already at $s=15$ and up until $s=70$.   

\subsection{Stability analysis} \label{sec:analysis_stab}
\begin{figure}[H]
\subcaptionbox*{}{\includegraphics[scale=0.5]{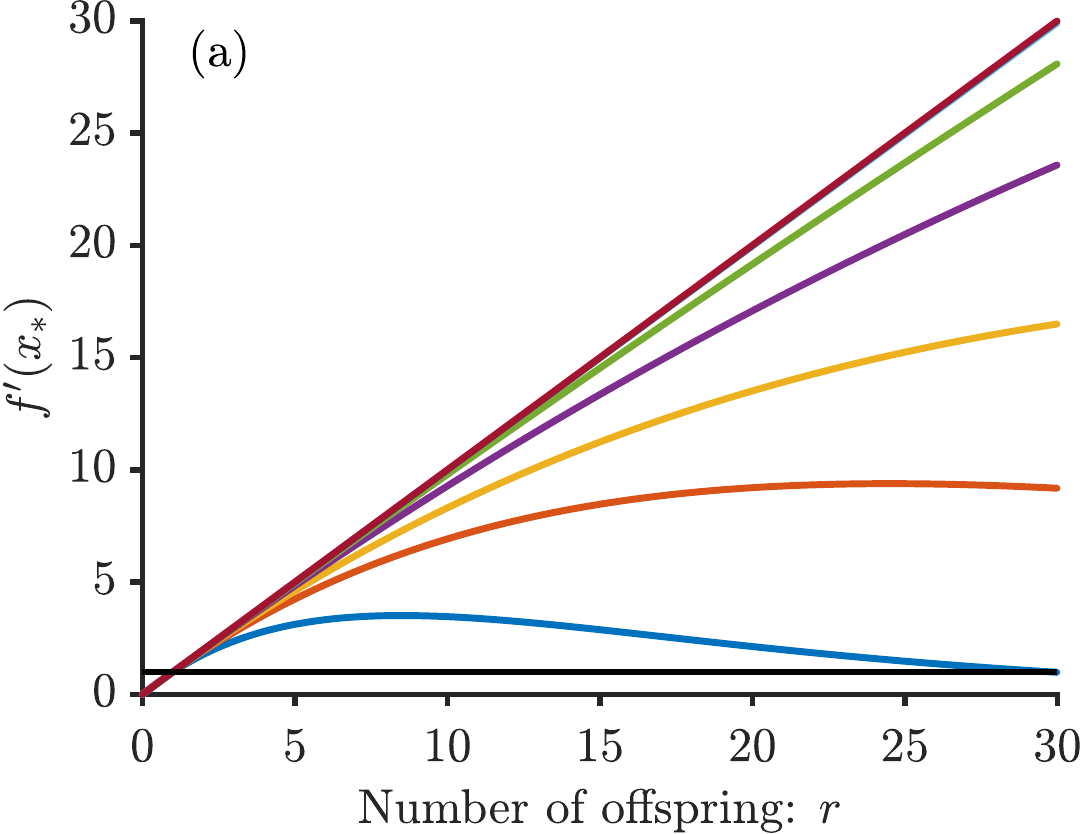}}
\subcaptionbox*{}{\includegraphics[scale=0.5]{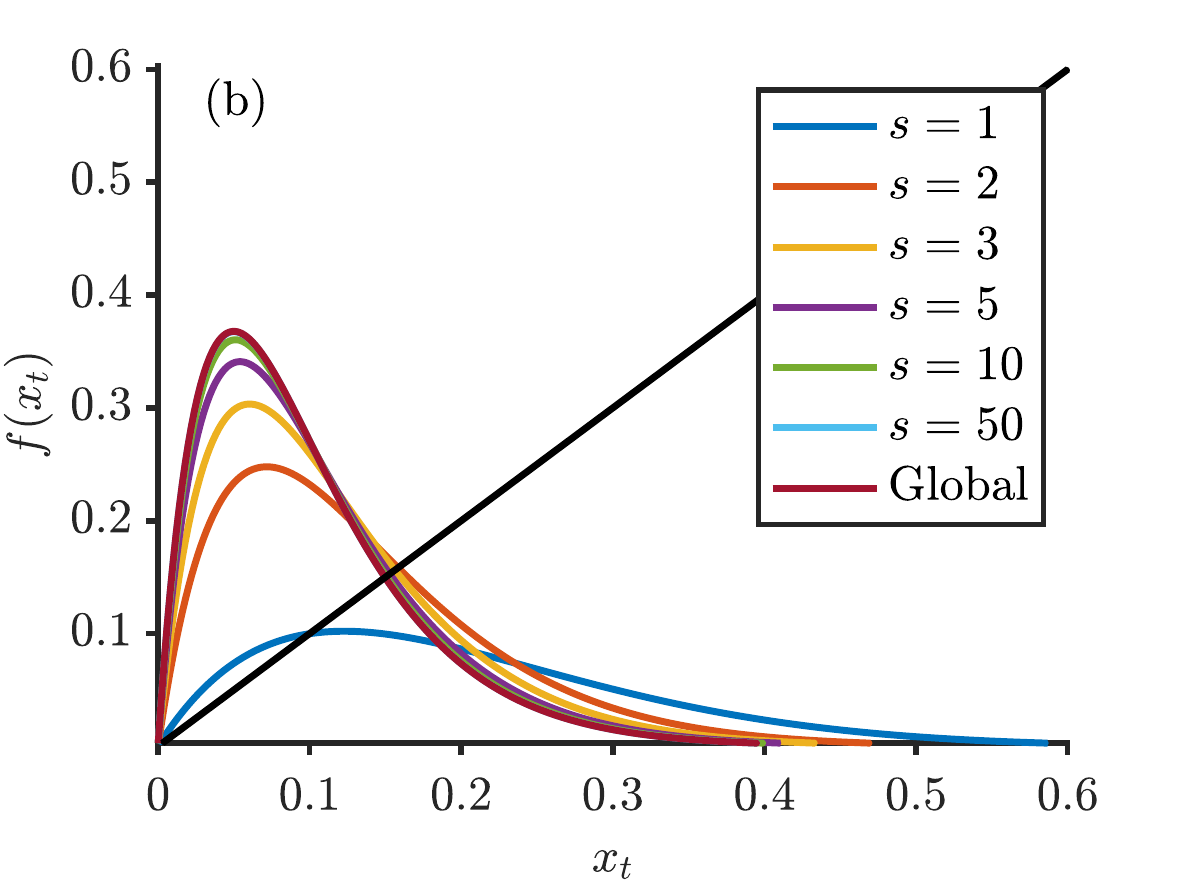}}
\caption{(a) Plot of $f'(x_*)$, for $x_*=0$ for various values of $s$ (see legend) . The model denoted global approximation is $f(x_t)=rx_t e^{-rx_t}$, i.e. equation \eqref{eq:meanfieldprevious}. The black line shows $f'(x_*)=1$.  (b) Plot of the approximation in equation \eqref{mainapprox} for various values of $s$ (see legend) and for $r=20$. The model denoted Global is $f(x_t)=rx_t e^{-rx_t}$, i.e. equation \eqref{eq:meanfieldprevious}). The black line shows $x_{t+1}=x_t$.}  \label{fig:f_prim}
\end{figure}

One important property of the individual-based model, that is not captured in the global approximation equation, is that the population goes extinct for some values of $r$, both when $s$ is very local ($s=1$) and when $s$ is global,as seen in Figure \ref{fig:bifmodel} (a) and (f). We want to find for which values of $r$ this happens for the local correlation approximation. To do so, we first observe that $x_*=0$ is always a steady state for equation \eqref{mainapprox}.  We then check the stability for the steady state $x_*=0$. Differentiating $f(x_*)$ we obtain: 
\begin{dmath}
f'(x_*)=r\left(\frac{(2s+1)^2-1}{(2s+1)^2}\right)^{r-1}\left(x_*\left(\left(\frac{(2s+1)^2-1}{(2s+1)^2}\right)^r-1\right)+1\right)^{(2s+1)^2-2}\left((2s+1)^2x_*\left(\left(\frac{(2s+1)^2-1}{(2s+1)^2}\right)^r-1\right)+1\right)
\end{dmath}
When  $x_*=0$,  we have
\begin{equation}
f'(0)=r\left(\frac{(2s+1)^2-1}{(2s+1)^2}\right)^{r-1}= r\left(1-\frac{1}{(2s+1)^2}\right)^{r-1}.
\end{equation} 
The stability condition $|f'(x_*)|<1$ for $x_*=0$ yields
\begin{equation} \label{extinct_ineq}
\left| r\left(1-\frac{1}{(2s+1)^2}\right)^{r-1} \right| <1,
\end{equation}
which, for $s=1$, is true for $1<r<30$. Thus, for our local correlation approximation, the population goes extinct when $r>30$, which is seen in Figure \ref{fig:bifapprox} (a). This is a slightly larger value than what is seen in bifurcation diagram of the individual-based model (Figure \ref{fig:bifmodel} (a)), where, after $t=5000$ time steps, extinction is observed already for $r=23$.  When $s$ increases, the extinction value for $r$ in the local correlation approximation increases as well. In Figure \ref{fig:f_prim} (a), we see how the value of $s$ affects $f'(x_*)$ when $x_*=0$. For $s=2$ e.g., $r$ needs to be larger than $117$ for the population to go extinct. For global dispersal, we have that 
\begin{equation}
\lim_{s \to \infty}f'(0) = \lim_{s \to \infty} r\left(1-\frac{1}{(2s+1)^2}\right)^{r-1}= r,
\end{equation}
and thus the inequality \eqref{extinct_ineq} is only fulfilled for $r<1$. This means that the local correlation approximation never result in extinction for global dispersal. In Figure \ref{fig:f_prim} below, we see a graphical illustration of this: when $s=1$, $f'(x_*)$ crosses the line $y=1$ when $r=30$. As $s$ increases, the slope of $f'(x_*)$ increases. For the global approximation equation, $f'(x_*)=r$, and will thus never cross the line $y=1$. 
In Figure \ref{fig:f_prim} (b) we see $f(x_t)$ as a function of $x_t$ when $r=20$ and for different values of $s$.  In this cobweb diagram,  we see that the population dynamics becomes unstable as $s$ increases.

Now for our local correlation approximation with $R_{ij}^t \sim \operatorname{Bin}(r/q,q)$, given by equation \eqref{eq:popdyn_rand}, $x_*=0$ will also be a steady state,  and $f'(x_*)|_{x^*=0}$ will be given by 

\begin{equation}
f'(x_*)|_{x^*=0}=r\left(1- \frac{q}{(2s+1)^2}\right)^{r/p-1}.
\end{equation}
The stability of $x_*=0$ of the local correlation approximation for stochastic reproductive rate will thus be affected by the value of $q$, and for smaller values, the population will become extinct for higher $r$ values. 

\subsection{General extinction probability} \label{sec:analysis_ext}
\begin{figure}[H]
\centering
\includegraphics[scale=0.5]{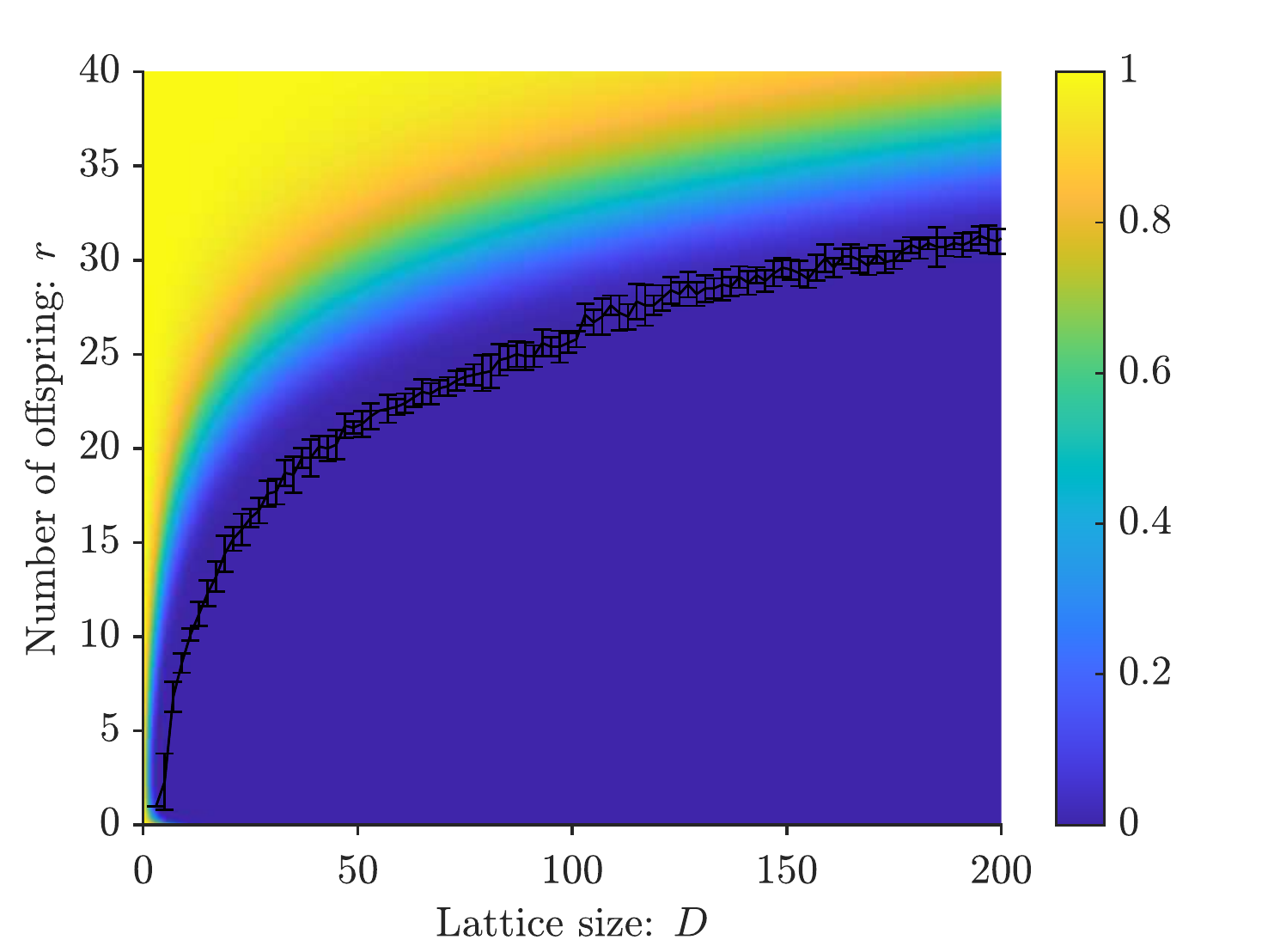}
\caption{Probability of going extinct in the next time step for different values of $D$ and $r$. The heat map is produced from equation \eqref{eq:prob}. The black line and the error bar show for each value of $D$,  the mean and standard deviation of the lowest value of $r$ for which the population was extinct after 5000 time steps for 10 simulation of each parameter set of the individual-based model.  \label{fig:extinctionprob}} 
\end{figure}

The individual-based model clearly displays population extinction for some values of $r$ when dispersal is global, as seen in the bifurcation diagram in Figure \ref{fig:bifmodel}(f). However, this extinction is not due to the stability of the stationary point $x_*=0$, but due to the stochasticity of the individual-based model: when $r$ is sufficiently large there is always a probability that all sites get overcrowded in one generation and the population goes extinct. We now want to find for which values of $r$ this happens. To find an expression for this, we make use of the different life-cycle stages within one time step: the population before competition, the population after competition but before reproduction, and the population after reproduction. Consider the individual-based model with  $n=D \times D$ resource sites. For the population to survive after competition, at least one resource site must contain exactly one individual (in case of scramble competition).  Let $x_{t}$ be the number of individuals before competition and reproduction at generation $t$, and $x_{t'}$ be the number of individuals after reproduction in the same time step. When dispersal is global,  the probability that a particular resource site contains exactly one individual, $P(X=1)$, will be a function of $x_t$, $p_1(x_t)$, since $X$ is Poisson distributed with mean $x_t/n$. We can thus write
\begin{equation}
p_1(x_t)= \frac{x_t}{n}e^{-\frac{x_t}{n}}.
\end{equation}
$p_1(x_t)$ will be maximized at $x_t=n$: 
\begin{equation}
p_1(n)= \frac{n}{n}e^{-\frac{n}{n}}=e^{-1}=p_1^*.
\end{equation}
Thus $p_1^*>p_1(x_t)$  $\forall x_t$.  When this is maximized,  there will be $e^{-1}\cdot n$ individuals that survived after competition,  meaning that, if the reproductive rate is constant,  there will be  
\begin{equation}
x_{t'}= n r e^{-1}
\end{equation}
individuals after reproduction. Thus, the probability for a particular resource site containing exactly one individual after reproduction
\begin{equation}
p_1(x_{t'})= \frac{n r e^{-1}}{n}e^{-\frac{n r e^{-1}}{n}}=  r e^{-1}e^{-r e^{-1}}.
\end{equation}
Then the probability that none of the resource sites has exactly one individual is given by
\begin{equation} \label{eq:prob}
p_{\text{extinct}}^* = (1-p_1(x_{t'}))^n=(1-  r e^{-1}e^{-r e^{-1}})^n.
\end{equation}
Since $p_1^*>p_1(x_t)$  $\forall x_t$,  $p_{\text{extinct}}^*<p_{\text{extinct}}(x_{t'})$  $\forall x_{t}$. Thus, this is the minimum probability of going extinct for each generation. In Figure \ref{fig:extinctionprob} we see how this probability is affected for different values of $D$ and $r$, and how this is connected to the actual simulations of the individual-based model (the black line and error bar). Recall that $p_{\text{extinct}}$ is the minimum probability of going extinct \textit{at each time step}, and thus when simulating over many generations (below for 5000 generations),  the population will go extinct when $p_{\text{extinct}}$ is quite low ($\approx 0.1$). 

\subsection{Convergence to the global approximation model} \label{sec:analysis_conv}

When deriving the local correlation approximation, we made the assumption that  $s<<D$. However Figures \ref{fig:bifapprox} and \ref{fig:f_prim} (a) suggest that as $s$ increases the equation \eqref{mainapprox} comes increasingly close to the global approximation (equation \eqref{eq:meanfieldprevious}).  Indeed, when $s=50$ in Figure \ref{fig:f_prim} (b), the curve defined by equation \eqref{mainapprox} is indistinguishable from the global approximation equation. These numerical results suggest that when $s$ is large, the local correlation approximation reproduces the dynamics of the global approximation, even though the $s << D$ assumption that was made for the approximation is not fulfilled. To rigorously determine whether this result holds, we need to check if our population dynamics model in equation \eqref{mainapprox} approaches the original global approximation as $s \to \infty$. This is equivalent to showing that the limit $\lim_{s \to \infty}Y \rightarrow \text{Po}(r x_t)$,  in distribution.  Here $Y$ is the number of offspring of all of the parents in the $(2s+1) \times (2s+1)$ Moore neighbourhood of a focal site that disperse to the focal site.  Thus, $\lim_{s \	to \infty}Y \rightarrow \text{Po}(r x_t)$,  means that when dispersal is global the number of offspring from the whole lattice that end up in a particular site, is Poisson distributed with mean $rx_t$. To show this limit, we can make use of the following theorem:

\begin{theorem} \label{theoremx} ({\bf Yannaros 1991})
Let $\xi_1, \xi_2, ...$ be i.i.d. Bernoulli variables with $P(\xi_j=1)=p$ and $N$ a non-negative integer valued random variable which is independent of the $\xi_j$'s. Let $S_N=\sum_{j=1}^N \xi_j$, then
\begin{equation}
d(S_N, U_{pa}) \leq \min \big\{\frac{p}{2\sqrt{1-p}}, p \Ex[ 1- e^{-pN}]\big\}+	\min\big\{p \Ex [|N- a|], \frac{1}{2}\sqrt{p\frac{\Ex [N-a]^2}{a}}\big\},
\end{equation}

for any $a>0$. Here $d$ is the total variation distance and $U_\lambda$ is the Possion distribution with mean $\lambda$. 
\end{theorem}  

The total variation distance is a measure of the closeness of two probability distributions.  Intuitively,  the total variation distance between two probability distribution is the largest possible difference that the two probability distributions can assign to the same event.  Thus, a total variation distance equal to zero,  indicates that the probability distributions are the same.

\begin{proposition} \label{convergeprop}
As $s \to \infty$,
$Y$ converges in distribution to a Poisson distribution with mean $r x_t$.
\end{proposition}

\begin{proof}
By choosing $a$ in the right way, we can use theorem \ref{theoremx}  to prove proposition \ref{convergeprop}. 
First, recall that  
\begin{equation}
Y=\sum_{j=1}^{F \cdot r} O_j.
\end{equation}
Now since $O_j$ are i.i.d. Bernoulli variables with $P(O_j=1)=\frac{1}{(2s+1)^2}$, we can equate these with  $\xi_j$, where $p=\frac{1}{(2s+1)^2}$. Likewise, we can equate $N =r \cdot F$, since it is an integer valued random variable which is independent on the $O_j$'s. Thus $Y$ is equal to $S_N$.  We can now find an $a>0$ such that $d(Y, U_{\frac{a}{(2s+1)^2}}) \to 0$ as $s \to \infty$.

Now choose $a$ to be the expected value of $r \cdot F$, i.e. $a= \Ex [r \cdot F]=r x_t (2s+1)^2$. Since $p=\frac{1}{(2s+1)^2}$, $U_{pa}=U_{r x_t}$.
We want to compute 
\begin{equation}
\lim_{s \to \infty}d(Y_{r\cdot F}, U_{r x_t}).
\end{equation}
By applying Theorem \ref{theoremx} we know that
\begin{equation} 
\begin{aligned}
\lim_{s \to \infty}d(Y_{r\cdot F}, U_{r x_t}) \leq  \lim_{s \to \infty} \Big( \min \big\{\frac{\frac{1}{(2s+1)^2}}{2\sqrt{1-(\frac{1}{(2s+1)^2})}}, \frac{1}{(2s+1)^2} \Ex [1- e^{-\frac{r \cdot F}{(2s+1)^2}}]\big\}
\\
+\min\big\{\frac{1}{(2s+1)^2} \Ex [|r \cdot F- a|], \frac{1}{2}\sqrt{p\frac{\Ex [r \cdot F-a]^2}{a}}\big\} \Big).
\end{aligned}
\end{equation}

Since $\frac{\frac{1}{(2s+1)^2}}{2\sqrt{1-(\frac{1}{(2s+1)^2})}} \to 0$ as $s \to \infty$, the first term, $\min \big\{\frac{\frac{1}{(2s+1)^2}}{2\sqrt{1-(\frac{1}{(2s+1)^2})}}, \frac{1}{(2s+1)^2} \Ex [1- e^{-\frac{r \cdot F}{(2s+1)^2}}]\big\}$, will approach zero as $s \to \infty$.

For the second term, note that by choosing $a= E (r \cdot F)=r x_t (2s+1)^2$ we get that $\Ex (r \cdot F-a)^2= \Ex [r \cdot F-\Ex [r \cdot F]]=\Var(r \cdot F)=(2s+1)^2x_t(1-x_t)$. So the second term will be 

\[\min\big\{\frac{1}{(2s+1)^2} \Ex [|r \cdot F- a|], \frac{1}{2}\sqrt{\frac{1}{(2s+1)^2}\frac{(2s+1)^2x_t(1-x_t)}{r x_t (2s+1)^2}}\big\}. \]

Since $\frac{1}{2}\sqrt{\frac{1}{(2s+1)^2}\frac{(2s+1)^2x_t(1-x_t)}{r x_t (2s+1)^2}}=\frac{1}{2}\sqrt{\frac{(1-x_t)}{r (2s+1)^2}} \to 0$ as $s \to \infty$, we get that the second term goes to zero as $s \to \infty$.  
\end{proof}


\section{Discussion}

With increasing computational capacity during the last 30 years,  there has been a shift from 'top-down' models in ecology,  describing the overall population structure, to 'bottom up' individual-based models capturing the local behaviour of the population. In light of this shift,  a question arises as to whether it is possible to approximate individual-based models with a small number of analytically tractable equations, describing the overall population dynamics.  For individual-based models in continuous time or space or in which evolution in space is 'smooth' \citep{touboul2014propagation, surendran2018spatial, omelyan2019spatially, wallhead2008spatially}, such as interacting particle systems and spatial point processes, there are a range of analytical approximations available \citep{patterson2020probabilistic,bolker1997using,oelschlager1989derivation}.  For models which are, what \cite{berec2002techniques} refer to D-space,  D-time, like the one we study here, analytical approaches have proved more limited. 

The current work has started to look at ways into this problem, by studying a specific model which has both spatial patterning and large oscillations over a single time step. The spirit of our work here is perhaps similar to how Levin first looked at  problems for spatially discrete, but smoothly changing processes \cite{durrett1994importance,levin1992problem}, which eventually led to the more rigorous treatments referenced above. 

We have identified two novel approaches here. The first is a local correlation approximation, which led to equation \ref{mainapprox}. This provided a one-dimensional dynamical system, with two parameters – $s$ for dispersal range and $r$ for reproduction number – which allows us to the draw a bifurcation diagram illustrating how these parameters determine stability. In this way, we show that our individual-based model can be reasonably approximated  for small dispersal distances (compare Figures \ref{fig:bifmodel} (a) and \ref{fig:bifapprox} (a)). The second in the long-range approximation, consisting of two coupled maps, that captures some aspects of the simulation when $s$ is large. Again, this allows us to draw a bifurcation diagram (Figure \ref{fig:bif-twopatch}). 

These two approximations shed light on a common discrepancy between theoretical and empirical ecology: theoretical models, as suggested by \cite{may1974biological} often show chaotic behaviour, but in real ecological time series chaos is infrequent. Often, two explanations are given for this inconstancy: populations stabilize due to demographic stochasticity \citep{jaggi2001incorporating}, or they are said to stabilize due to dispersal.  With our local correlation approximation, we show that a deterministic difference equation have the same stabilizing effect on the population dynamics, thus  demonstrating local dispersal as the stabilizing effect in this case. 

The long-range approximation establishes a link between our model and two patch metapopulations discussed by \cite{hastings1993complex} and \cite{gyllenberg1993does}. Indeed, when dispersal is large in the individual-based model, the population dynamics can be well approximated with a two patch system, as seen in Figures \ref{fig:IBM_bif_s} and \ref{fig:bif-twopatch}.  Long-range dispersal has been shown to play an important role for seed dispersal and is key for understanding plant population dynamics and community composition \citep{levey2008modelling}.


The work presented here can be extended in several natural directions. First, our characterization of the spatial dynamics could be extended to other survival functions, like those examined by \cite{brannstrom2005role} for the non-spatial case. 
Second, our two approximation techniques can be considered also for situation with two sexes, or several populations. A better approximation should also result if correlations are tracked also between generation, as opposed to currently within a single generation. The possibility to approximate spatially explicit individual-based models with analytically tractable dynamical systems does not only facilitate systematic exploration of parameter space by speeding up numerical investigations but can potentially reveal deeper insights about the role of space in population dynamics. Also, as we showed here, this approximation also helped us establish an extinction probability curve a result which could be extended further. We believe that the results presented here has contributed to this understanding and that future efforts will help to elucidate how and why spatial structure affects the dynamics of populations.

It is also worth pausing to think about the limits which our results point towards. Our starting question was the degree to which we could analyse the complex dynamics generated by a relatively simple individual-based model. The answer is that, through our two approximations and statistical measurements, we can get some insight in to those dynamics, but this is not a complete mathematical explanation of the phenomena that arise, i.e. of the patterns we see in Figure \ref{fig:snapshots} and \ref{fig:D201_snapshot}. In particular, neither of the approximations work particularly well for $s$ between 3 and 10 in Figure \ref{fig:snapshots} . This is in itself an interesting observation. Our model is far from being the most complex bottom-up model, yet there seems to be a limit to what we can currently say about it using mathematical approaches. Our approach goes beyond, for example, mean-field (well-mixed) equations, yet ultimately it still fails to capture the richness of the individual-based model. While we have little doubt that approximating stochastic individual-based models with analytical approaches is a useful contribution, it might also be that there are quite strong limits to what can readily be achieved with a mathematical analysis.


\newpage

%

\newpage



%
%

\bibliography{references_approximation}

%
\newpage

\setcounter{figure}{0}
\renewcommand{\figurename}{Fig.}
\renewcommand{\thefigure}{S\arabic{figure}}
\renewcommand{\theequation}{S\arabic{equation}}
\renewcommand{\thetable}{S\arabic{table}}
\renewcommand{\thefigure}{S\arabic{figure}}
\setcounter{equation}{0}
\setcounter{table}{0}
\setcounter{figure}{0}

\subsection*{S1: Moran's Index} 

The patterns of the individual-model leads us to explore more thoroughly the role of spatial structure in the model.  Here, we study the spatial autocorrelation. This can be measured by the global Moran's Index. The global Moran's $I$ is defined by the equation 	
\begin{equation}
I = \frac{N}{W}\frac{\sum_i \sum_j w_{ij}(x_i-\bar{x})(x_j-\bar{x})}{\sum_i(x_i-\bar{x})^2},
\end{equation}
where $N$ is the total number of sites in the lattice, indexed by $i$ and $j$; $\bar{x}$ is the average population density;  $x_i$ is the number of individuals at site $i$; $w_{ij}$ is the connectivity matrix; and $W$ is the total number of connections between sites. More specifically $w_{ij}$ is an $N \times N$ matrix, stating which sites are connected to each other in a von Neumann neighbourhood: if $i \neq j$ and site $i$ is in the von Neumann neighbourhood of site $j$, then $w_{ij}=1$, otherwise $w_{ij}=0$. Since each site is connected to exactly four neighbouring sites, $W$ equals 4 times $N$.  Moran's $I$ takes values between -1 and 1 and its expected value in the absence of spatial autocorrelation is $E(I)= -1/(N-1)$, which is close to zero when $N$ is large. $I < E(I)$ indicates negative spatial autocorrelation (dispersion) and $I> E(I)$ indicates positive spatial autocorrelation (clustering). Examples of how the spatial configuration affects Moran's $I$ is seen in Figure \ref{fig:moran_ex} below.

\begin{figure}[H] \centering
  \definecolor{mycolor}{RGB}{238,233,233}
\definecolor{mygreen}{RGB}{0,0,0}
\definecolor{myred}{RGB}{0,0,0}
\begin{tikzpicture}

\draw [fill=gray, opacity=0.7] (0,1.5) -- (0,2) -- (0.5,2) -- (0.5,1.5) -- (0,1.5);
\draw [fill=gray, opacity=0.7] (1,1.5) -- (1,2) -- (1.5,2) -- (1.5,1.5) -- (1,1.5);

\draw [fill=gray, opacity=0.7] (0.5,1) -- (0.5,1.5) -- (1,1.5) -- (1,1) -- (0.5,1);
\draw [fill=gray, opacity=0.7] (1.5,1) -- (1.5,1.5) -- (2,1.5) -- (2,1) -- (1.5,1);

\draw [fill=gray, opacity=0.7] (0,0.5) -- (0,1) -- (0.5,1) -- (0.5,0.5) -- (0,0.5);
\draw [fill=gray, opacity=0.7] (1,0.5) -- (1,1) -- (1.5,1) -- (1.5,0.5) -- (1,0.5);

\draw [fill=gray, opacity=0.7] (0.5,0) -- (0.5,0.5) -- (1,0.5) -- (1,0) -- (0.5,0);
\draw [fill=gray, opacity=0.7] (1.5,0) -- (1.5,0.5) -- (2,0.5) -- (2,0) -- (1.5,0);
\draw[step=0.5cm,color=black] (0,0) grid (2,2);

\draw [fill=gray, opacity=0.7] (3,1.5) -- (3,2) -- (3.5,2) -- (3.5,1.5) -- (3,1.5);
\draw [fill=gray, opacity=0.7] (4.5,1.5) -- (4.5,2) -- (5,2) -- (5,1.5) -- (4.5,1.5);

\draw [fill=gray, opacity=0.7] (4,1) -- (4,1.5) -- (4.5,1.5) -- (4.5,1) -- (4,1);

\draw [fill=gray, opacity=0.7] (3,0.5) -- (3,1.0) -- (3.5,1) -- (3.5,0.5) -- (3,0.5);
\draw [fill=gray, opacity=0.7] (4,0.5) -- (4,1) -- (4.5,1) -- (4.5,0.5) -- (4,0.5);

\draw [fill=gray, opacity=0.7] (3,0) -- (3,0.5) -- (3.5,0.5) -- (3.5,0) -- (3,0);
\draw [fill=gray, opacity=0.7] (4.5,0) -- (4.5,0.5) -- (5,0.5) -- (5,0) -- (4.5,0);
\draw[step=0.5cm,color=black] (2.999,0) grid (5,2);

\draw [fill=gray, opacity=0.7] (7,0) -- (7,2) -- (8,2) -- (8,0) -- (7,0);
\draw[step=0.5cm,color=black] (5.999,0) grid (8,2);

\draw[thin,<->] (-0.2,-0.3) -- (8.2,-0.3);

\node (1) at (1, -0.5) {\scriptsize Moran's $I<0$};
\node (2) at (4, -0.5) {\scriptsize Moran's $I \approx 0$};
\node (2) at (7, -0.5) {\scriptsize Moran's $I>0$};

\end{tikzpicture}
  \caption{Spatial structure indicated by Moran's $I$. Negative Moran's $I$ indicates regularly spaced individuals and positive Moran's $I$ indicates clustering. Moran's $I$ close to zero indicates random a configuration.} 
  \label{fig:moran_ex}	
\end{figure}

\begin{figure}[H] \centering
\subcaptionbox*{}{\includegraphics[scale=0.5]{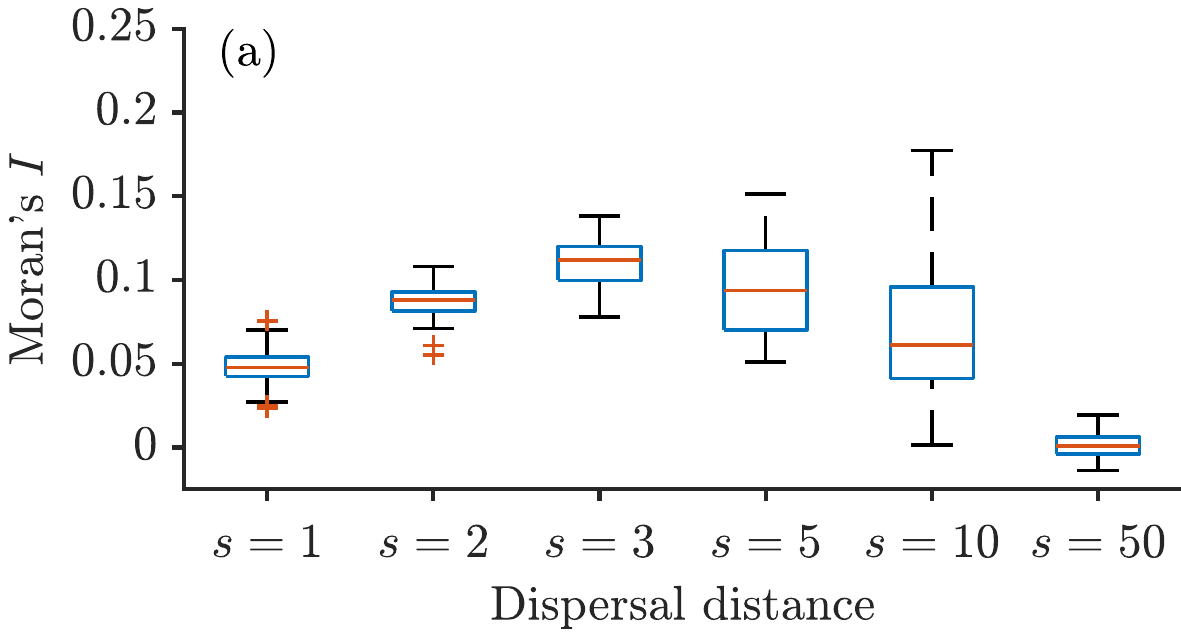}}
\subcaptionbox*{}{\includegraphics[scale=0.5]{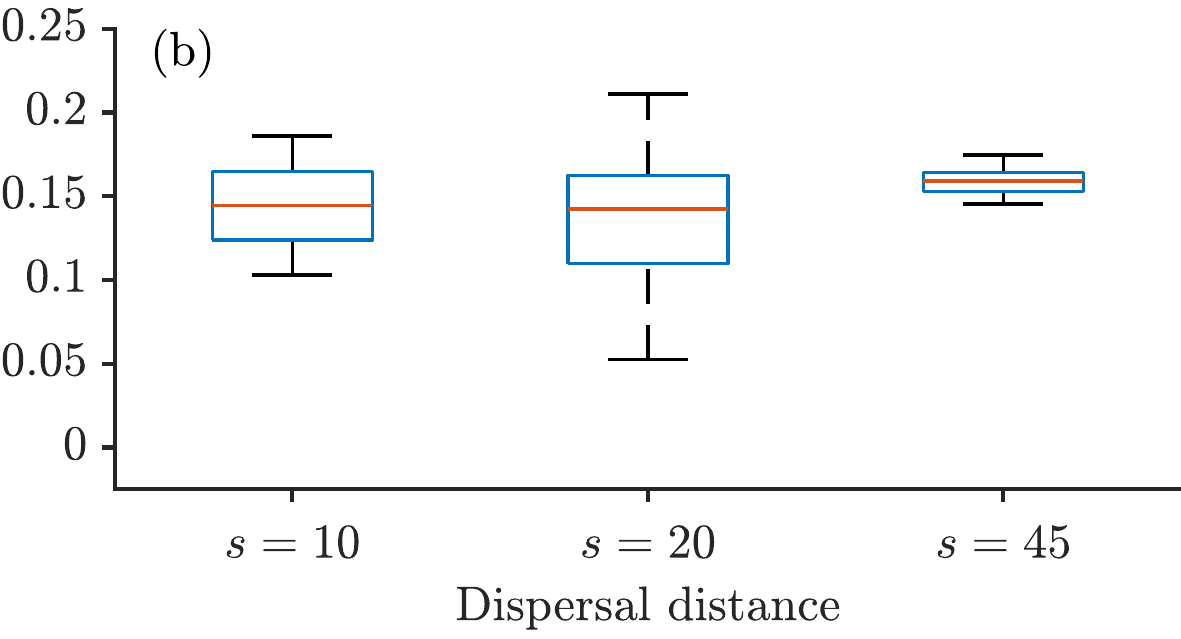}}
\caption{Box plot Moran's $I$ for 100 consecutive snapshots of the individual-based model simulated on a a) $101 \times 101$ lattice  with $r=20$, (the same parameter values as in Figure 2 in the main article) and b) $201 \times 201$ lattice with $r=30$, for different values of $s$ (the same parameter values as in Figure 4 in the main article).} \label{fig:morans}
\end{figure}

Figure \ref{fig:morans} (a) shows that, for small dispersal distances ($s=1$ and $s=2$) Moran's $I$ lies around 0.05 and 0.1, and the variation is small. This suggests that the population is close to randomly distributed, but there is a small indication of spatial clustering. There are not any big fluctuations in spatial figuration between the time steps. As $s$ increases, both the maximum value of the Moran's $I$ and the variation increases, indicating more evident spatial patterns, but also higher variations between time steps. 
When dispersal is global, i.e. $s=50$, the mean Moran's $I$ is close to zero, and the variations are small. Analysing the snapshots of the $201 \times 201$ lattice in Figure \ref{fig:morans} b), the median Moran's $I$ is roughly the same ($\approx 0.15$) for all three dispersal distances, but the variation is very high for $s=20$ and very low for $s=45$. 

\subsection*{S2: Illustration of 4TLQV method}

\begin{figure}[H] \centering
  \definecolor{mycolor}{RGB}{238,233,233}
\definecolor{mygreen}{RGB}{0,0,0}
\definecolor{myred}{RGB}{0,0,0}
\begin{tikzpicture}

\draw [thick] (0.5,0.5) -- (4.5,0.5) -- (4.5,4.5) -- (0.5,4.5) -- (0.5,0.5);

\draw [thick] (2.5,0.5) -- (2.5,4.5);

\draw [thick] (0.5,2.5) -- (4.5,2.5);

\draw [thick] (0.5,5.5) -- (4.5,5.5) -- (4.5,9.5) -- (0.5,9.5) -- (0.5,5.5);

\draw [thick] (0.5,7.5) -- (4.5,7.5);

\draw [thick] (2.5,5.5) -- (2.5,9.5);

\draw [thick] (5.5,0.5) -- (9.5,0.5) -- (9.5,4.5) -- (5.5,4.5) -- (5.5,0.5);

\draw [thick] (7.5,0.5) -- (7.5,4.5);

\draw [thick] (5.5,2.5) -- (9.5,2.5);

\draw [thick] (5.5,5.5) -- (9.5,5.5) -- (9.5,9.5) -- (5.5,9.5) -- (5.5,5.5);

\draw [thick] (5.5,7.5) -- (9.5,7.5);

\draw [thick] (7.5,5.5) -- (7.5,9.5);

\node (1) at (0.15, 1.5) {\Large b};
\node (2) at (0.15, 3.5) {\Large b};

\node (3) at (1.5, 4.85) {\Large b};
\node (4) at (3.5, 4.85) {\Large b};

\node (5) at (1.5, 1.5) {\Large -3};
\node (6) at (1.5, 3.5) {\Large +1};

\node (7) at (3.5, 1.5) {\Large +1};
\node (8) at (3.5, 3.5) {\Large +1};

\node (1) at (0.15, 6.5) {\Large b};
\node (2) at (0.15, 8.5) {\Large b};

\node (3) at (1.5, 9.85) {\Large b};
\node (4) at (3.5, 9.85) {\Large b};

\node (5) at (1.5, 6.5) {\Large +1};
\node (6) at (1.5, 8.5) {\Large -3};

\node (7) at (3.5, 6.5) {\Large +1};
\node (8) at (3.5, 8.5) {\Large +1};

\node (1) at (5.15, 6.5) {\Large b};
\node (2) at (5.15, 8.5) {\Large b};

\node (3) at (6.5, 9.85) {\Large b};
\node (4) at (8.5, 9.85) {\Large b};

\node (5) at (6.5, 6.5) {\Large +1};
\node (6) at (6.5, 8.5) {\Large+1};

\node (7) at (8.5, 6.5) {\Large +1};
\node (8) at (8.5, 8.5) {\Large -3};

\node (1) at (5.15, 1.5) {\Large b};
\node (2) at (5.15, 3.5) {\Large b};

\node (3) at (6.5, 4.85) {\Large b};
\node (4) at (8.5, 4.85) {\Large b};

\node (5) at (6.5, 1.5) {\Large +1};
\node (6) at (6.5, 3.5) {\Large +1};

\node (7) at (8.5, 1.5) {\Large -3};
\node (8) at (8.5, 3.5) {\Large +1};

\end{tikzpicture} 
  \caption{Illustration of the calculation for 4TLQV: the number of individuals in the '+1' blocks are summed and added together and the number of individuals in the '-3' block are summed, multiplied by 3 and subtracted. The result is then squared and an average of all four possibilities is taken.}
  \label{4tlqv_illustration}	
\end{figure}

\subsection*{S3: Stochastic reproductive rate}

In the individual based model in the main text we used a fixed $r$. Here we instead let $R_{ij}^t \sim  \operatorname{Bin}(r/q,q)$ (thus $\Ex[R_{ij}^t]=r$). We simulate this version of the individual based model for $q=0.1, 0.2$ and $0.5$ in the figures below.

\begin{figure}[H] \centering
\subcaptionbox*{}{\includegraphics[scale=0.48]{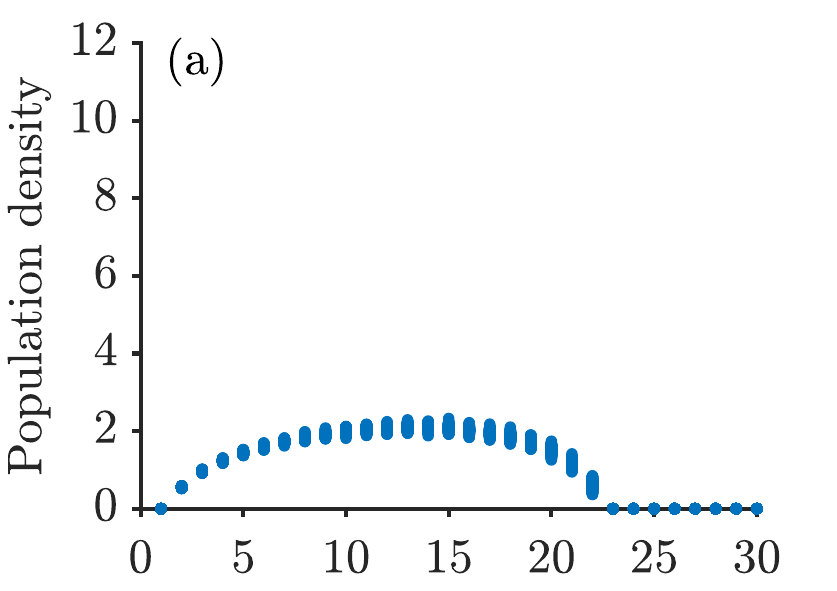}}
\subcaptionbox*{}{\includegraphics[scale=0.48]{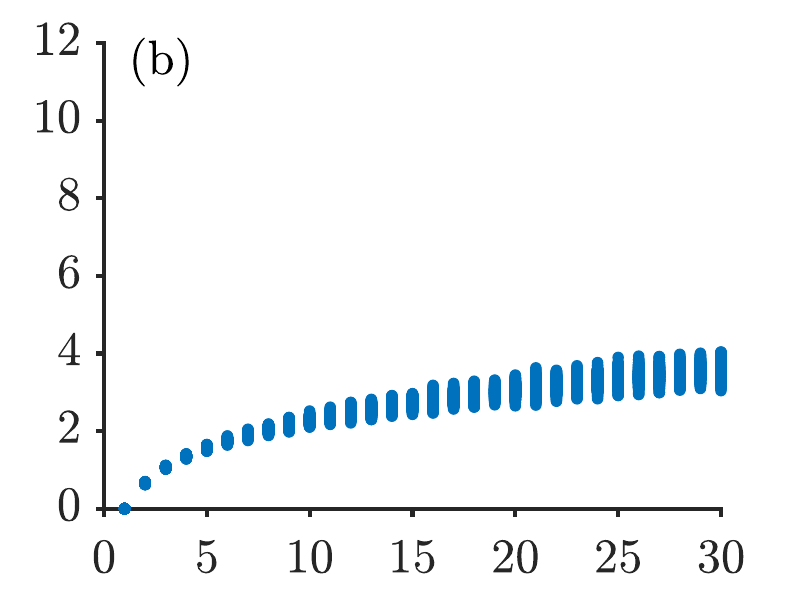}}
\subcaptionbox*{}{\includegraphics[scale=0.48]{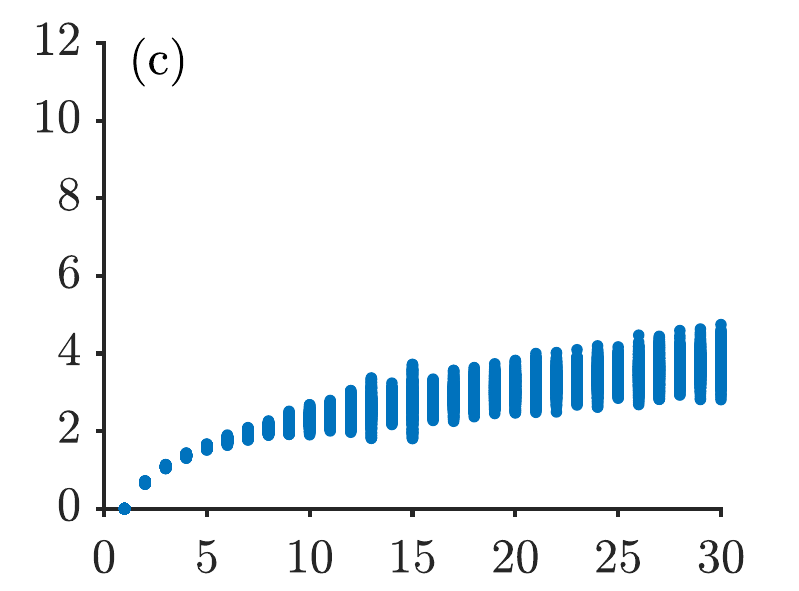}}
\subcaptionbox*{}{\includegraphics[scale=0.48]{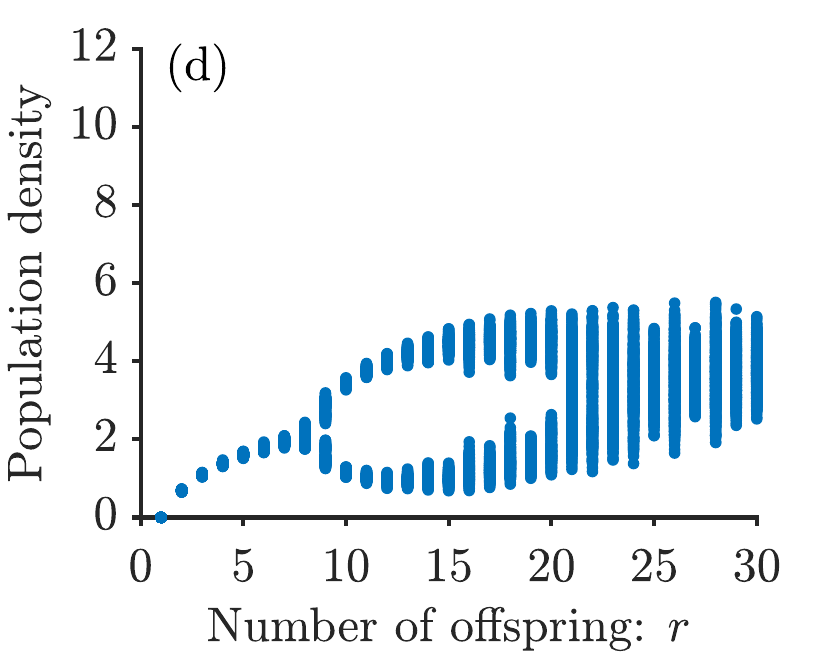}}
\subcaptionbox*{}{\includegraphics[scale=0.48]{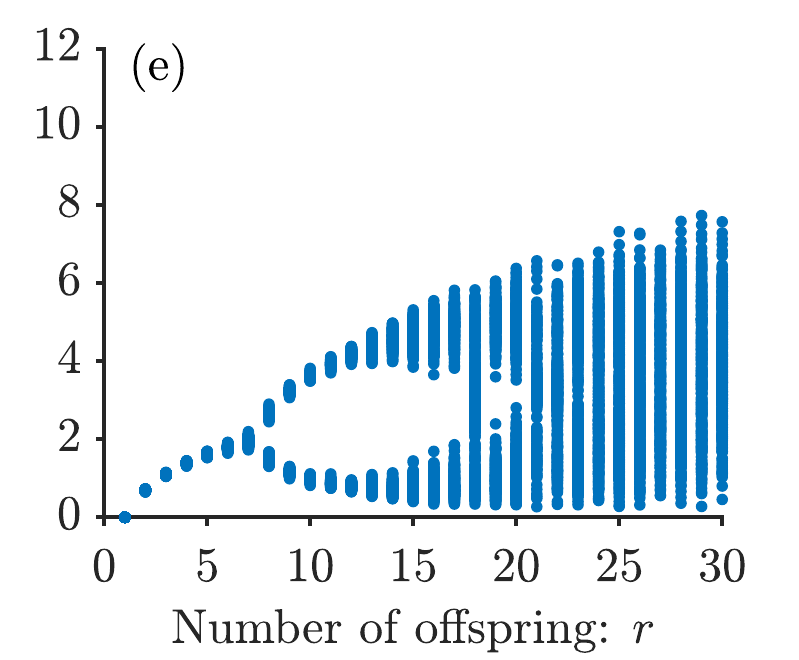}}
\subcaptionbox*{}{\includegraphics[scale=0.48]{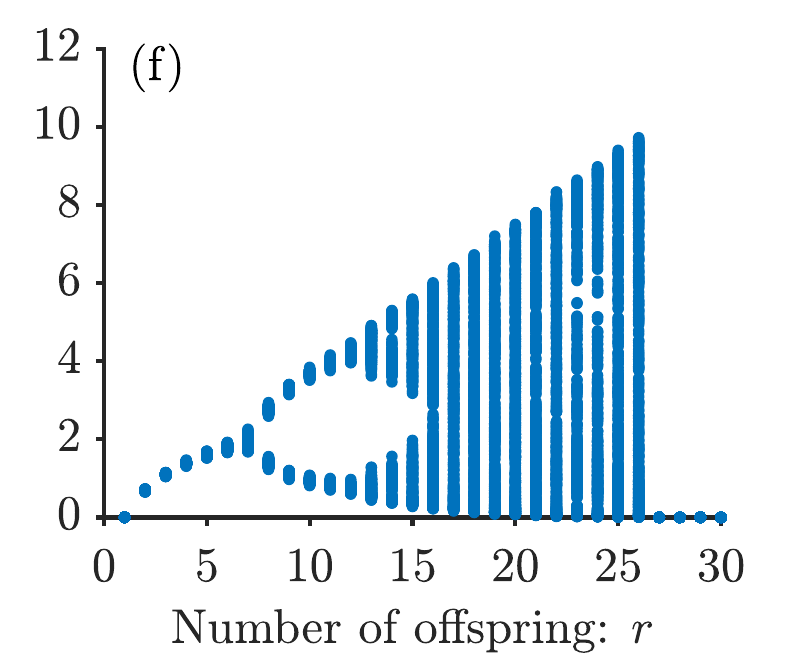}}
\caption{Bifurcation diagram for the individual-based model, with $\phi(k)$ given as Equation \eqref{eq:scramble_phi} in the main manuscript and $R_{ij}^t \sim  \operatorname{Bin}(r/q,q)$,  where $q=0.1$. In each subplot we vary $r$ and simulate the model for 5000 time steps and plot last 500 time steps. The population density is given by the number of individuals divided by the number of resource sites, and is counted after reproduction. We give results for (a) $s=1$ (very local dispersal) (b) $s=2$, (c) $s=3$, (d) $s=5$, (e) $s=10$ and (f) $s=50$  (global dispersal).  \label{fig:bifmodel_p01}}
\end{figure}

\begin{figure}[H] \centering
\subcaptionbox*{}{\includegraphics[scale=0.48]{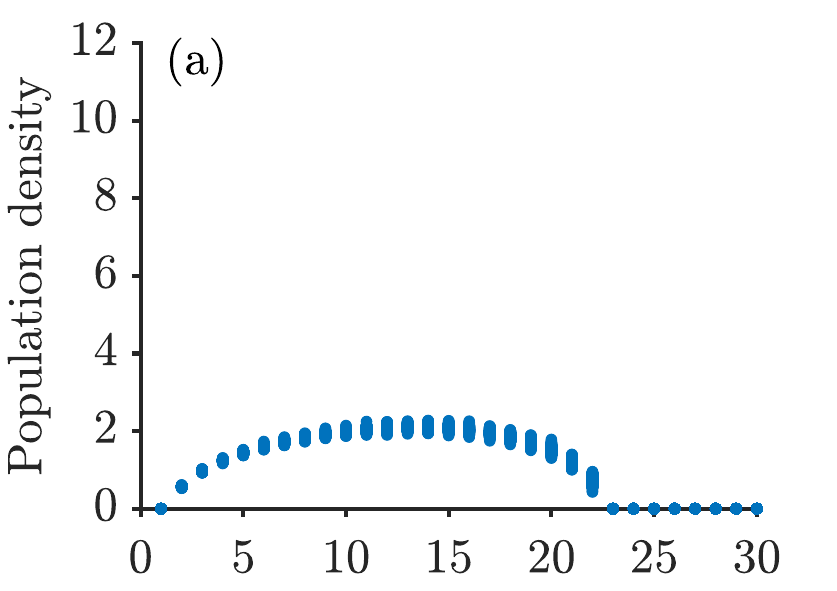}}
\subcaptionbox*{}{\includegraphics[scale=0.48]{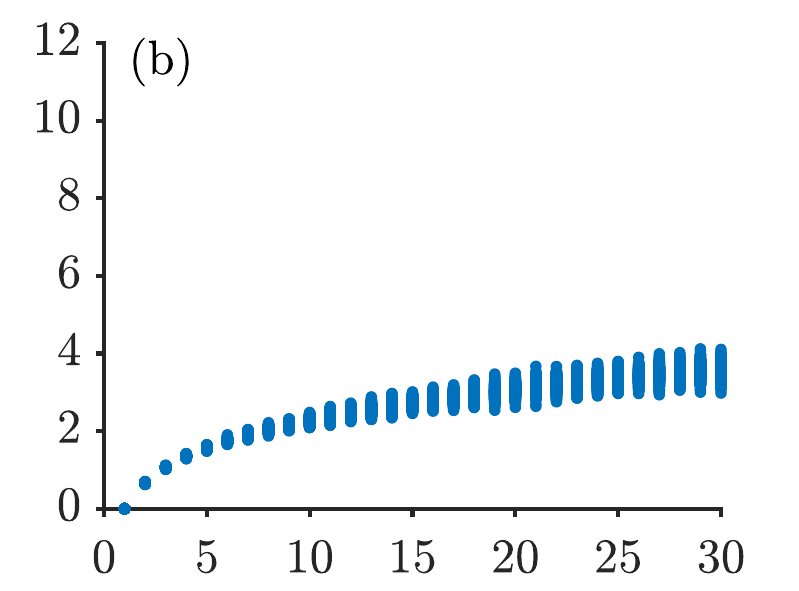}}
\subcaptionbox*{}{\includegraphics[scale=0.48]{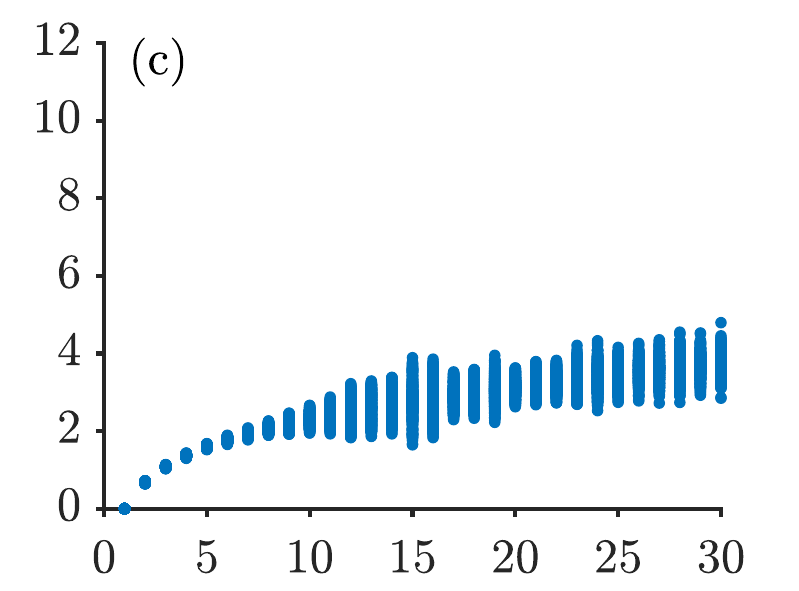}}
\subcaptionbox*{}{\includegraphics[scale=0.48]{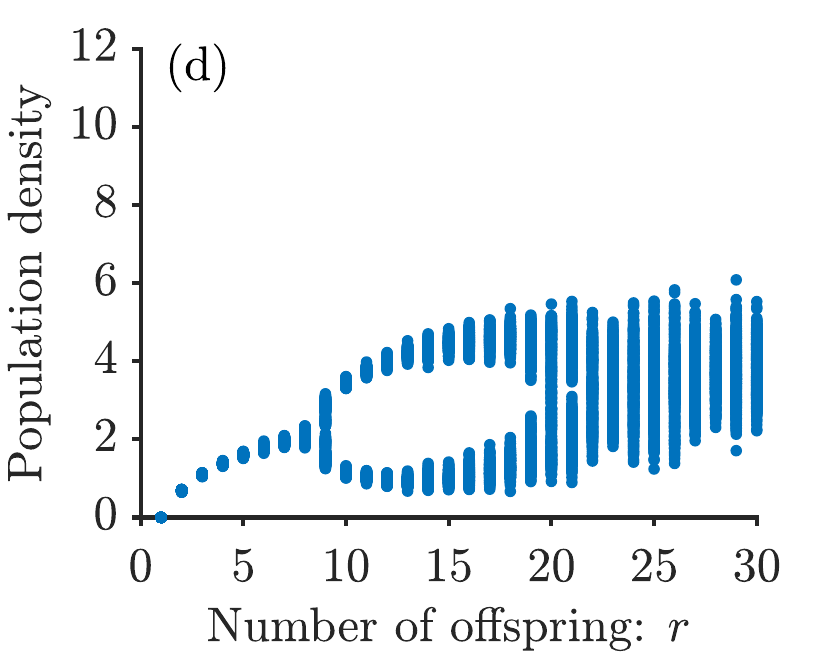}}
\subcaptionbox*{}{\includegraphics[scale=0.48]{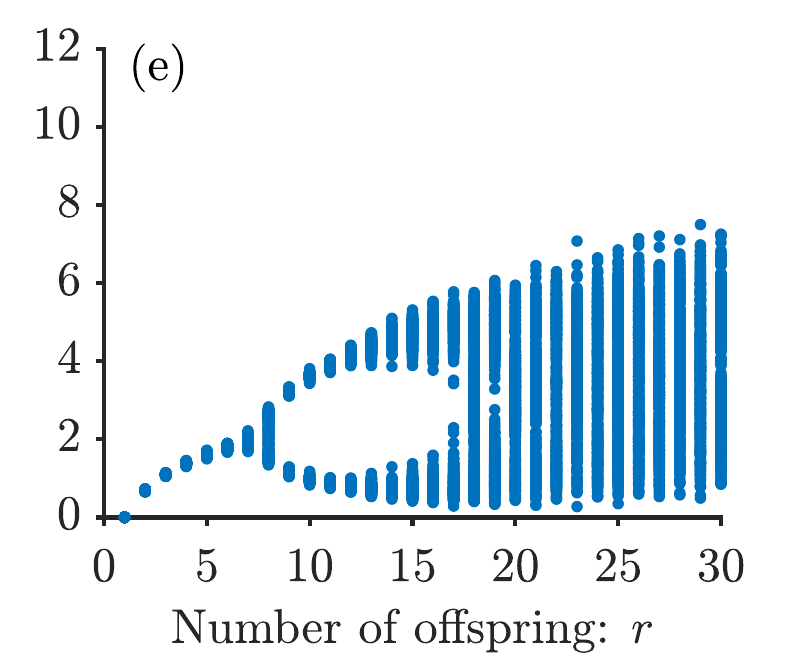}}
\subcaptionbox*{}{\includegraphics[scale=0.48]{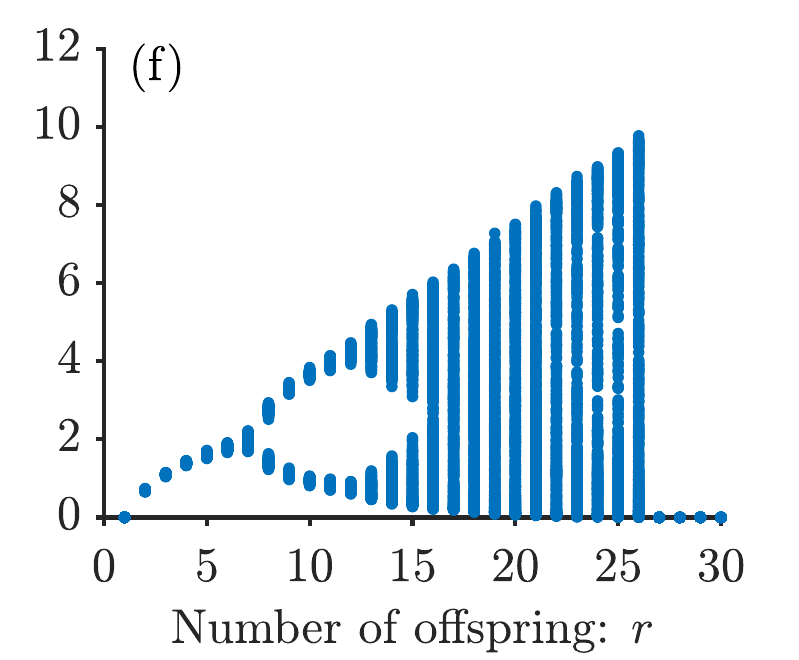}}
\caption{Bifurcation diagram for the individual-based model, with $\phi(k)$ given as equation \eqref{eq:scramble_phi} in the main manuscript  and $R_{ij}^t \sim  \operatorname{Bin}(r/q,q)$,  where $q=0.2$. In each subplot we vary $r$ and simulate the model for 5000 time steps and plot last 500 time steps. The population density is given by the number of individuals divided by the number of resource sites, and is counted after reproduction. We give results for (a) $s=1$ (very local dispersal) (b) $s=2$, (c) $s=3$, (d) $s=5$, (e) $s=10$ and (f) $s=50$  (global dispersal).  \label{fig:bifmodel_p02}}
\end{figure}

\begin{figure}[H] \centering
\subcaptionbox*{}{\includegraphics[scale=0.48]{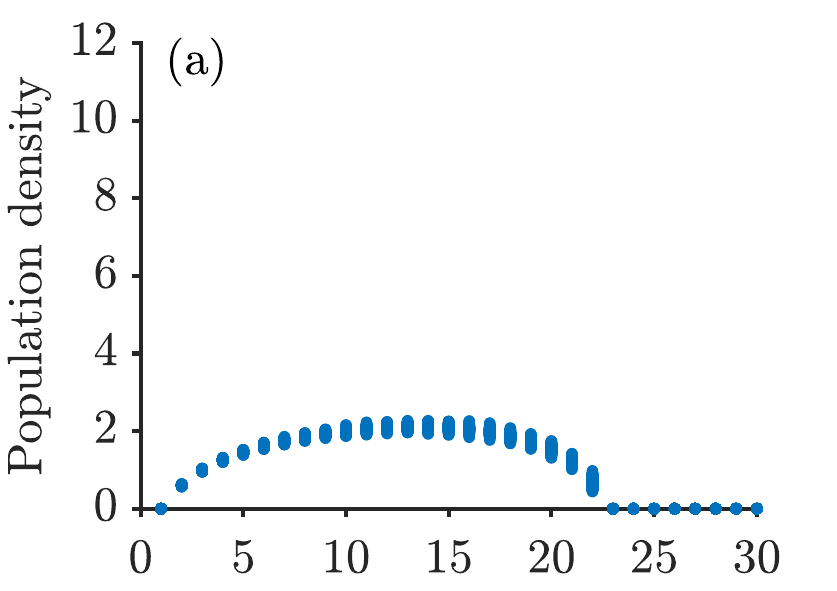}}
\subcaptionbox*{}{\includegraphics[scale=0.48]{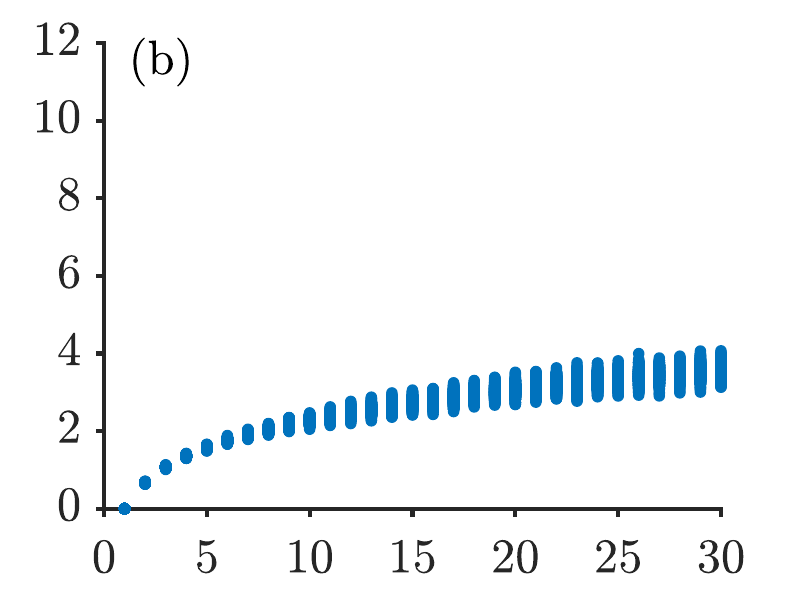}}
\subcaptionbox*{}{\includegraphics[scale=0.48]{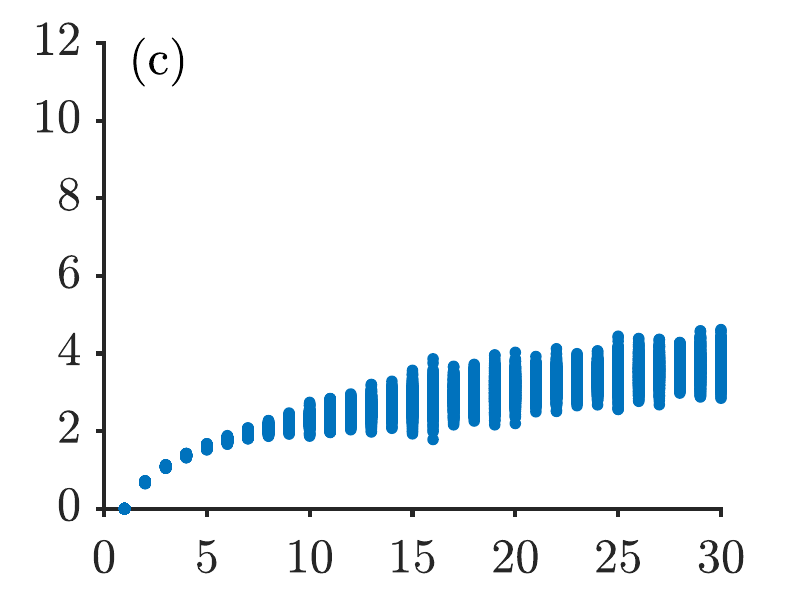}}
\subcaptionbox*{}{\includegraphics[scale=0.48]{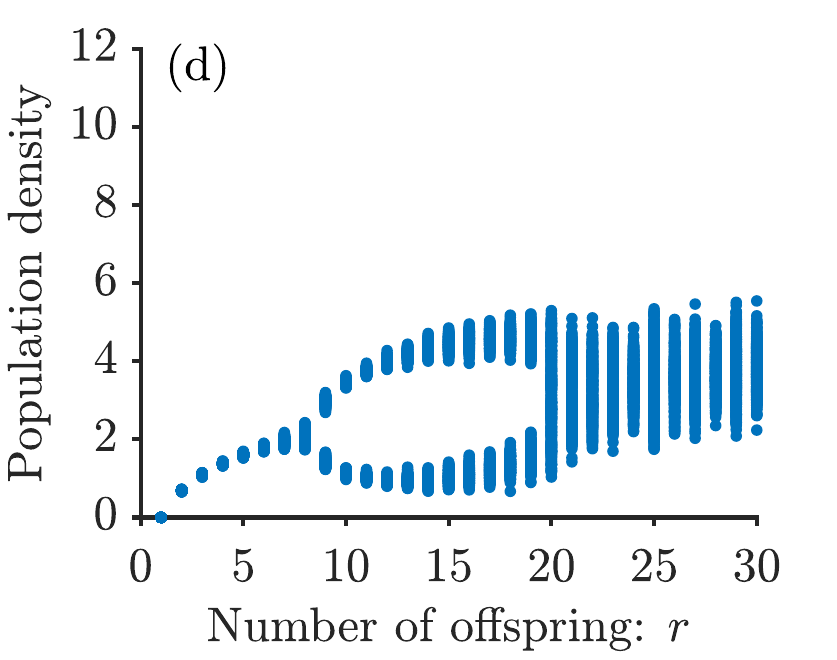}}
\subcaptionbox*{}{\includegraphics[scale=0.48]{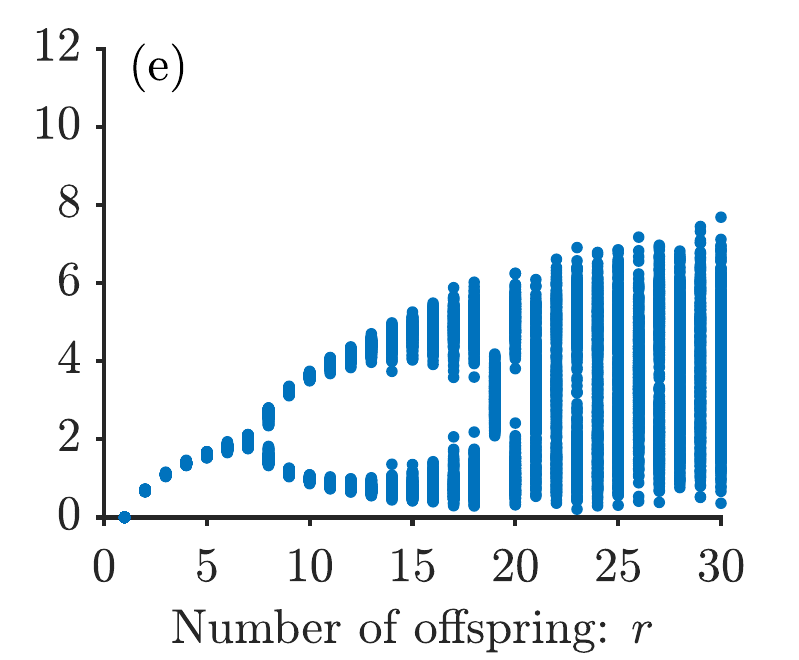}}
\subcaptionbox*{}{\includegraphics[scale=0.48]{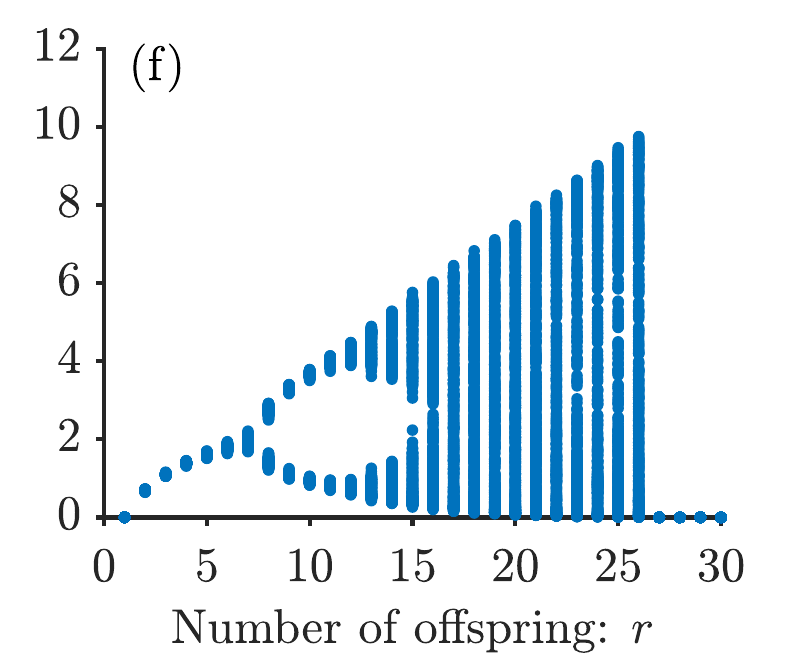}}
\caption{Bifurcation diagram for the individual-based model, with $\phi(k)$ given as equation \eqref{eq:scramble_phi} in the main manuscript  and $R_{ij}^t \sim  \operatorname{Bin}(r/q,q)$,  where $q=0.5$. In each subplot we vary $r$ and simulate the model for 5000 time steps and plot last 500 time steps. The population density is given by the number of individuals divided by the number of resource sites, and is counted after reproduction. We give results for (a) $s=1$ (very local dispersal) (b) $s=2$, (c) $s=3$, (d) $s=5$, (e) $s=10$ and (f) $s=50$  (global dispersal).  \label{fig:bifmodel_p05}}
\end{figure}

We see that letting the reproductive rate be a random variable does not change the qualitative behaviour of the model.  

\subsection*{S4: General derivation of the local approximation}
In our main text we derived the distribution of the number of offspring ending up at our focal site, $Y$, using the law of total probability.  Another approach to derive the distribution of $Y$ is to use probability generating functions.  This approach is especially useful when the reproductive rate, $r$, is not constant, but given by i.i.d random variables  $R_{ij}^t$. 
Let $R_{ij}^t$ be i.i.d with $R_{ij}^t \sim  \operatorname{Bin}(r/q,q)$, assuming $r/q$ is an integer.  In the case of constant reproductive rate, each parent produce $r$ offspring, meaning that there are $r \cdot F$ offspring in the Moore neighbourhood of our focal site. Now, the number of offspring of all of the parents in the neighbourhood of the focal site will instead be given by the sum
\begin{equation}
S_F=\sum_{m=1}^F R_m
\end{equation}
where $R_i\sim \operatorname{Bin}(r/q,q)$.
Thus, 
\begin{equation}
Y = \sum_{j=1}^{S_F} O_j =\sum_{j=1}^{\sum_{m=1}^F R_m} O_j 
\end{equation}

Let $g_Y(z)$ be the probability generating function of $Y$, and $g_O(z)$, $g_R(z)$ and $g_F(z)$ be the probability generating functions of $O$, $R$ and $F$ respectively.  $g_Y(z)$ will then be given by

\begin{equation}
g_Y(z)=g_F\left(g_R\left(g_O\left(z\right)\right)\right),
\end{equation}
Recall that $O_j \sim \operatorname{Ber} (\frac{1}{(2s+1)^2} )$, $F \sim \operatorname{Bin} ((2s+1)^2, x_t)$ and $R \sim  \operatorname{Bin}(r/q,q)$.  The corresponding probability generating functions will thus be given by
\begin{equation}
g_O(z)=1-\frac{1}{(2s+1)^2}+\frac{1}{(2s+1)^2}z,
\end{equation}
\begin{equation}
g_R(z)=\left(1-q+qz\right)^{r/q},
\end{equation}
and
\begin{equation}
g_F(z)=\left(1-x_t+x_tz\right)^{(2s+1)^2}.
\end{equation}
Thus $g_Y(z)$ can be written as
\begin{equation}
g_Y(z)=\left(1-x_t+x_t\left(1-q+q\left(1-\frac{1}{(2s+1)^2}+\frac{1}{(2s+1)^2}z\right)\right)^{r/q}\right)^{(2s+1)^2}.
\end{equation}

From the probability generating function we can recover $P(Y=k)$ from 
\begin{equation}
P(Y=k)=\left(\frac{1}{k!}\right) g^{(k)}_Y(0),
\end{equation}
where $g^{(k)}_Y(0)$ is the $k$-th derivative of $g_Y$ evaluated at 0. 
Thus, $p_1$ will be given by 
\begin{dmath}
p_1= g'_Y(0)=rx_t\left(\left(1-\frac{1}{(2s+1)^2}\right)q-q+1\right)^{r/q-1}\\
\left(x_t  \left(\left(1-\frac{1}{(2s+1)^2}\right)q-q+1\right)^{r/q}-x_t+1\right)^{(2s+1)^2-1},
\end{dmath}
which means that, for scramble competition, the population dynamics will be given by
\begin{dmath} \label{eq:popdyn_rand}
x_{t+1}=rx_t\left(\left(1-\frac{1}{(2s+1)^2}\right)q-q+1\right)^{r/q-1}\\
\left(x_t \left(\left(1-\frac{1}{(2s+1)^2}\right)q-q+1\right)^{r/q}-x_t+1\right)^{(2s+1)^2-1} = f(x_t).
\end{dmath}

Having $r$ fixed is the same as setting $q=1$, so that $R_i=r$ with probability 1. By setting $q=1$ in equation \eqref{eq:popdyn_rand} we obtain 
 \begin{eqnarray}
x_{t+1}  & =rx_t\left(1-\frac{1}{(2s+1)^2}\right)^{r-1}\left(x_t \left(1-\frac{1}{(2s+1)^2}\right)^{r}-x_t+1\right)^{(2s+1)^2-1},
\end{eqnarray}
which is our local correlation approximation in Section 3.2.

To get a better understanding of the dynamics of equation \eqref{eq:popdyn_rand} we produced bifurcation plots for the local correlation approximation for stochastic reproductive rate, iterating equation \eqref{mainapprox} for different values of $r$,  for three different values of $q$: 0.1,0.2 and 0.5.   In these plots we see that the overall population dynamics is not affected by having a stochastic reproductive rate.  However,  in Figures \ref{fig:bifmodel_p01}-\ref{fig:bifmodel_p05}(a), we see that the population is not extinct when $r=30$ and $s=1$, in contrast to the approximation for deterministic $r$.  By differentiating $h(x_t)$ (equation \ref{eq:popdyn_rand}) we see that the stability of the steady state $x_*=0$ is affected by $q$: 

\begin{equation}
f'(0)=r\left(1-\frac{q}{(2s+1)^2}\right)^{r/q-1},
\end{equation}

which will be larger than 1 when $q \lessapprox0.839$, for $s=1$, $r=30$. 

\begin{figure}[H] \centering
\subcaptionbox*{}{\includegraphics[scale=0.48]{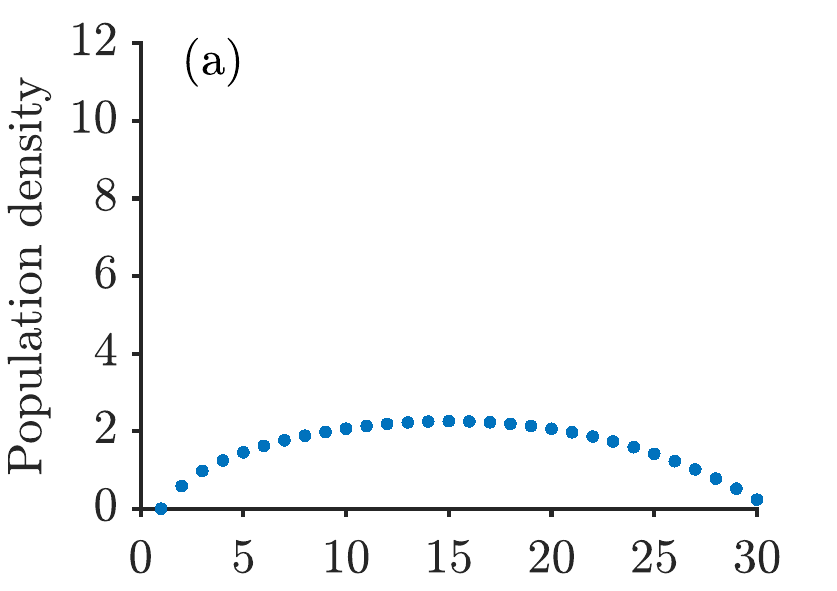}}
\subcaptionbox*{}{\includegraphics[scale=0.48]{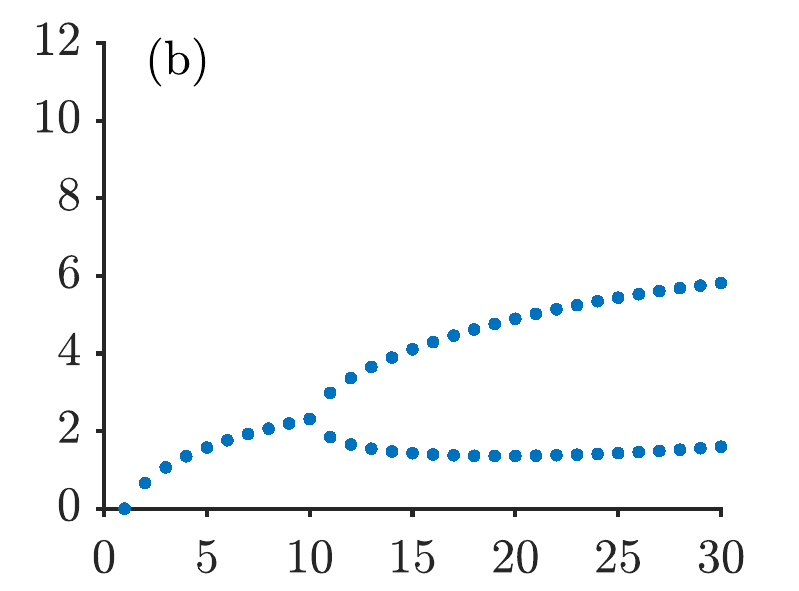}}
\subcaptionbox*{}{\includegraphics[scale=0.48]{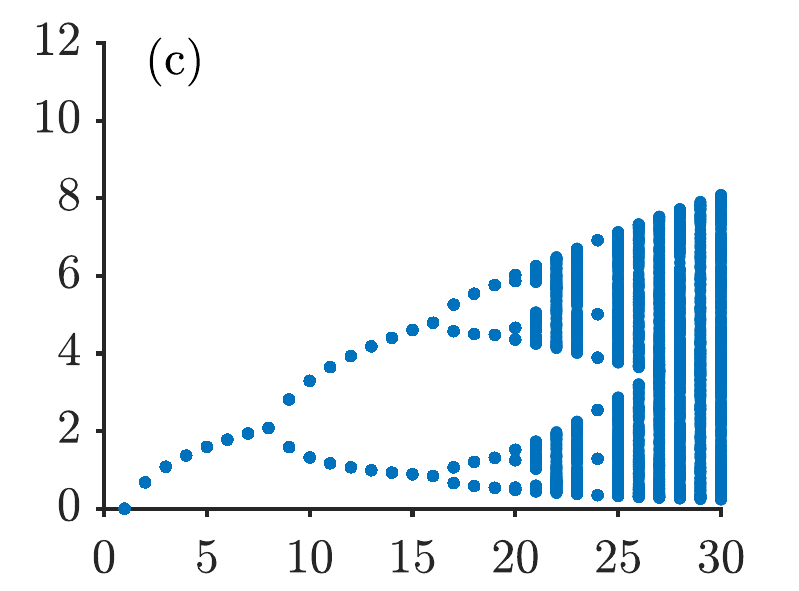}}
\subcaptionbox*{}{\includegraphics[scale=0.48]{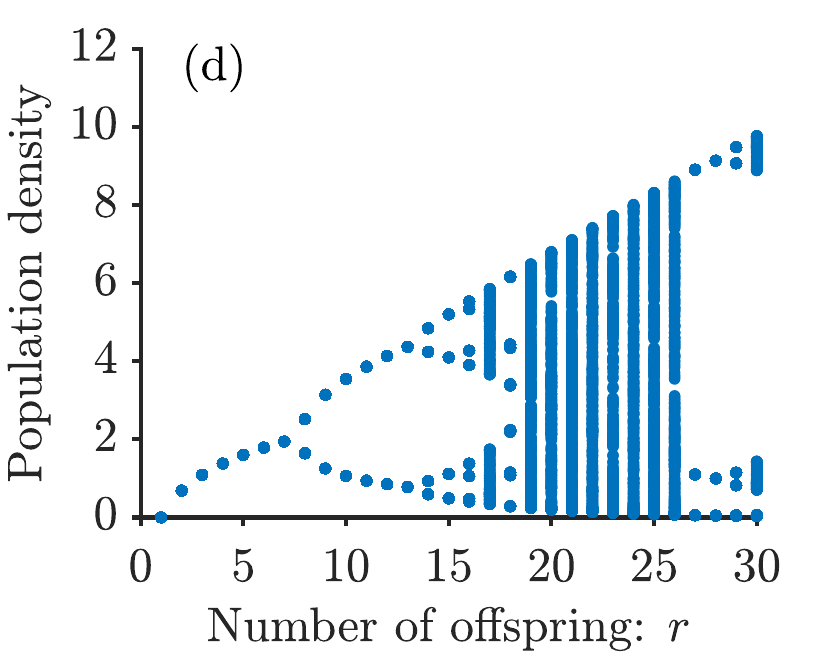}}
\subcaptionbox*{}{\includegraphics[scale=0.48]{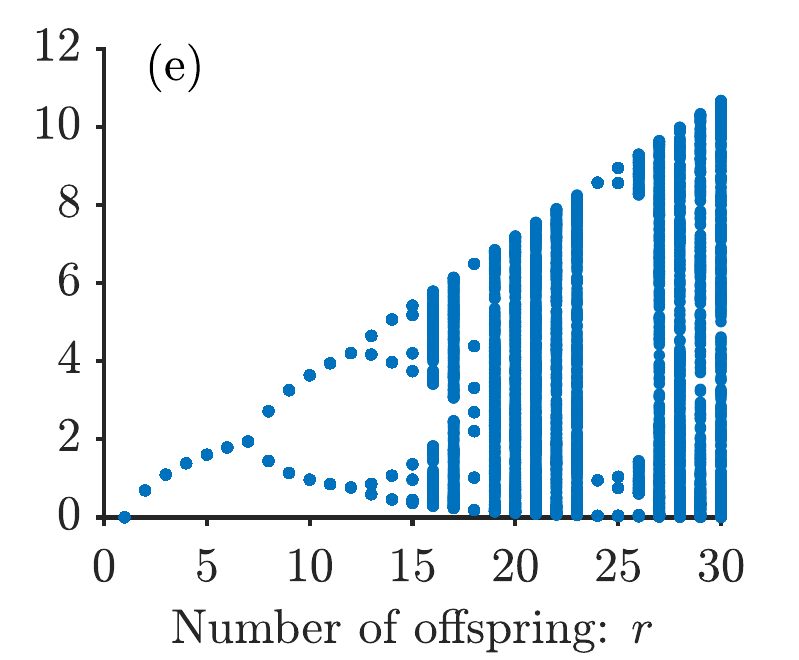}}
\subcaptionbox*{}{\includegraphics[scale=0.48]{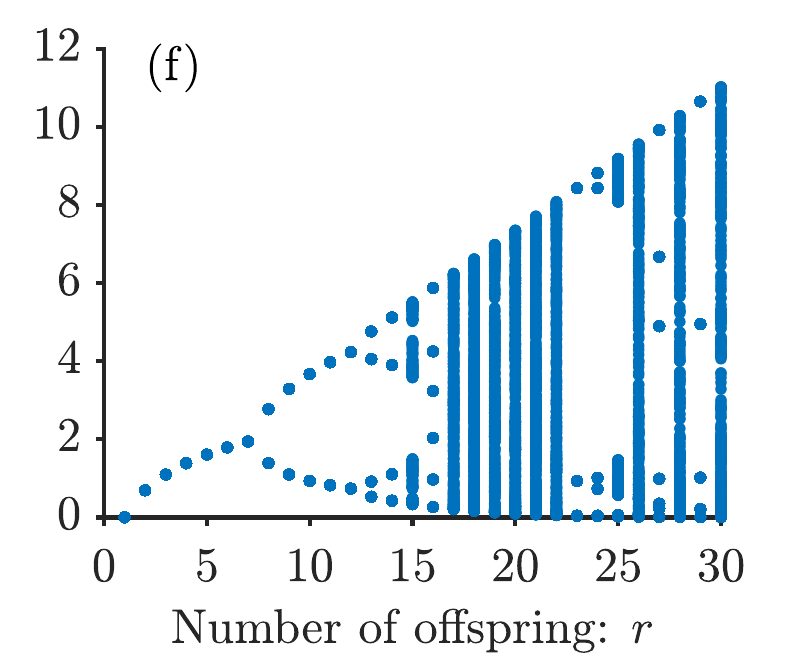}}
\caption{Bifurcation diagram for the local correlation approximation for stochastic reproductive rate,  given given by equation \eqref{eq:popdyn_rand}, when $q=0.1$. In each subplot we vary $r$ and iterate through $x_t$ for 5000 time steps and plot last 500 time steps. We give results for (a) $s=1$ (very local dispersal) (b) $s=2$, (c) $s=3$, (d) $s=5$, (e) $s=10$ and (f) $s=50$ (close to global dispersal).  \label{fig:bifapprox_p01}}
\end{figure}

\begin{figure}[H] \centering
\subcaptionbox*{}{\includegraphics[scale=0.48]{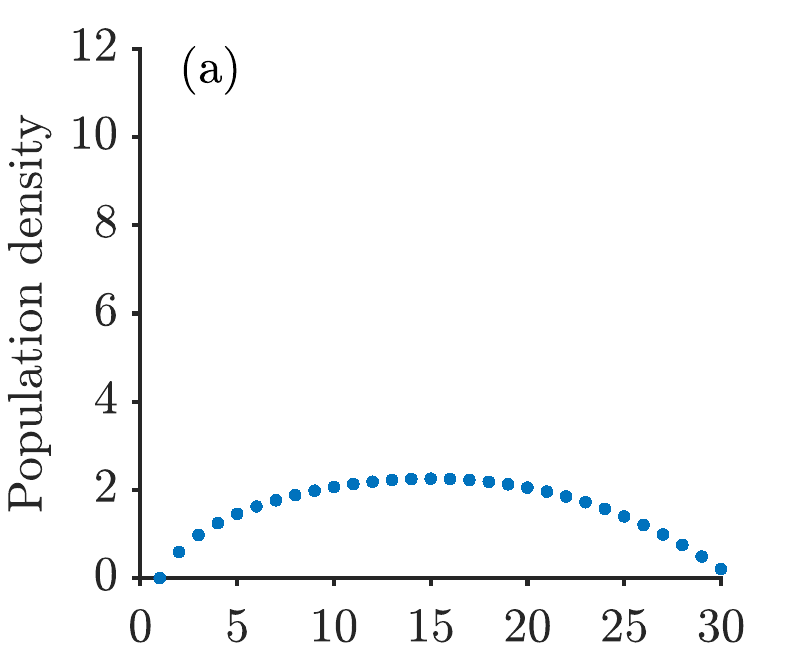}}
\subcaptionbox*{}{\includegraphics[scale=0.48]{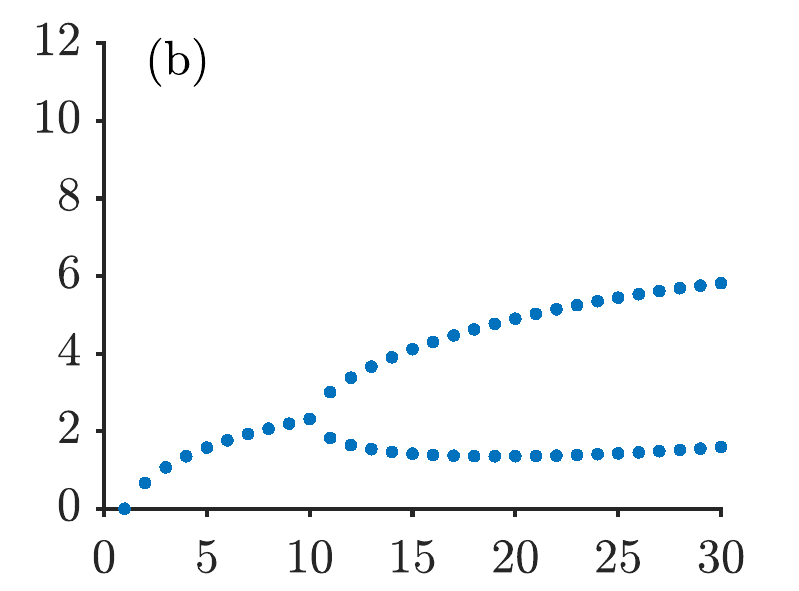}}
\subcaptionbox*{}{\includegraphics[scale=0.48]{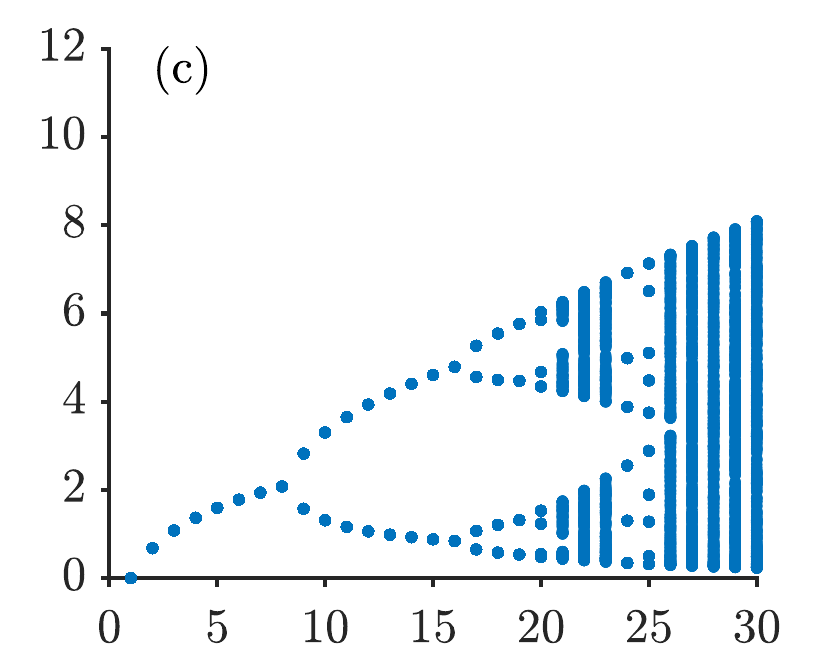}}
\subcaptionbox*{}{\includegraphics[scale=0.48]{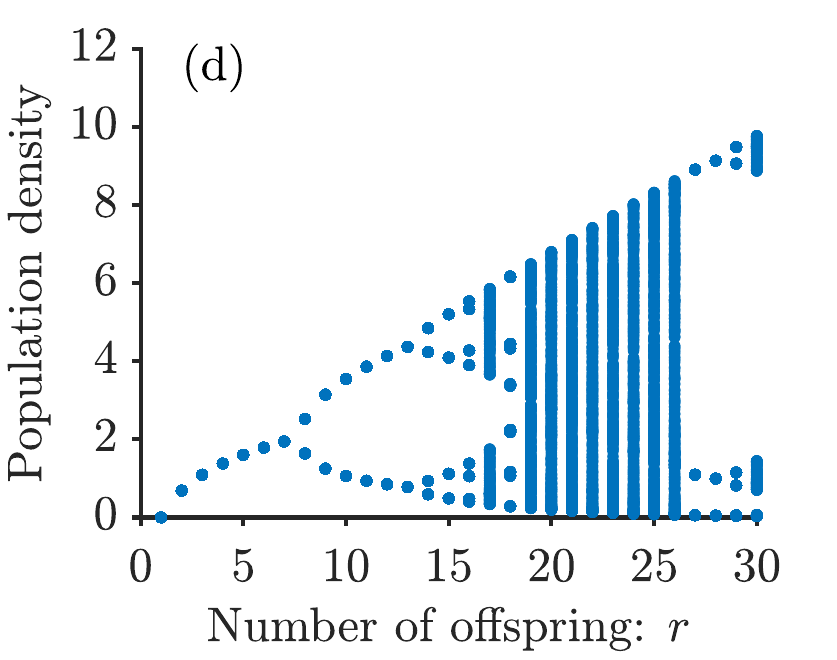}}
\subcaptionbox*{}{\includegraphics[scale=0.48]{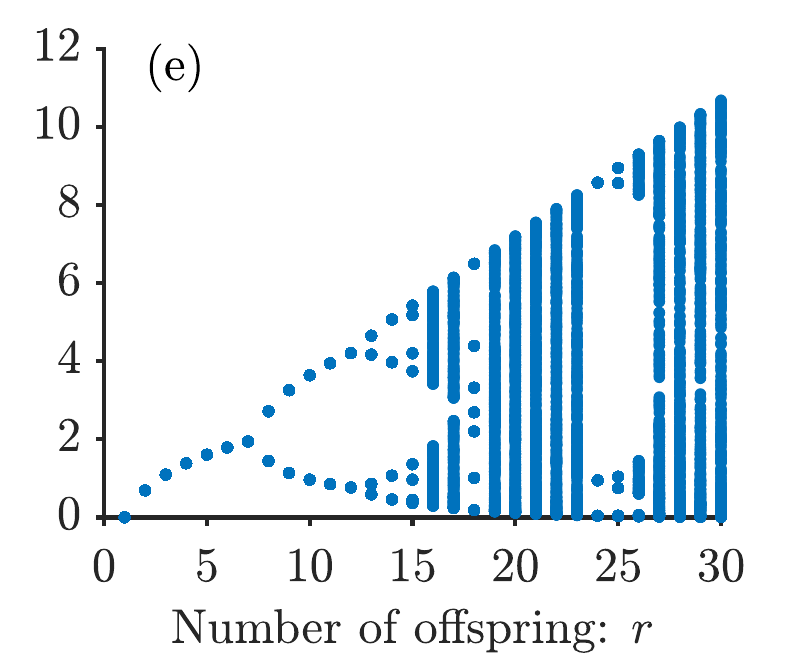}}
\subcaptionbox*{}{\includegraphics[scale=0.48]{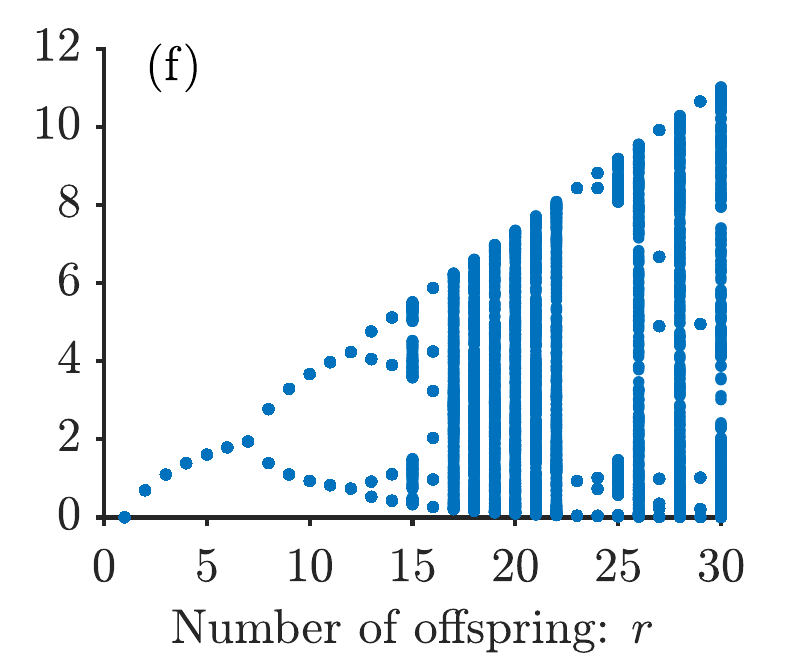}}
\caption{Bifurcation diagram for the local correlation approximation for stochastic reproductive rate,  given given by equation \eqref{eq:popdyn_rand},  when $q=0.2$. In each subplot we vary $r$ and iterate through $x_t$ for 5000 time steps and plot last 500 time steps. We give results for (a) $s=1$ (very local dispersal) (b) $s=2$, (c) $s=3$, (d) $s=5$, (e) $s=10$ and (f) $s=50$ (close to global dispersal).  \label{fig:bifapprox_p02}}
\end{figure}

\begin{figure}[H] \centering
\subcaptionbox*{}{\includegraphics[scale=0.48]{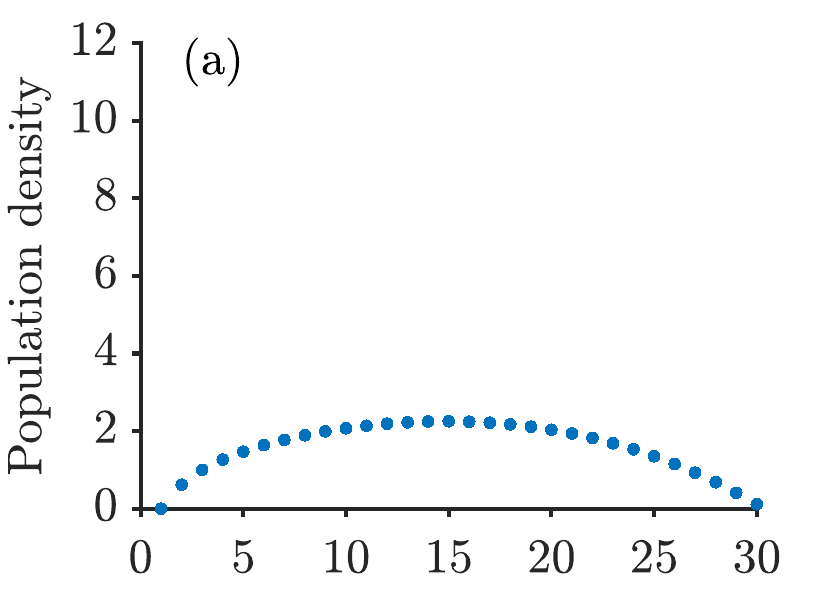}}
\subcaptionbox*{}{\includegraphics[scale=0.48]{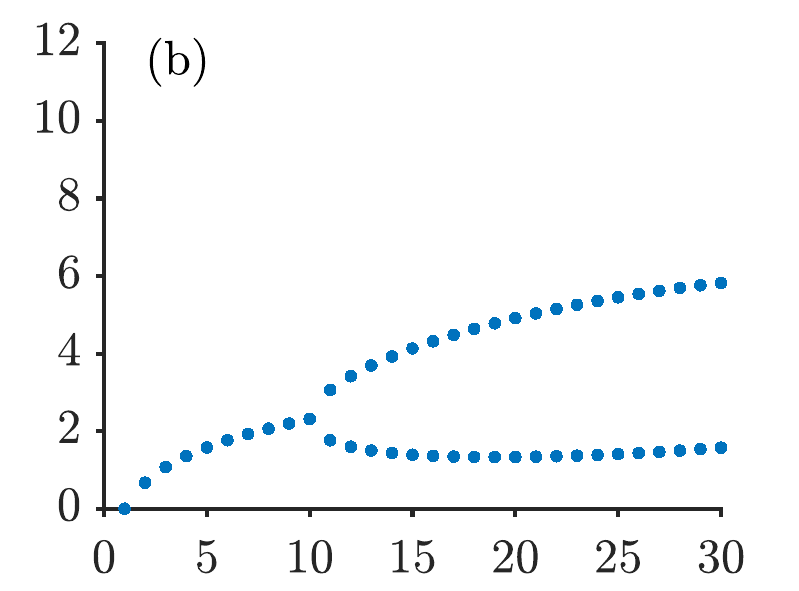}}
\subcaptionbox*{}{\includegraphics[scale=0.48]{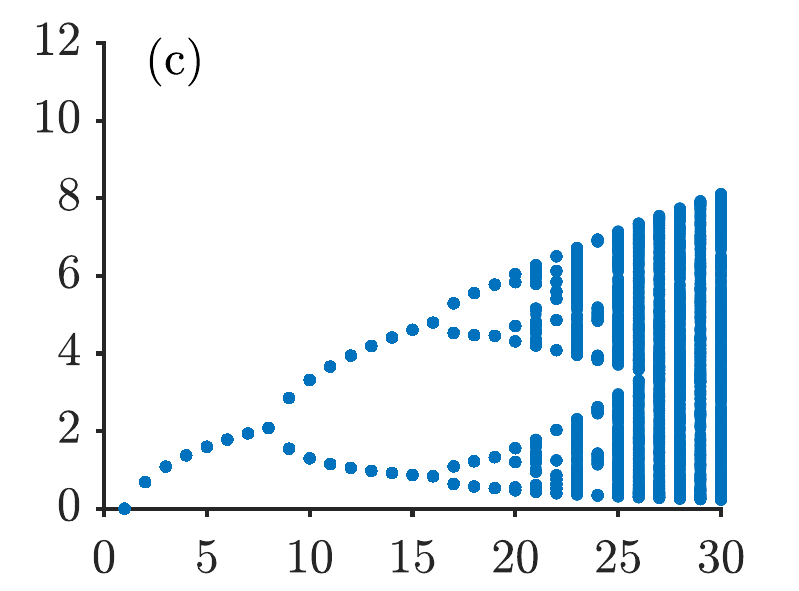}}
\subcaptionbox*{}{\includegraphics[scale=0.48]{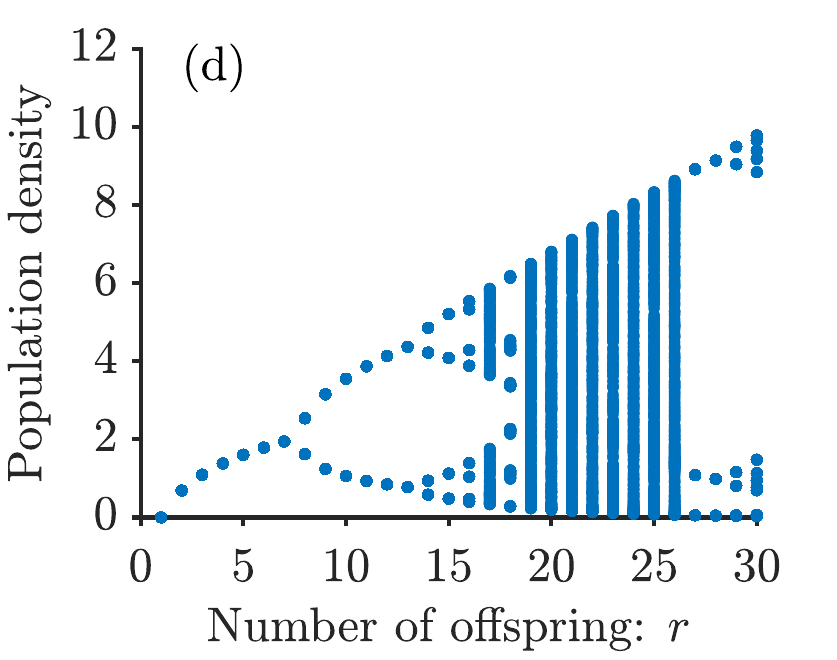}}
\subcaptionbox*{}{\includegraphics[scale=0.48]{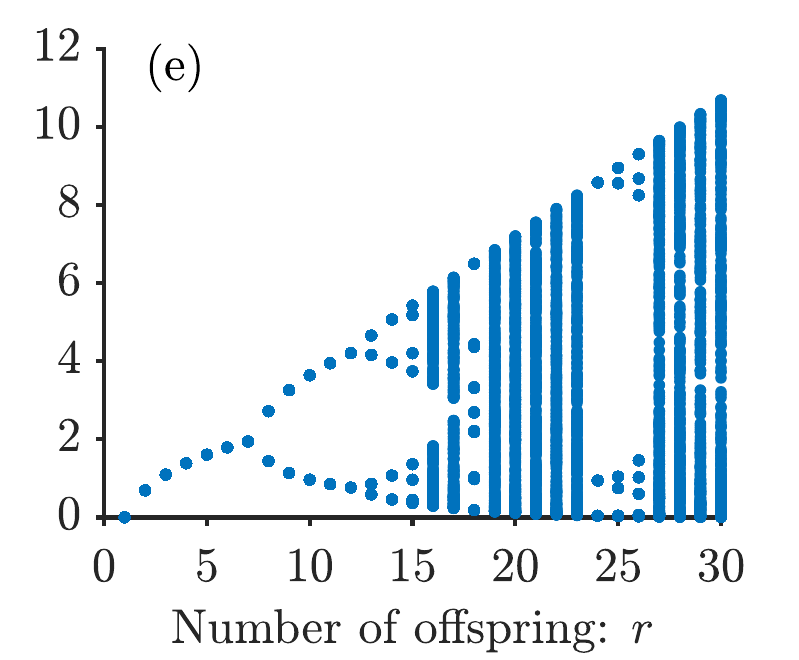}}
\subcaptionbox*{}{\includegraphics[scale=0.48]{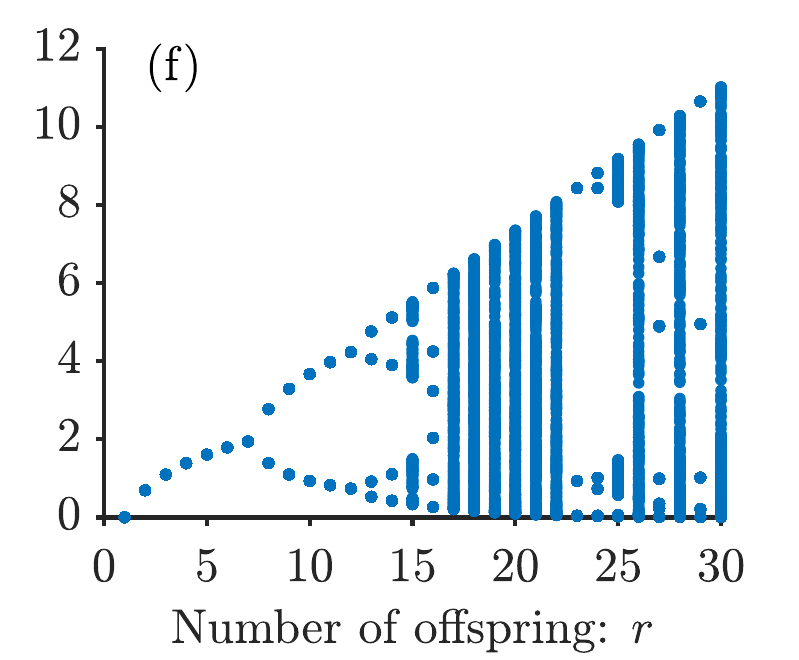}}
\caption{Bifurcation diagram for the local correlation approximation for stochastic reproductive rate,  given given by equation \eqref{eq:popdyn_rand},  when $q=0.5$. In each subplot we vary $r$ and iterate through $x_t$ for 5000 time steps and plot last 500 time steps. We give results for (a) $s=1$ (very local dispersal) (b) $s=2$, (c) $s=3$, (d) $s=5$, (e) $s=10$ and (f) $s=50$ (close to global dispersal). \label{fig:bifapprox_p05}}
\end{figure}

\end{document}